\documentstyle[epsfig,aaspp4]{article}

\newcommand{\Msun}{\mbox{$M_{\sun}$}}

\newcommand{\Mdot}{\mbox{\.{M}}}
\newcommand{\eflux}{\mbox{ergs s$^{-1}$ cm$^{-2}$}}
\newcommand{\pflux}{\mbox{photons s$^{-1}$ cm$^{-2}$}}
\newcommand{\ps}{\mbox{s$^{-1}$}}
\newcommand{\pcmsq}{\mbox{cm$^{-2}$}}

\newcommand{\x}{\mbox{$\times$}}
\newcommand{\dg}{\mbox{$^\circ$}}

\newcommand{\pnn}{\phn\phn}
\let\tnm=\tablenotemark

\begin{document}

\title{THE PROPERTIES OF X-RAY AND OPTICAL LIGHT CURVES OF X-RAY NOVAE}

\author{Wan Chen$^{1,2}$, C. R. Shrader$^{1,3}$, and Mario Livio$^{4}$}

\affil{$^{1}$ Laboratory for High Energy Astrophysics, NASA/Goddard Space
Flight Center, Greenbelt, MD 20771 \\
 $^{2}$ Department of Astronomy, University of Maryland, College Park, MD 
20742 \\
 $^{3}$ Universities Space Research Association \\
 $^{4}$ Space Telescope Science Institute, 3700 San Martin Drive,
Baltimore, MD 21218 \\
chen@milkyway.gsfc.nasa.gov, shrader@grossc.gsfc.nasa.gov, mlivio@stsci.edu
 \\
\hspace{2in} \\
(Accepted on 10 July 1997 for publication in the Astrophysical Journal,
Part 1)} 

\begin{abstract}

We have collected the available data from the literature and from public
data archives covering the past two decades for the long-term X-ray and
optical light curves of X-ray nova (XN) outbursts. XN outbursts are due to
episodic accretion events, primarily in low-mass X-ray binaries normally
characterized by low mass transfer rates. Dynamical studies indicate that
most XNs contain a black hole. The soft X-ray emission during outburst
traces the accretion rate through the inner edge of the accretion disk,
while the optical light curve traces the physical conditions at the outer
disk -- thus collectively they contain information on the time-dependent
behavior of accretion processes through the disks.

In this paper we carry out for the first time a systematic, statistical
study of XN light curves which are classified into 5 morphological types.
Basic light curve parameters, such as the outburst peak flux, amplitude,
luminosity, rise and decay timescales, the observed and expected outburst
durations, and total energy radiated, are tabulated and discussed. We find
that the rise timescales have a flat distribution while the decay
timescales have a much narrower and near-Gaussian distribution, centered
around 30 days and dominated by the strongest outbursts. The peak
luminosity is also distributed like a Gaussian, centered around 0.2 in
Eddington units, while the total energy released has a much broader
distribution around $10^{44}$ ergs. We find no intrinsic difference between
black hole and neutron star systems in their distribution of peak
amplitudes.

We identify and discuss additional light curve features, such as
precursors, plateaus, and secondary maxima. The plateaus exhibited in the
light curves of black hole sources are found to have, on average, longer
durations and they are followed by longer decays. The identified secondary
maxima seem to occur mostly in black hole systems. For the frequency of
outbursts, we find that the average XN outburst rate is about 2.6 per year
for events $>0.3$ Crab, and that the mean recurrence time between outbursts
from a single source is 6 years. The spatial and $\log(N) -\log(S)$
distribution of the XN sources, with limited statistics, agrees with a
source population in the Galactic disk, as observed from a point at a
distance of 8.5 kpc from the Galactic center. Finally, we point out that
the observed XN light curve properties can in general be explained by a
disk thermal instability model, although some important problems still
remain.

\end{abstract}

\section{INTRODUCTION}

On 4 April 1967 an X-ray rocket experiment detected a new X-ray source
almost as bright as Sco X-1 in the 2--5 keV range (\cite{hmff67}). The new
source, Cen XR-2, was not detected on 28 October 1965 (\cite{ghst66}), thus
indicating a flux increase by a factor of $>100$ during 1967. This was
unprecedented among about a dozen known X-ray sources at the time. A series
of follow-up rocket flights found that the X-ray flux of Cen X-2 peaked on
April 10, then declined {\it exponentially} with an e-folding time of
$\sim30.5$ days, disappearing by late September. This behavior led its
discoverers to name Cen X-2 an {\em X-ray nova}, i.e., a transient X-ray
source with a light curve analogous to that of a classical nova in the
optical (\cite{cmrs68}).

In the 1970's and early 1980's, many bright X-ray transient events were
discovered by scanning instruments or all-sky monitors (ASMs) aboard the
{\it UHURU, Vela, Ariel 5, OSO 7}, and {\it Hakucho} satellites, and also
by the pointing instruments on the {\it Ariel 5, SAS-3, HEAO-1}, and {\it
EXOSAT\/} satellites. These transients were classified according to their
distinctive X-ray spectral properties near the peak of the outbursts (e.g.,
\cite{wks84} and references therein). Sources with a characteristic
bremsstrahlung temperature $kT_{\rm b}<15$ keV were called soft X-ray
transients (SXTs) while sources with $kT_{\rm b}>15$ keV were called hard
X-ray transients (HXTs). 

The HXTs were later found to be Be binary pulsars in highly eccentric
orbits with periods of a few weeks to a few months (e.g., \cite {mtvdh76}).
An X-ray outburst may occur in these systems when the neutron star (NS)
comes close to the Be star and plunges into its highly flattened dense
wind. Because Be star winds are episodic, not every periastron passage
involves an outburst. The HXTs usually display a light curve roughly
symmetric in the rise and decay phases. 

The SXTs, on the other hand, occur in low-mass X-ray binaries (LMXBs), in
which a Roche-lobe overflowing, main-sequence or subgiant star of typically
$<1$ \Msun\ orbits a NS or a black hole (BH), with an orbital period of a
few hours to a few days. SXTs undergo episodic X-ray and optical outbursts
with light curves similar to that of the first X-ray nova Cen X-2, and they
are the subject of this paper. Several SXTs exhibit Type-I X-ray bursts,
but no pulsations have ever been detected. About $\sim50$\% of the SXTs
were called ``ultra-soft" (\cite{wks84}), due to the fact that in their
peak, a soft excess of $kT_{\rm brem} \lesssim$ a few keV has been observed
on top of a power-law hard tail. The spectra usually harden as the flux
decreases. This property is reminiscent of the bi-modal spectral behavior
of BH candidates (BHCs) such as Cyg X-1, and thus suggests a possible high
incidence of BH primaries in SXTs. 

The terms {\em X-ray novae} and {\em soft X-ray transients} have been used
interchangeably in the literature. Several recent events (e.g., GRS
1716--249, GS 2023+338, and GRO J0422+32), however, showed pure power-law
spectra without the soft excess, while having low mass companions and
displaying light curves similar to other SXTs. They are not SXTs in the
traditional sense but belong to the SXT family. For clarity, we use the
term X-ray nova (XN) throughout this paper.\footnote {One should not
confuse XNs with classical novae in which the accreting object is a white
dwarf. For example, the name ``Nova Oph 1993'' has been attributed to both
X-ray Nova GRS 1716--249 and a classical nova which occurred in the same
year. We suggest that such X-ray transients should always be referred to as
an {\it X-ray} nova, e.g., ``X-ray Nova Ophiuchi 1993'' or ``XN Oph
1993''.} 

XNs offer a significant advantage over persistent LMXBs in terms of the 
potential to perform dynamical
studies of the binary system. After the outburst, an XN eventually returns
to a quiescent state with its optical light dominated by the secondary
star, thus allowing for detailed photometry and spectroscopy to determine
the system parameters of the binary, notably the mass function. XN
0620--00 (XN Mon 1975) was the first low-mass binary proven in this manner
to contain a compact object of mass definitively greater than 3 \Msun\
(\cite{mr86}), the widely accepted mass upper limit for a stable NS. Thus,
XN 0620--00 may contain a BH (see also \cite{hrhsa93}; \cite{snc94}).

Since 1988, the ASMs on the {\it Ginga} and {\it GRANAT} satellites and on
the {\it Mir-Kvant} space station, and the BATSE instrument on the {\it
Compton Gamma Ray Observatory} have been detecting bright XNs at a rate of
1--2 per year. Many of them have subsequently been studied in great detail
both in X-rays and in the UV, optical, and radio wavelengths. Remarkable
progress has been achieved in identifying 7 sources as BHCs on the basis of
orbital dynamical studies (Table~\ref{tb:BH}; see also \cite{wvp96}). 
Five of these are
sources discovered since the {\it Ginga} mission was flown. Currently, the
majority of the XNs are classified as BHCs, at least on the basis of their
X-ray spectral properties (\cite{wne94}; \cite{cap92}; \cite{tl95}). XNs
are thus ideal sources for the study of accretion disks around BHs over a
large range of accretion rates, and for studying the formation and
evolution of high mass-ratio BH binary systems. 

\begin{table}[tbp]
\dummytable\label{tb:BH}
\end{table}

A review of historical XNs was presented in van Paradijs \& Verbunt (1984),
and their spectral properties were summarized by White, Kaluzienski, \&
Swank (1984). A catalog of 34 X-ray transients, including 17 XNs, was
compiled by Amnuel \& Guseinov (1979). Owing to the improved all-sky
coverage since the late 1980's, the number of cataloged XNs has by now more
than doubled. In the recent catalog of X-ray binaries (\cite{vpj95},
hereafter vP95), 41 transients were included among 124 LMXBs. A few
additional sources have recently appeared (see Tanaka \& Shibazaki (1996)
and Tanaka \& Lewin (1995) for recent reviews). Some of the fundamental
outburst properties reported previously are by now well established, and more
detailed features of the XN light curves and spectra have emerged. This
trend is likely to continue, in light of the recently launched {\it Rossi
X-ray Timing Explorer (RXTE)}. In the meantime, substantial progress has
been made in our understanding of the physical mechanisms involved (e.g.,
\cite{ccl95} and references therein; \cite{nbm97}; \cite{nr97} and
references therein). This seems, therefore, to be an appropriate time for a
systematic study of all the known XNs and the properties of their light
curves using all the data available, as we attempt in this paper. 

The paper is organized as follows. In \S~\ref{sc:data} we discuss the
source selection criteria, the methods and scope of our data collection,
and the procedures we followed to present the data in a uniform manner. In
\S~\ref{sc:m} we discuss general light curve morphologies and
identify different types of light curve shapes and features. Before going
into details of the light curves, in \S~\ref{sc:q} we discuss the
properties of quiescent emission of the XN sources. In \S~\ref{sc:main} we
present our analysis of the light curve properties, including the basic
parameters which can be defined for almost all the X-ray and optical light
curves (\S~\ref{ssc:b}), and several features which are seen only in a
subset of the light curves, but which are nevertheless of interest and
importance towards a complete understanding of the XN phenomenon
(\S~\ref{ssc:f}). In \S~\ref{sc:d} we discuss the frequency of XN
outbursts, recurrence times, the spatial distribution, and population of XN
sources in the Galaxy. In \S~\ref{sc:model} we touch briefly on the
applicability of several broad classes of theoretical models. A summary
and conclusions follow in \S~\ref{sc:s}. 

\section{SOURCE AND DATA COLLECTION}
\label{sc:data}

\subsection{Source Selection}
\label{ssc:s-s}

In the literature, the name transient is loosely applied to a wide range of
phenomena, including X-ray and $\gamma$-ray bursters which may or may not
show detectable fluxes outside the bursting periods, which are usually a
few seconds to a few minutes long (\cite{wse84}). To study X-ray transients
which have intrinsically similar {\em long-term} X-ray temporal behavior,
we define an XN to be a source which satisfies at least 4 of the following
5 conditions: \newline
\indent (1) The source had at least one X-ray outburst lasting more than 10
days; \newline
\indent (2) It is not identified with a Be HXT; \newline
\indent (3) Its X-ray light curve has the typical fast-rise and
exponential-decay profile; \newline
\indent (4) Its highest peak X-ray flux is at least a factor of 10 higher
than its quiescent flux; \newline
\indent (5) Its quiescent phase lasts at least 10 times longer than the
outburst phase. \newline
In Table~\ref{tb:source} we list all the sources we have identified so far
under these criteria. 

\begin{table}[tbp]
\dummytable\label{tb:source}
\end{table}

Our sample of 24 sources includes 21 of the 41 low-mass transients listed
in the vP95 catalog, for which the data collection ended in December 1992,
and three new XNs: GRS 1009--45, GRO J1655--40, and GRS 1716--249. Of the
20 low-mass X-ray transients listed as possible XNs
(Table~\ref{tb:possible}), Cir X-1 is probably a high-mass X-ray binary
with high and low states similar to Cyg X-1 (\cite{ma92}). MX 1730--335 is
the Rapid Burster, for which large flux variations are probably caused by
type I and type II bursts (\cite{gg77}). The remaining 18 sources, many of
which are located in the crowded Galactic center region, are of unclear
transient nature.

\begin{table}[tbp]
\dummytable\label{tb:possible}
\end{table}

It may be difficult to distinguish an XN from an extremely variable but
persistent source. Certain sources may straddle a stability line which
depends on the amount of X-ray irradiation of the disk -- above it the
system will be stabilized (\cite{vpj96}). For example, 4U 1608--522 is
classified as an XN in the literature, but it has exhibited fairly high
flux levels between many moderate-amplitude outbursts (\cite{lrd94}), and
it is sometimes called a persistent X-ray source. On the other hand, the
BHC GX 339--4 is listed as a transient in vP95, but is often regarded as a
persistent, highly variable source possessing three distinctive (high, low,
and off) intensity states (e.g., \cite{mthea73}; \cite{tl95}). We included
4U 1608--522 in our sample and GX 339--4 in our list of possible sources,
but we note that they may belong to the same subclass located in the
transition region between XN and persistent sources. 

Table~\ref{tb:general} lists some general information for known XNs
(hereafter we will refer to a source by its celestial coordinates name,
such as XN 0620--00 or simply XN~0620). Among the 24 XN sources, there are
7 dynamically confirmed and 10 X-ray spectroscopic BHCs (71\%) and 7 NS
(basically all those not considered to be BHCs); 19 (80\%) sources are
optically identified but only 9 (37\%) of them have the companion star and
orbital information available. For 19 (80\%) XN sources we have distance
estimates from either the X-ray absorption column or optical measurements. 

\begin{table}[tbp]
\dummytable\label{tb:general}
\end{table}

\subsection{Light Curve Database}
\label{ssc:db}

For the 24 XNs in our sample, there are 66 recorded outbursts. For 49 of
the outbursts, we have collected a total of 58 X-ray and 10 optical light
curves, including light curves of the same outburst in different energy
bands. Most of these light curves are taken from the literature by
digitizing the published light curve plots. Some light curves were kindly
provided by the original observers. We have searched the IAU circulars from
the late 1960's up to the present and the {\it Ariel 5, Vela 5B, Ginga},
and {\it UHURU} databases at the High Energy Astrophysics Science Archive
Research Center (HEASARC) for information on additional, weak outbursts and
data points that complement the published light curves. 

The XN light curves were reported in a variety of time units such as the
calendar date, the Julian date, days of the outburst year, or days after
the outburst. X-ray fluxes were also reported in a variety of units such as
instrument count rates, UHURU flux units (UFU), energy flux (\eflux),
photon flux (\pflux), or Crab units. The last case is particularly useful
in that it allows a comparison of source brightnesses despite the lack of
precise knowledge regarding spectral and instrument response properties.
Still, the quantitative comparisons we conduct in this paper will be
limited in their precision by the absence of reliable instrument cross
calibrations. 

To ease comparison, we report all the light curve measurements in a common
set of units. The time axis in both the X-ray and optical plots will be
given by the days after the {\em peak date} of the X-ray outburst. When
there is no data for the rising phase, we use the first data point as the
peak date. For clarity, the peak calendar date is marked and the truncated
Julian date (JD-2440000) is used as the upper abscissa. The intensity axis
is given in Crab units or magnitudes in the reported energy or photometric
band, while we also plot the energy flux in units of \eflux\ as the right
ordinate. 

\section{LIGHT CURVE MORPHOLOGIES}
\label{sc:m}

\begin{figure}
\psfig{figure=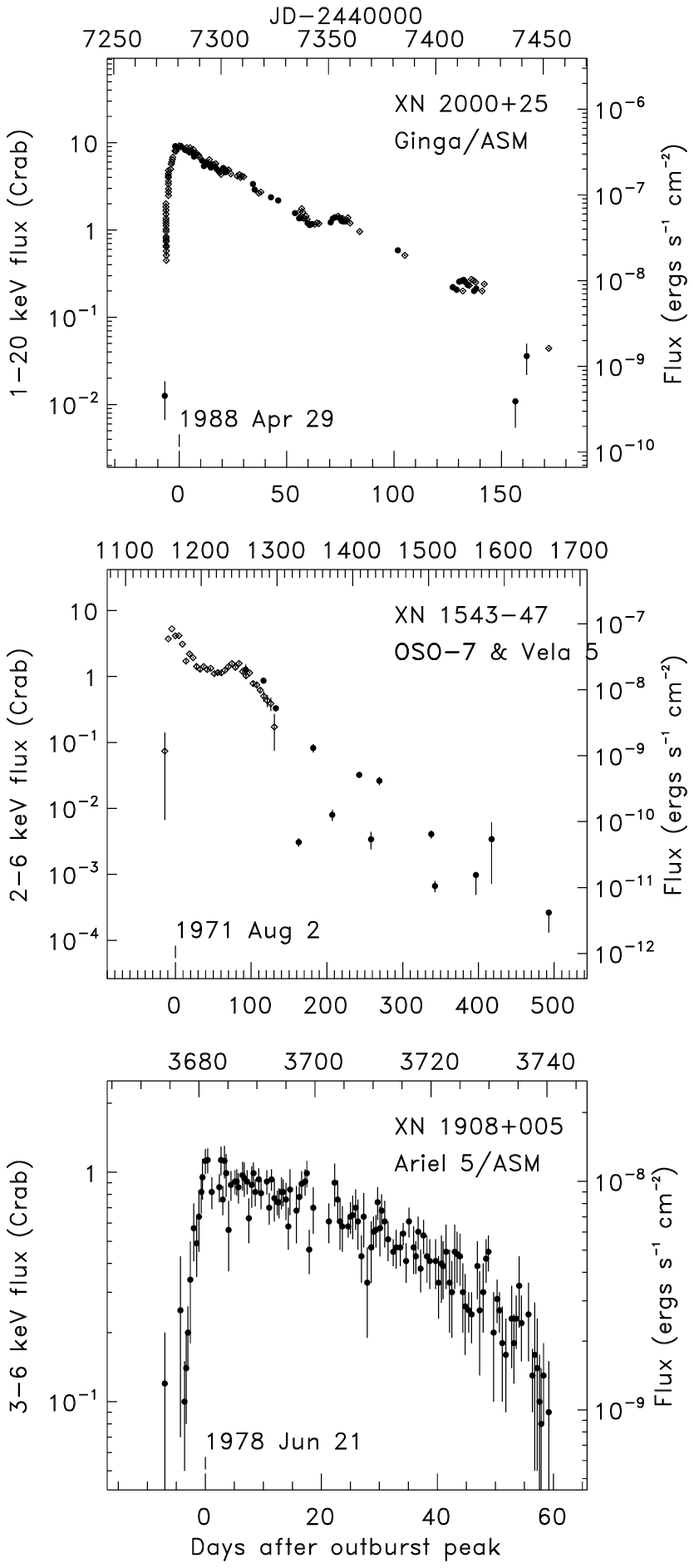}
\caption{ Examples of the FRED-type X-ray light
curves.  (a) The 1988 outburst of GS 2000+25 (Tsunemi et al.\ 1989), a
BHXN; (b) the 1974 outburst of 4U 1543--47 (Li et al.\ 1976), also a BHC;
(c) the 1978 outburst of 4U 1908+005 (Charles et al.\ 1980), a NSXN. Note
that both (a) and (b) exhibit a `glitch'-type secondary maximum (see
\S~4.3.3). \label{fg:F} }
\end{figure}

\begin{figure}
\psfig{figure=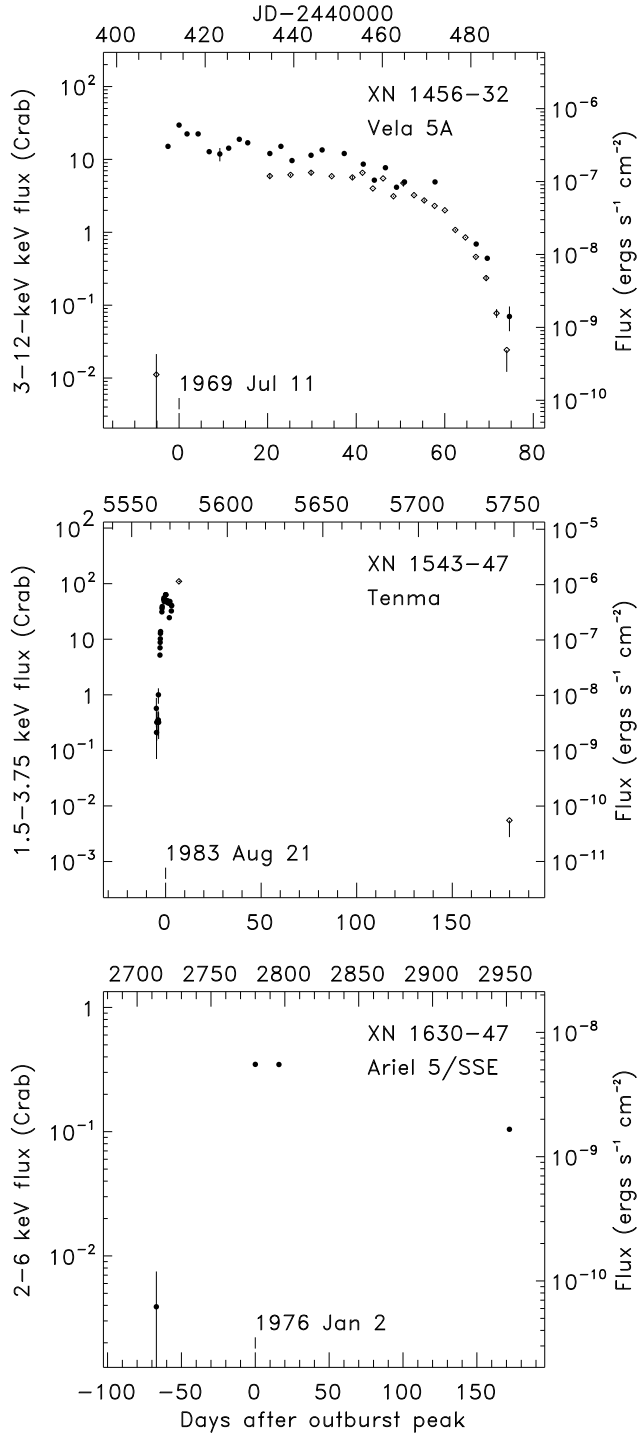}
\caption{ Examples of possible FRED light
curves. (a) The 1969 outburst of 4U 1456--32 (Evans et al.\ 1970); (b) the
1983 outburst of 4U 1543--47 (Kitamoto et al.\ 1984); (c) the 1976
outburst of 4U 1630--472 (Jones et al.\ 1976). \label{fg:PF} }
\end{figure}

The most prominent morphological trend we see in the collected light curves
is that a large fraction of them have a similar shape: a fast rise followed
by an exponential decay (Fig.~\ref{fg:F}). This is consistent with previous
findings (\cite{vpv84}; \cite{wks84}). In some cases, the exponential decay
lasts for more than 200 days and over 3 orders of magnitude in dynamic
range. The {\em fast-rise-exponential-decay} light curves are also seen in
X-ray and $\gamma$-ray bursts on much shorter time scales, where they are
given the acronym FRED\footnote{This term was coined by Chip Meegan and
has now been widely used in the $\gamma$-ray literature.}. We adopt this
acronym to describe the canonical XN light curves. In cases such as the
1977 outburst of XN 1543--47 (Fig.~\ref{fg:F}b), the flux decay later
becomes erratic -- but the {\em upper envelope} of the light curve
maintains an exponential form. We classify these light curves as
FREDs also. The best examples of the FREDs are XN 0620--00, XN 1124--683,
XN 2000+25, and XN J0422+32, where the first three are recorded in soft
X-rays and the last one in hard X-rays. Light curves which have either a
fast rise or an exponential decay profile but not both are classified as
possible FREDs -- examples are shown in Fig.~\ref{fg:PF}.

\begin{figure}
\psfig{figure=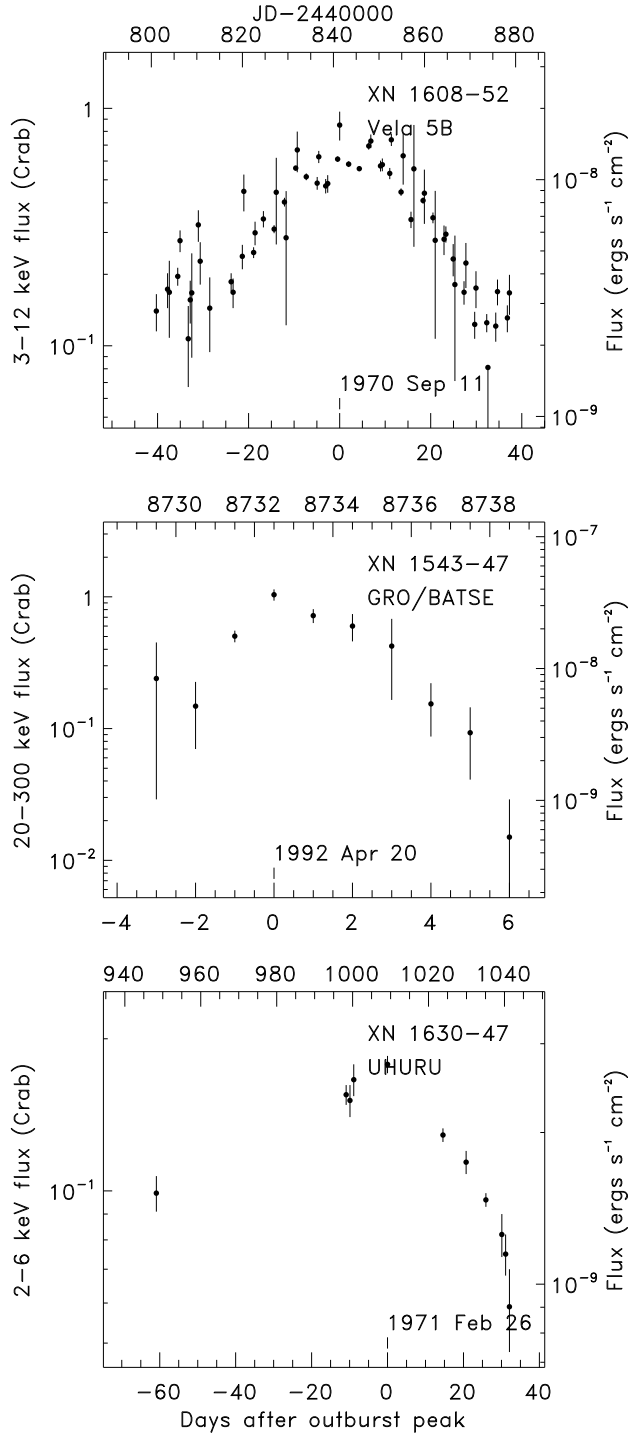}
\caption{ Examples of triangular light curves.
(a) The 1970 outburst of 4U 1608--52 (Lochner \& Roussel-Dupr\'e 1994); (b)
the 1992 outburst of A 1543--47 (Harmon et al.\ 1994) in hard X-rays; (c)
the 1971 outburst of 4U 1630--472 (Forman et al.\ 1976) in which the rise
timescale may be longer than the decay timescale. \label{fg:T} }
\end{figure}

\begin{figure}
\psfig{figure=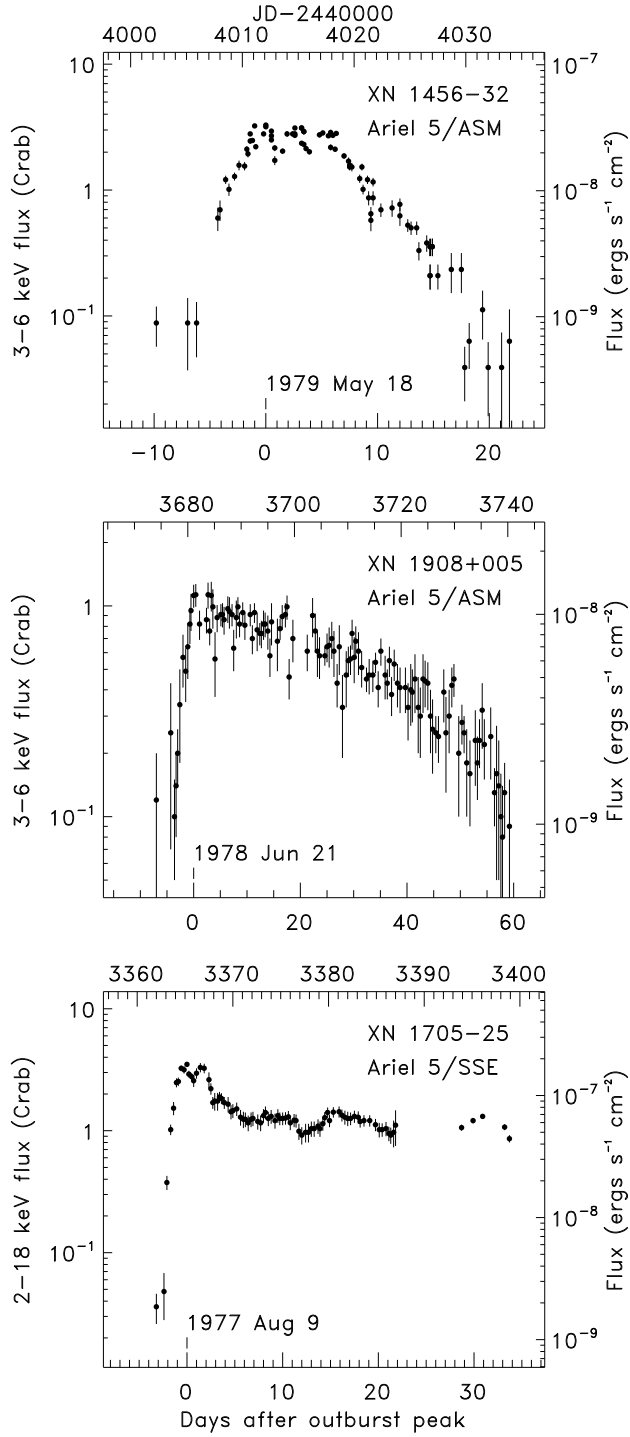}
\caption{ Examples of short plateaus which
are usually followed by normal decay. (a) The 1979 outburst of 4U 1456--32
(Kaluzienski et al.\ 1980); (b) the 1978 outburst of 4U 1908+005 (Charles
et al.\ 1980); (c) the 1977 outburst of 1H 1705--25 (Griffiths et al.\
1978; Share et al.\ 1978). The late phase of this outburst is uncertain.
\label{fg:SP} }
\end{figure}

\begin{figure}
\psfig{figure=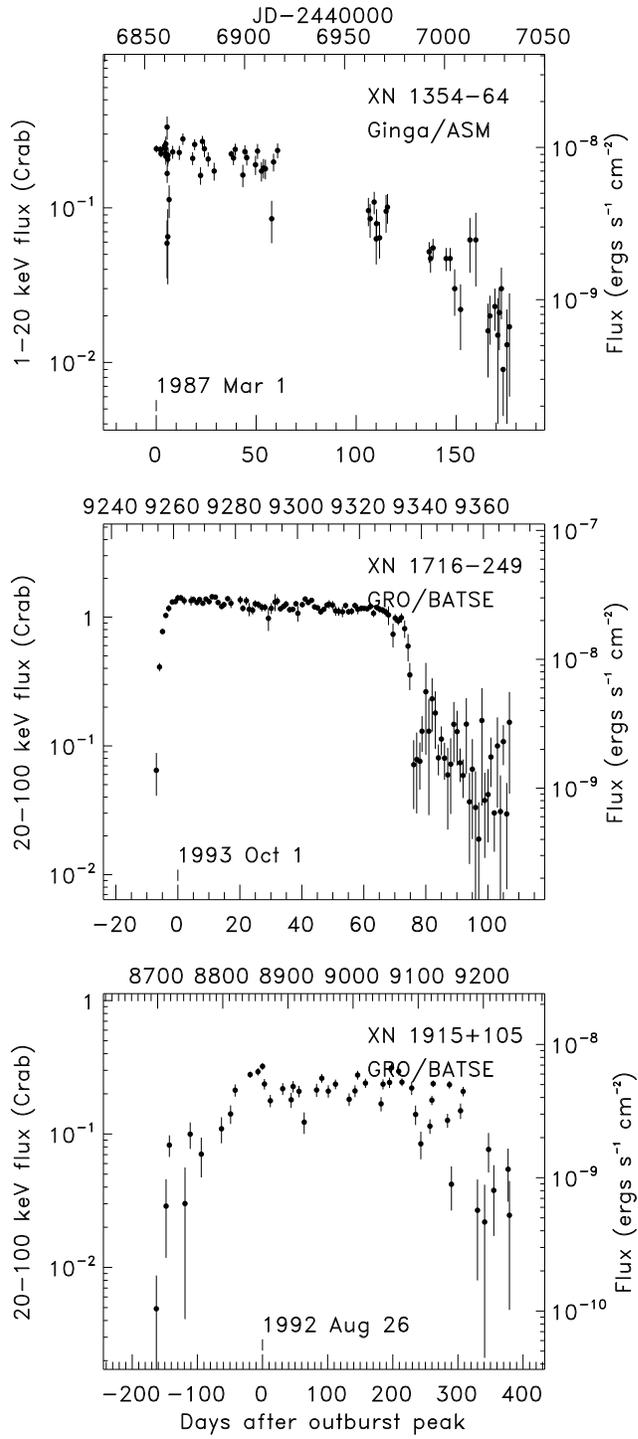}
\caption{ Examples of long plateau X-ray
light curves. (a) The 1987 outburst of A 1354--64 (Kitamoto et al.\ 1990),
a plateau of 50 days followed by normal decay; (b) the 1993 outburst of GRS
1716--249 (Harmon et al.\ 1994) which has a fast rise and a sudden cutoff;
(c) the 1992 outburst of GRS 1915+105 (Harmon et al.\ 1994), the longest
plateau to date, which has a very slow rise. \label{fg:LP} }
\end{figure}

\begin{figure}
\psfig{figure=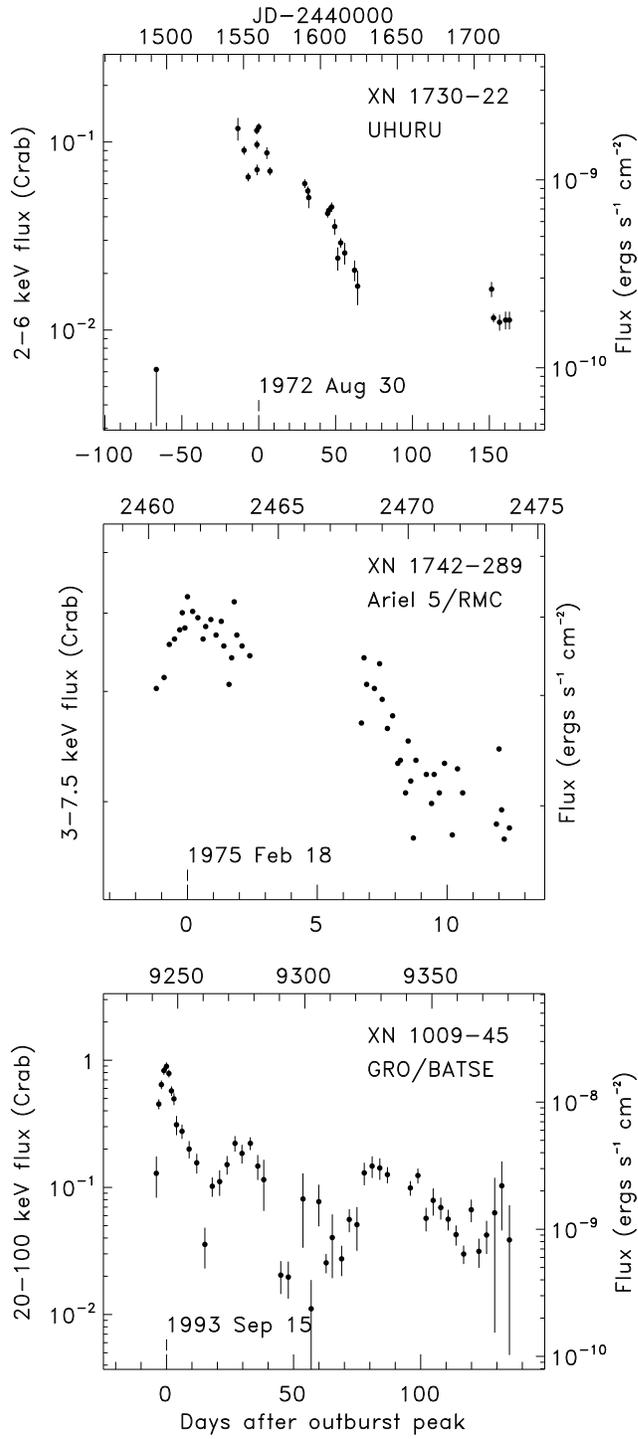}
\caption{ Examples of light curves with
variable decay constants. (a) The 1972 outburst of 4U 1730--22 (Cominsky et
al.\ 1978); (b) the 1975 outburst of A 1742--289 (Eyles et al.\ 1975;
Branduardi et al.\ 1976); (c) the 1993 outburst of GRS 1009--45 (Harmon et
al.\ 1994) in hard X-rays. \label{fg:V} }
\end{figure}

\begin{figure}
\psfig{figure=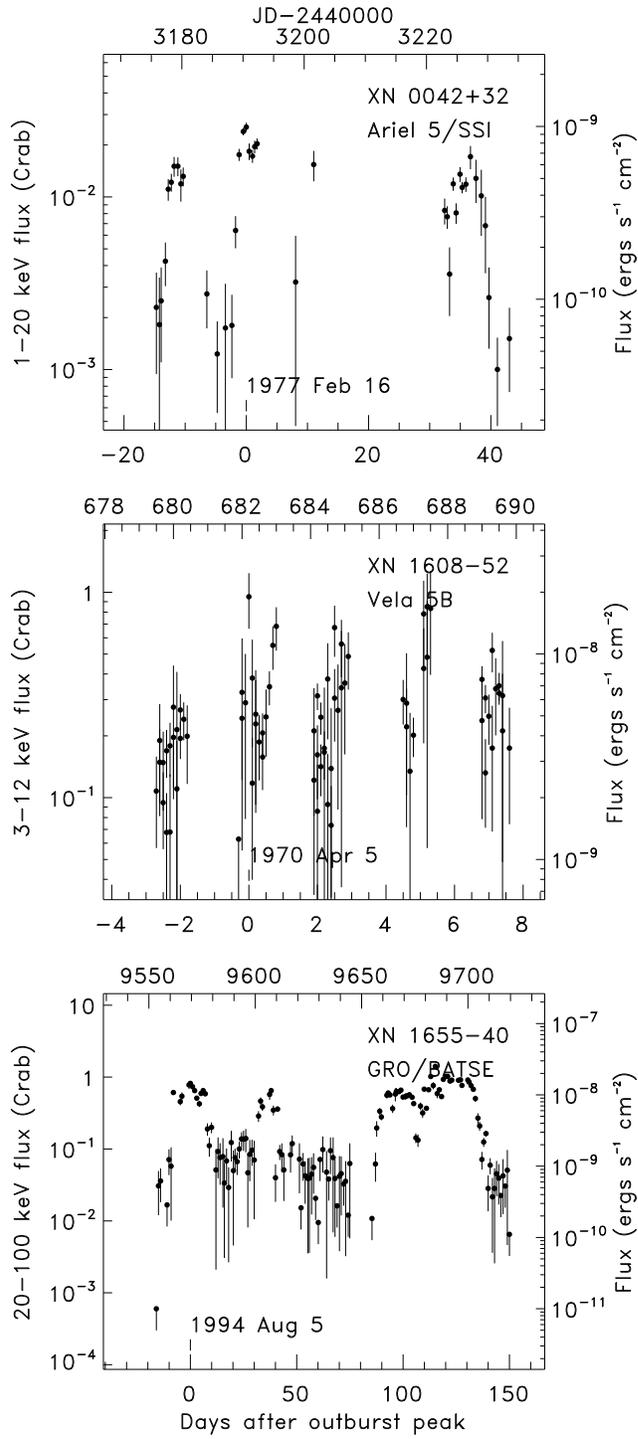}
\caption{ Examples of light curves with
multiple peaks of comparable strength. (a) The 1977 outburst of 3U 0042+32
(Watson \& Ricketts 1978); (b) the 1970 outburst of 4U 1608--52 (Lochner \&
Roussel-Dupr\'e 1994); (c) the 1994 outburst of GRO J1655--40 (Harmon et
al.\ 1995) in hard X-rays. \label{fg:MP} }
\end{figure}

In addition to the FRED case, we have identified four more general light
curve morphologies, {\em triangular, plateau, variable decay}, and {\em
multi-peak}, which were not previously recognized. They are illustrated in
Figs.~\ref{fg:T}-\ref{fg:MP} and listed in Table~\ref{tb:type}.

\begin{table}[tbp]
\dummytable\label{tb:type}
\end{table}

The {\em triangular} profile is one in which the rise time is similar to,
or even longer than the decay time (Fig.~\ref{fg:T}). The dynamic range of
the triangular outbursts is always small so that it is unclear if the rise
and decay are exponential or linear. 

The {\em plateau} light curves (Figs.~\ref{fg:SP}-\ref{fg:LP}) are those in
which the source stays at the peak ($\pm10\%$) for an extended period of
time ($>3$ days). In some cases, the plateau phase is relatively short
($<30$ days) and it is followed by a more or less normal exponential decay
(Fig.~\ref{fg:SP}). In other cases, the plateau lasts much longer but
eventually decays normally (Fig.~\ref{fg:LP}a). In a few extreme cases,
however, a very long plateau is terminated by a sudden cutoff
(Fig.~\ref{fg:LP}b), thus bypassing the exponential decay phase
altogether. No plateaus have been seen in the optical light curves.

The {\em variable-decay} light curves show an otherwise smooth decay
profile broken into several segments, each having a considerably different
decay timescale and duration (Fig.~\ref{fg:V}a, b). In some cases they have
a complicated substructure (Fig.~\ref{fg:V}c). Although the data in
Fig.~\ref{fg:V}a and b are not sufficiently sampled, they clearly show that
the decay in the late stages does not follow the same pattern as in earlier
stages. 

The last morphological type is the {\em multi-peak} light curve seen only
in a few cases (Fig.~\ref{fg:MP}). Different from all other types which
have a main outburst, these events undergo several consecutive, {\em
similar} outbursts with later ones sometimes stronger than the first. The
three different sources show different timescales in both peak intervals
and individual peaks. An important point is that the multi-peak light
curves have been seen in both soft and hard X-rays, therefore, they are
{\em not} an exclusive feature of a given energy band.

\begin{figure}
\psfig{figure=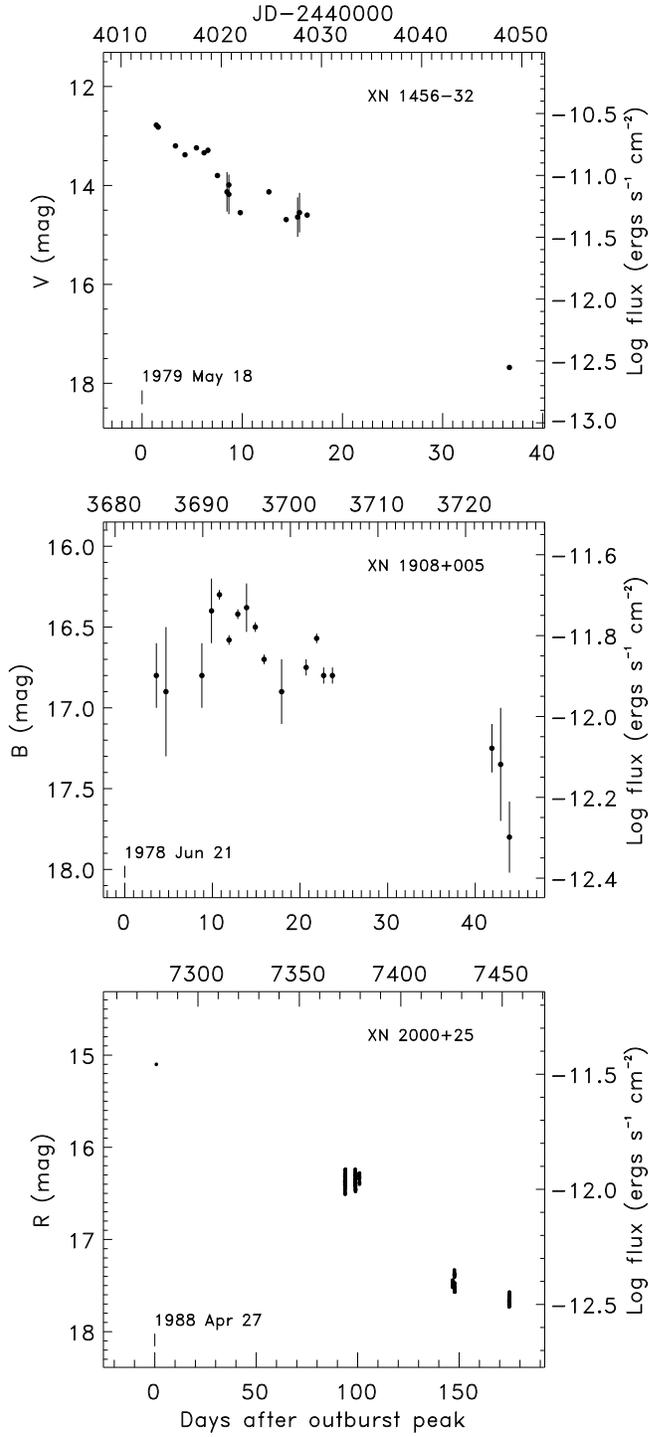}
\caption{ Examples of optical light curves. (a)
The 1979 outburst of 4U 1456--32 (Canizares, McClintock, \& Grindlay 1980);
(b) the 1978 outburst of 4U 1908+005 (Charles et al.\ 1980); (c) the 1988
outburst of GS 2000+25 (Chevalier \& Ilovaisky 1990). \label{fg:opt} }
\end{figure}

We have only a very limited collection of optical light curves. Most of them
can be classified as FREDs or possible FREDs, as illustrated in
Fig.~\ref{fg:opt}.

The light curve morphology can vary not only from source to source but also
from outburst to outburst of the same source. For some recurrent sources,
different outbursts show similar peak fluxes and durations, e.g.,
XN 1908+005 (\cite{cpaea80}), while for others, different outbursts evolve
very differently, e.g., XN 1630--472 (\cite{paw95}) and XN 1608--52
(\cite{lrd94}). 

In addition to these major morphological classes, a number of light curves
display distinct features, namely the {\em precursors} and {\em secondary
maxima}, which are superposed on their otherwise canonical profiles. During
the rising phase, a small number of outbursts exhibit a weak precursor peak
prior to the main peak (e.g., XN 1524--62 and XN 0620--00). In other cases,
both the X-ray and optical light curves exhibit one or more {\em secondary
maxima\/} in the form of a `glitch' (e.g., XN 0620--00 and XN 1124--683),
a `bump' (e.g., XN 0620--00), or a series of relatively large amplitude
`mini-outbursts' (e.g., XN J0422+32 in the optical). These light curve
features will be discussed in more detail in \S~\ref{ssc:f}.

Finally, we emphasize that: (1) the shapes of the light curves are generally
energy dependent with more variability and less uniformity in hard X-rays
($>10$ keV) than in soft X-rays, but (2) none of the light curve shapes
or features exclusively belong to any given energy band. For example, the
{\em Ginga\/} light curve of XN 1124--683 in the 1-37 keV band, which is
dominated by the soft component ($<9$ keV) during the first 150 days, has a
proto--typical FRED profile, but the light curve in the 9-37 keV band does
not comply with any of the above specified categories (Fig.~\ref{fg:1124}).
On the other hand, the 20-100 keV light curve of the 1992 outburst of
XN~J0422+32 is the best example of a smooth hard X-ray FRED
(Fig.~\ref{fg:0422}). 

\section{Quiescent Emission}
\label{sc:q}

Before we discuss the outbursts in detail, let us first examine the
properties of quiescent XN emission. Quiescent XNs are usually very faint. 
Most XNs listed in vP95 have not been detected in quiescence, although the
upper limits from previous X-ray instruments are not highly constraining.
Recent {\it ROSAT} and {\it ASCA} observations of 14 quiescent XNs detected
7 sources and set stringent upper limits on the others. In
Table~\ref{tb:q}, we list for all XNs their quiescent X-ray fluxes or upper
limits in col.~(5), in units of milli-Crab in the observed energy band.
The fluxes are then converted to 0.4-10 keV fluxes in units of ergs \ps\
\pcmsq, and listed in col.~(6), and to the corresponding luminosities
listed in col.~(7). In the calculation, we used the spectral information
from either the quiescent or the outburst spectrum. When neither was
available or not applicable, we assumed a power law of photon index 2. 

\begin{table}[tbp]
\dummytable\label{tb:q}
\end{table}

Table~\ref{tb:q} shows that only 9 out of the 24 sources have been detected
in quiescence so far. Among them 4 are NSs and 5 are BHCs. We note that the
quiescent luminosities of XN 0836--429 and XN 1354--64 seem somewhat high;
thus they may have not been seen in their true quiescent state.
Table~\ref{tb:q} also indicates that the quiescent X-ray luminosity spans
several orders of magnitude, particularly for the BHCs. XN~0620--00, for
example, has a flux of only 4.3 $\mu$Crab in 0.4--1.4 keV (\cite{mhr95}),
which leads to a record low X-ray luminosity of $6.2\times 10^{30}$ ergs
\ps\ in 0.4-10 keV. While XN~2000+25 has the lowest flux upper limit of
only 0.36 $\mu$Crab, its luminosity limit is still higher than that of XN
0620--00, due to its larger distance and absorption column (\cite{vbjkl94}).
XN~2023--338, on the other hand, is 4 times brighter and 3 orders of
magnitude more luminous than XN~0620 (\cite{wshkh94}). For the NS systems,
the spread is much less pronounced. The quiescent luminosities of the 4 NS
systems detected so far (XN~0748, XN~1456, XN~1608 and XN~1908) all cluster
around $10^{33}$ ergs \ps\ with a small dispersion of $\Delta L_{\rm q} /
<L_{\rm q}> =0.65$. For the 3 detected BHCs (XN~0620, XN~1655, and
XN~2023), the dispersion is 1.61.

The quiescent mass accretion rates listed in col.~(8) of Table~\ref{tb:q}
are derived from ${\dot M}_{\rm q} = L_{\rm q} / \eta c^2$ where we have
assumed a radiation efficiency of $\eta = 0.1$. These numbers should be
taken with caution, since it is possible that many of these systems involve
an advection dominated accretion flow (ADAF), which radiates with a very low
efficiency (e.g., \cite{nmy95}; \cite{nbm97}). If in fact $\eta$ is much
smaller, the corresponding ${\dot M}_{\rm q}$ will be proportionally
higher. We note that much insight into the outburst mechanism and the
quiescent state of the system may be gained by comparing the quiescent mass
accretion rate derived here with the average mass transfer rate from the
secondary star, ${\dot M}_2$. The latter has only been inferred for a few
sources from the average mass accretion rate required to power the observed
outbursts during the period since last outburst (\cite{vpj96}): $1.3\times
10^{-11} \Msun$ yr$^{-1}$ for XN~0620, $1.0\times 10^{-11}\Msun$ yr$^{-1}$
for XN~1456, $1.6\times 10^{-10}\Msun$ yr$^{-1}$ for XN~1908, and
$2.0\times 10^{-10}\Msun$ yr$^{-1}$ for XN~2023 (see also
\S~\ref{sc:model}). 

\section{PROPERTIES OF XN LIGHT CURVES}
\label{sc:main}

\subsection{Basic Light Curve Properties}
\label{ssc:b}

While the XN light curves display several different morphologies, most of
them can still be characterized by a common set of parameters. An ideal
FRED-type X-ray light curve is completely describable in terms of the peak
flux $F_{\rm p}$ or amplitude $F_{\rm p}/F_{\rm q}$, the rise timescale
$\tau_{\rm r}$, and the decay timescale $\tau_{\rm d}$. This is more or
less true for most other types of light curves. Even for the multi-peak
light curves, we can use these parameters for a single peak to represent
the entire light curve. Another important quantity is the duration of the
outburst which however, except in a few cases, is instrument dependent.
In Table~\ref{tb:b} we list these basic observational properties of the XN
light curves. Col.~(2) gives the year and month of the outburst peak.
Col.~(3) gives the light curve morphology type, as defined in \S~\ref{sc:m}
and Table~\ref{tb:type}; Col.~(4) gives the satellite/instrument and its
energy band for which the peak X-ray flux in Col.~(5) is reported in Crab
units; Col.~(6) gives the luminosity in the same energy band; Cols.~(7)-(9)
give the X-ray rise timescale $\tau_{\rm r}$, decay timescale
$\tau_{\rm d}$, and total duration $T_{\rm obs}$ respectively.
Cols.~(10)-(14) give the corresponding optical data, with Col.~(11) giving
the brightness change (or lower limit) in magnitudes. Col.~(15) gives the
main references from which the data were obtained. We will discuss each of
these quantities separately following the time sequence of an outburst.

\begin{table}[tbp]
\dummytable\label{tb:b}
\end{table}

\subsubsection{Rise timescales}
\label{sssc:rise}

A complete empirical picture of how an XN outburst rises from quiescence to 
the peak has not yet been established. From Figs.~\ref{fg:F}--\ref{fg:LP},
however, we see that a large portion of most of the {\it observed} rising
phases takes an approximately exponential form, which in some cases holds
well for more than 2 orders of magnitude. Moreover, the steepest flux
increase seems always to be exponential and it dominates the last decade of
flux increase, although the flux increase usually slows down significantly
near the peak. In this paper we define the rise timescale $\tau_{\rm r}$ to
be the e-folding time over the time period which experiences the fastest
flux increase. This quantity is readily measurable for most light curves,
is instrument independent, and is directly comparable with the commonly
used e-folding decay timescale. The results are listed in col.~(7) of
Table~\ref{tb:b}. In cases where we have only two data points between the
first detection and the peak, we assume an exponential rise between these
two points. The corresponding rise timescale will be an upper limit to its
true value.

\begin{figure}
\psfig{figure=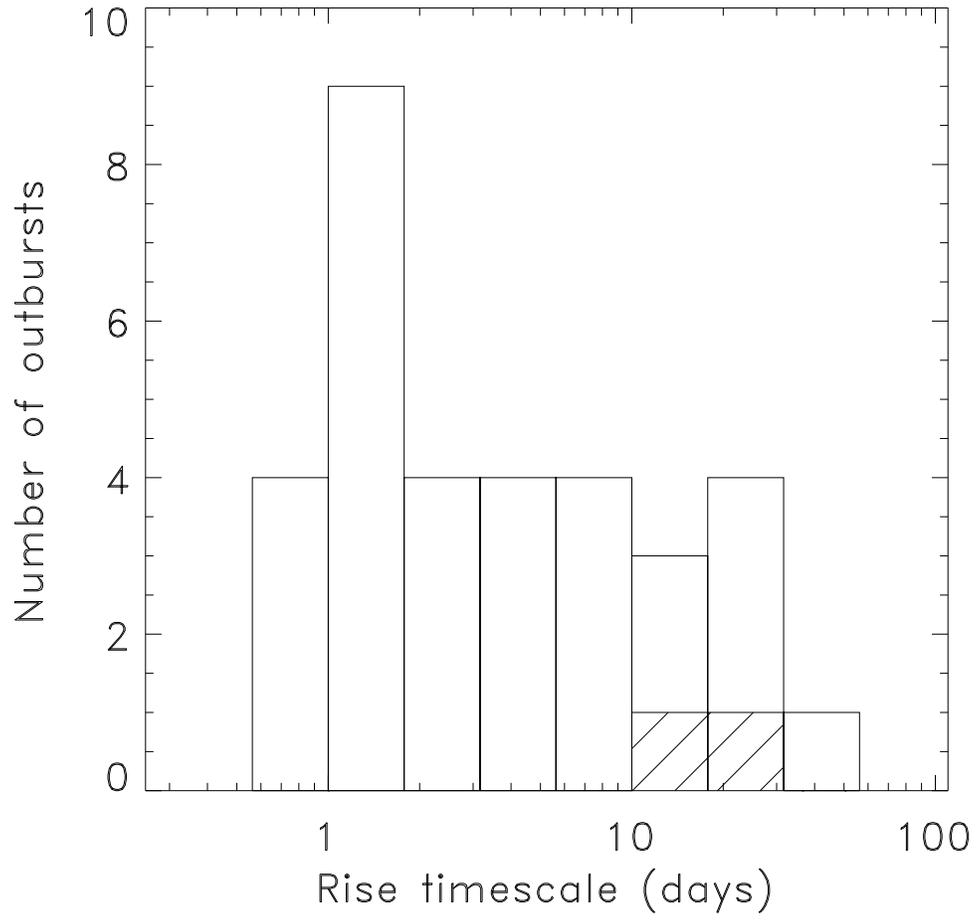}
\caption{ The distribution of rise timescales,
$\tau_{\rm r}$ in logarithmic spacing. The shaded areas are the {\em
upper} limits.  Notice the almost flat distribution and the single peak at
1--2 days. \label{fg:rise} }
\end{figure}

In Table~\ref{tb:b} we see that: (1) the rise timescales are generally of
the order of a few days but with a large spread, $<$$\log(\tau_{\rm r})$$>
= 0.533 \pm 0.621$ which corresponds to a mean $\tau_{\rm r}$ of 3.4 days
and the 1-$\sigma$ upper and lower bounds at 14.3 and 0.8 days.
\footnote{Due to large dynamical ranges and asymmetric distribution
properties, the mean and standard deviation for $\tau_{\rm r}$ and most
other light curve parameters in the following sections are calculated in
logarithmic scale; for convenience, we also report the corresponding mean
value and 1-$\sigma$ boundary in their natural units, in the format of {\it
mean (upper bound, lower bound)}.} Fig.~\ref{fg:rise} shows that
$\tau_{\rm r}$ in logarithmic scale is, surprisingly, evenly distributed
between 0.6 and 30 days with a narrow peak at the 1--2 day bin (or it is at
least a very broad peak if the true values of the two upper limits were
smaller by a few bins). Due to limited statistics, we do not know if this
indicates two different types of flux increase mechanisms; (2) about 25\%
of the outbursts have $\tau_{\rm r}>10$ days. This is partly due to the
lack of pre-peak coverage (e.g., XN 1354 in 1967 and XN 1630 in 1971) which
always makes $\tau_{\rm r}$ longer. Several outbursts with long
$\tau_{\rm r}$, however, do have adequate rise phase coverage (e.g., XN
1354 in 1987, XN~1915 in 1992, and XN~J1655 in 1996), and therefore, are
truly slow risers. 

The rise timescale so defined characterizes the rate of flux increase but
not the {\it duration} of the rising phase. If a major portion of the flux
increase follows a single exponential form, which has been seen in many
cases such as XN~1124 and XN~2000, we can define the expected duration of
the rising phase to be the rise timescale multiplied by the total number of
e-folds from quiescence to peak, i.e., 
\begin{equation}
T_{\rm r,exp} = \tau_{\rm r} \ln A_{\rm p},
\end{equation}
where $A_{\rm p} = L_{\rm p}/L_{\rm q}$ is the outburst amplitude which we
will discuss later. Therefore, an outburst with $\tau_{\rm r} =3$ days and
an amplitude of $10^3$, would likely have a rising phase lasting at least
about 7 e-folding times, or 3 weeks. 

We note that $T_{\rm r,exp}$ as defined serves only as a general gauge or a
lower limit to the true duration of the rising phase, since the real rising
phase may be more complicated than a single exponential form so that a
small $T_{\rm r,exp}$ does not necessarily imply a short rising phase. 
The rising history of XN~1524 in 1974 (\cite{kljea75}) consists of two
periods with different e-folding times, separated by a precursor peak
(\cite{kljea75}). The 2.9-day rise timescale listed in Table~\ref{tb:b}
is for the second, main period. In this case, the last decade of flux
increase took 22 days instead of 6.7 days expected from a single
exponential. 

\begin{figure}
\psfig{figure=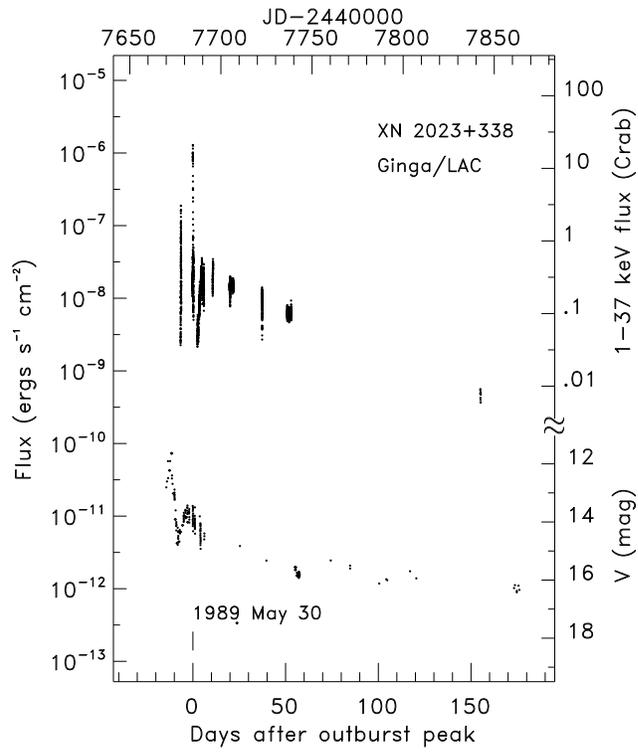}
\caption{ The {\it Ginga}/LAC 1-37 X-ray and V-band
optical light curves of XN 2023+338 in 1989. Notice that the optical may
have peaked a few days before the X-rays. \label{fg:2023} }
\end{figure}

Although there has been no complete optical coverage during the rising
phase of any XN outburst, there is evidence that the optical light may rise
and peak before the X-rays do by at least a few days. For the 1996 outburst
of XN J1655--40, Orosz et al.\ (1997) reported for the first time optical
observations fortuitously taken 6 days before the recorded rise of the
X-rays, which clearly show that the optical rises {\it before} the X-rays.
In that particular event it is also the case that the soft X-ray rise
precedes the hard X-rays (Hynes et al. 1997). However, there is no 
follow-up data to confirm that the optical actually {\it peaks} before the
X-rays do. For the 1989 outburst of XN 2023, the optical light does peak
before the X-rays (Fig.~\ref{fg:2023}), although we do not have a complete
optical rise profile due to large variability. There is also weak evidence
in the early phase of the 1991 outburst of XN~1124 suggesting that the
optical might have peaked 3 days before the X-rays (\cite{dvjw91}).

\subsubsection{Peak flux, luminosity and amplitude}
\label{sssc:amplitude}

At the peak of an outburst, the observed X-ray flux measures the apparent
brightness and the luminosity gives the intrinsic energy output. However,
the fluxes and luminosities listed in Table~\ref{tb:b} have to be converted
into common energy bands before meaningful comparisons can be made. Another
useful quantity is the outburst amplitude, i.e., the fractional change in
the bolometric luminosity from the quiescence to the peak, which reflects
the strength of the outburst. Since for most historical XNs, the spectral
information that is available is often insufficient for a reliable
bolometric correction to be made, we opt to compare the luminosities in a
{\em common} energy band which contains a major portion of the radiated power
during both the outburst and quiescence.

We first have to determine the best choice for this common energy band.
Historical XN outbursts were recorded by different instruments but they
often overlap more or less at the 2-10 keV band (except {\it
Compton}/BATSE, see Table~\ref{tb:b}). More importantly, a large fraction
of the power is radiated in this band for a thermal spectrum of $kT\sim1$
keV or a power law spectrum of photon index $>2$ or a combination of both.
In quiescence, on the other hand, many sources observed by {\it Rosat}
(0.4-2 keV) show a thermal component of $kT \sim 0.1$ keV. Therefore, a
broad band of 0.4-10 keV is likely to encompass much of the total radiated
power of XNs in both outburst and quiescence. Converting the observed
fluxes obtained by most X-ray instruments to this common energy band does
not require large extrapolation and thus should not introduce large
uncertainties.

Using the published spectral parameters and peak fluxes (Table~\ref{tb:b})
we calculated, in the 0.4-10 keV band for each outburst, the calibrated
outburst peak X-ray flux $F_{\rm p}$ in units of ergs s$^{-1}$ cm$^{-2}$,
peak luminosity $L_{\rm p}$ in units of both ergs s$^{-1}$. We also
computed the Eddington luminosity, $L_{\rm Edd}\sim 1.3 \times 10^{38}$
ergs s$^{-1}$ ($M/\Msun$), assuming solar composition, and peak amplitude
$A_{\rm p}=L_{\rm p}/L_{\rm q}$. The results are listed in
Table~\ref{tb:E}. It is seen that the peak flux, $F_{\rm p}$, spans 4
orders of magnitude ranging from $1.22 \times 10^{-10}$ (XN 1630--472 in
1979 March) to $1.35 \times 10^{-6}$ ergs s$^{-1}$ cm$^{-2}$ (XN 1456--32
in 1969 July), with $<$$\log(F_{\rm p})$$> = -7.84 \pm 0.82$ with the
corresponding the mean and 1-$\sigma$ boundary for $F_{\rm p}$ at $1.4\,
(9.5, 0.2) \times 10^{-8}$ ergs s$^{-1}$ cm$^{-2}$. The peak luminosity, on
the other hand, varies from $8.6 \times 10^{35}$ ergs s$^{-1}$ (XN 0748,
1985 February) to $8.63 \times 10^{39}$ ergs s$^{-1}$ (XN 1742, 1975
February), and $<$$\log(L_{\rm p})$$> = 37.72 \pm 0.79$ or $0.52\, (2.22,
0.08) \times 10^{38}$ ergs s$^{-1}$.

\begin{table}[tbp]
\dummytable\label{tb:E}
\end{table}

\begin{figure}
\psfig{figure=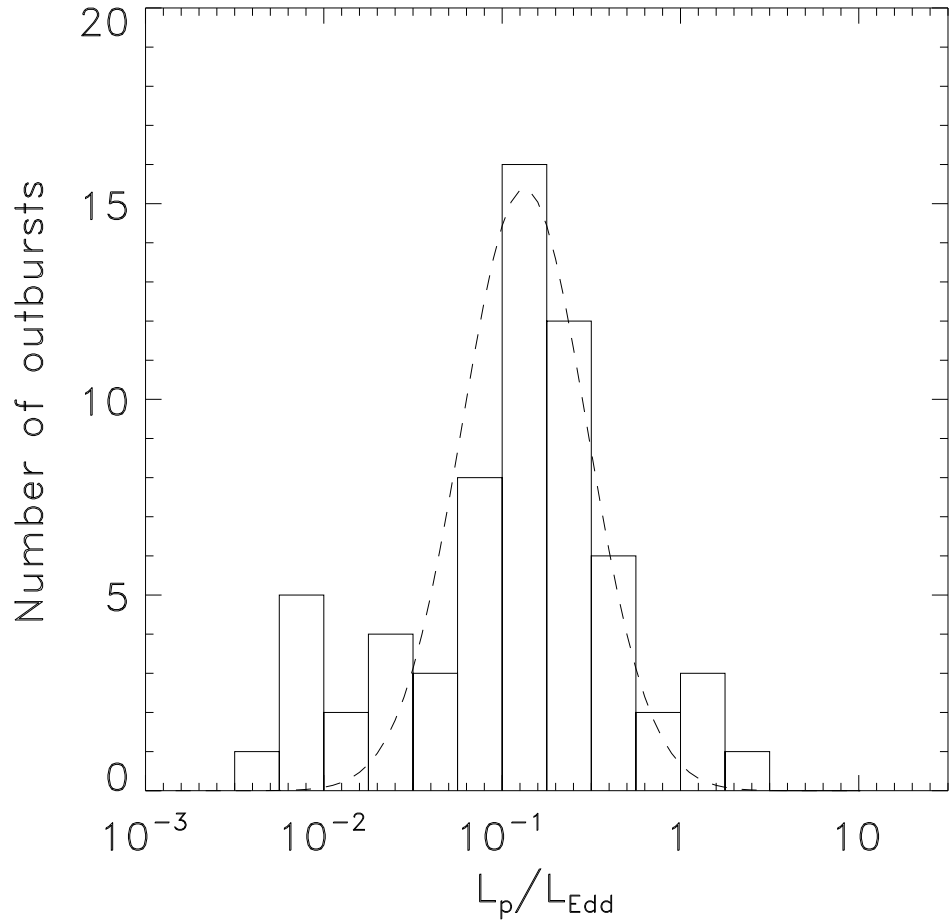}
\caption{ The peak luminosity distribution in the
units of Eddington limit in logarithmic scale. The dashed line is a Gaussian
fit. \label{fg:LpLE} }
\end{figure}

In deriving $L_{\rm p}/L_{\rm Edd}$ we have assumed a mass of $1.4\Msun$
for a NS and $10\Msun$ for a BH for sources whose compact object mass is
not available.  We see that $L_{\rm p} / L_{\rm Edd}$ ranges from 0.0031
(XN 0042+32, 1970 February) to 6.6 (XN 1742--289, 1975 February). The mean
is $<$$\log(L_{\rm p}/L_{\rm Edd})$$> = -1.00 \pm 0.68$ or equivalently the
mean $L_{\rm p}/ L_{\rm Edd}$ and its 1-$\sigma$ boundary at 0.10 (0.48,
0.02). This result deviates from the general belief that the peak
luminosity of XN outbursts is always near the Eddington limit. Changing the
unknown BH masses from $10\Msun$ to $5\Msun$ only increases this value by a
factor of $<2$. When we count only the most luminous outburst from each
source, the average luminosity increases to $0.13\, (0.97, 0.02)
L_{\rm Edd}$. In Fig.~\ref{fg:LpLE} we plot the distribution of the peak
luminosity in Eddington units. It is roughly a Gaussian centered at $0.2
L_{\rm Edd}$ with a FWHM of 0.82 in logarithmic scale, if we disregard the
large excess at $<0.03$. These values may have been biased by the
uncertainties in the distances, masses, and spectral shapes, especially in
the case of XN 0042+32, XN 0836-429, and XN 1918+146. The cases with
$L_{\rm p} / L_{\rm Edd} >1$ are probably also biased by these
uncertainties.

\begin{figure}
\psfig{figure=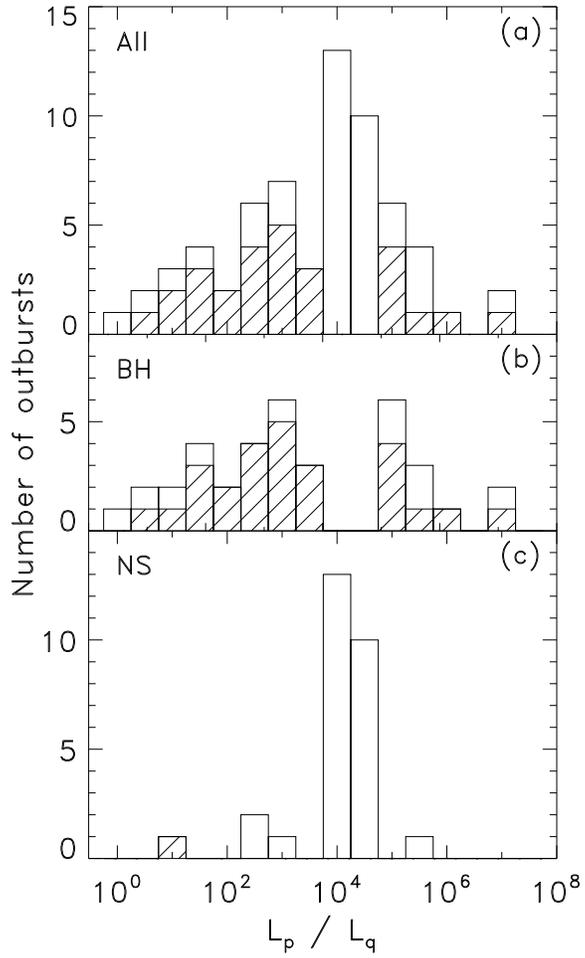}
\caption{ The peak amplitude distribution of XN
outbursts in logarithmic spacing. (a) All sources, (b) BH events, (c) NS
events. The shaded areas are the {\em lower limits} derived from the peak
outburst flux and the upper limits of the quiescent flux. \label{fg:Amp} }
\end{figure}

Next, we examine the outburst amplitude $L_{\rm p} / L_{\rm q}$, listed in
col.~(5) of Table~\ref{tb:E}. For sources only having upper limits for
their quiescent fluxes, we calculate the lower limits of their outburst
amplitudes. We see that the peak amplitude varies from only a factor of
less than 2 for XN 0836--429 in 1990 May to an astonishing $2\times10^7$
for XN 0620--00 in 1975 August. The average amplitude is $(7.17 \pm 0.78)
\times 10^3$. If we exclude the lower limits, the average increases by a
factor of $\sim2$ to $(1.54 \pm 0.08) \times 10^4$; if we take only the
highest amplitude from each of the 9 positively detected sources, the
average amplitude jumps by another factor of $\sim2$ to $(3.73 \pm 0.16)
\times 10^4$. Fig.~\ref{fg:Amp}a shows that the amplitude distribution in
logarithmic scale has a broad peak at $\sim 10^4$. Future detections of
quiescent fluxes will make the distribution move to the right.

One may ask if there is a difference in the amplitude distribution between
the BHs and NSs. Fig.~\ref{fg:Amp}b-c shows that the NS events have a
narrow distribution around $ \sim 10^4$, dominated by the recurrent events
of XN 1608--522 and XN 1908+005. If we take only one outburst from each
source, however, the amplitude distribution of NS events is similar to that
of the BHs, i.e., a broad hump between 10 and $10^7$. We note that most BH
events have only a lower limit on their amplitude and the true distribution
may be different from the one shown here. For the same reason, the lack of
BHCs events in the bins of $10^4$ and $3\times10^4$ seems to be accidental.

\subsubsection{Decay timescales}
\label{sssc:decay}

For most XN light curves, we can define the decay timescale $\tau_{\rm d}$
as simply the e-folding time computed over a time period long enough to
smooth out small scale variations. Generally speaking, $\tau_{\rm d}$ is a
well determined parameter, because the overall shape of most light curves,
or the upper envelope of variable light curves, is approximately
exponential in both the X-rays and optical. When the observed flux range is
small, we may not be able determine with confidence if the decay is
exponential, linear, or a power law. In these cases, we assume it is
exponential. For plateau or variable decay profiles, we calculate
$\tau_{\rm d}$ for the dominant phase of the decay. For example, for a
short plateau profile we use the decay timescale of the normal decay phase
which is usually longer than the plateau phase. For a long plateau we list
the quantity $\tau_{\rm d}$ which characterizes the decay timescale
{\it during} the plateau phase. 

\begin{figure}
\psfig{figure=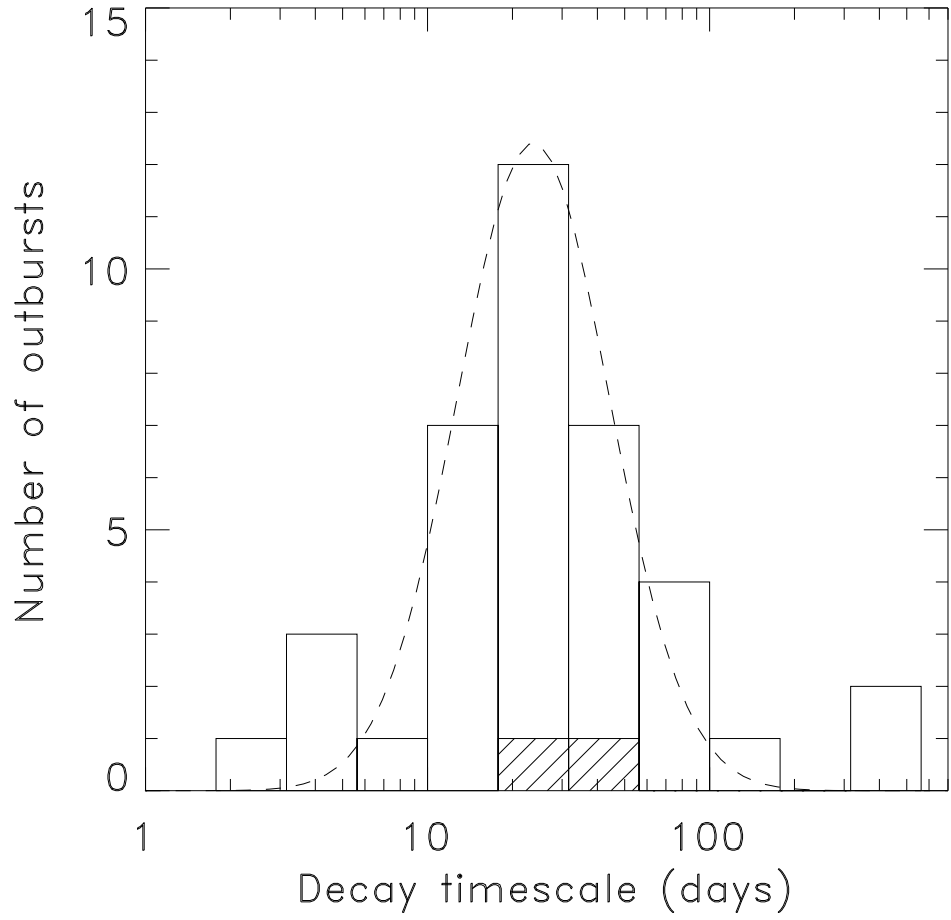}
\caption{ The distribution of decay timescales,
$\tau_{\rm d}$ in logarithmic spacing. The shaded areas are the {\em
upper limits}. \label{fg:decay} }
\end{figure}

A prominent trend found in historical XN light curves is a universal decay
timescale of approximately 30 days (\cite{wks84}). This feature still holds
in our enlarged sample, $<$$\log(\tau_{\rm d})$$> = 1.24\pm 0.36$ or
equivalently $17.4\, (39.8, 7.8)$ days, after excluding the plateau events.
Although the spread is still large, Fig.~\ref{fg:decay} shows that the
distribution of $\tau_{\rm d}$ is quite different from that of
$\tau_{\rm r}$. Here, we see an approximately Gaussian profile in the
logarithmic scale, which peaks at $\sim 24$ days, and with a FWHM of 0.65
in $\log \tau_{\rm d}$. The large spread is mainly caused by significant
residues at both short and long timescales. 

\begin{figure}
\psfig{figure=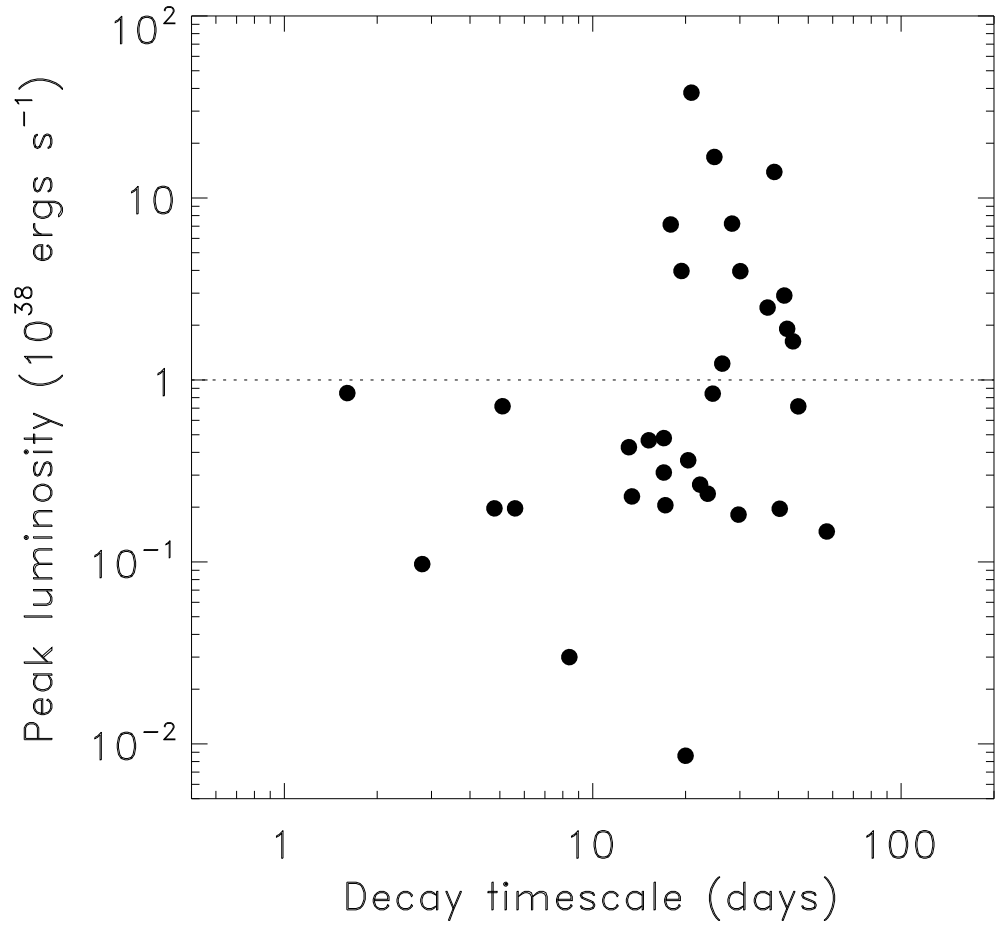}
\caption{ The correlation between outburst peak
luminosity and decay timescale of the normal outbursts (excluding the
plateau events). \label{fg:TdLp} }
\end{figure}

Considering the large intrinsic differences among these systems, the
observed universality of the exponential decay and its time constant are
remarkable and likely to be the results of some fundamental physical
processes involved. From Table~\ref{tb:b} we also find that for recurrent
sources, the universality holds for multiple events of {\em similar
intensity} (e.g., XN 1908+005 in 1975-1978), but not for multiple events of
{\em different intensity} (e.g., XN~1456--32, XN~1608--522, XN~1630--472).
Looking for a possible dependence of the decay timescale on some other 
physical quantity, we notice that, as shown in Fig.~\ref{fg:TdLp},
$\tau_{\rm d}$ of high luminosity outbursts (excluding the plateau events)
seems to be confined to a much narrower range than that of the low
luminosity outbursts.

The limited statistical information available on optical decay timescales
in our sample allows only for crude conclusions to be drawn. First, the
flux decay is generally a factor of 2.2 slower in the optical than in
X-rays. From Table~\ref{tb:b} we derive an average decay time of 67.6 days
in the optical and 34.9 days in X-rays (excluding the long plateaus). For
individual outbursts, this ratio is 1.5 for XN 1908+005 in 1978 and 4.8 for
XN J0422+32. 

\subsubsection{Duration and total energy}
\label{sssc:duration}

The observed duration of the outbursts, listed in Table~\ref{tb:b} for the
X-rays ($T_{\rm obs}$) and optical ($T_{\rm obs,o}$), is the total elapsed
time in days between the first and the last detections of the source. This
is clearly dependent on the instrumentation and on the conditions under
which the observations were made. Similarly to both the rise and decay
timescales, the observed duration has a large spread in time ranging from
$\sim10$ days to $>100$ days with a mean value of $<$$\log(T_{\rm obs})$$>
= 2.05\pm0.42$ which corresponds to the mean and 1-$\sigma$ boundary of
$112\, (294, 43)$ days. For the small number of observed $T_{\rm obs,o}$,
these values are $2.02 \pm 0.50$ and $104\, (327, 33)$ days. We have
searched for, but found no correlation between the duration and either the
peak flux or the amplitude. 

Since $T_{\rm obs}$ is highly instrument dependent, we define an expected
duration to be more reflective of the intrinsic event. First, similarly to
the rise phase duration $T_{\rm r,exp}$, we define the expected duration
for the decay phase to be the number of days required for the outburst to
decay from the peak to quiescence, i.e., 
\begin{equation}
T_{{\rm d,exp}} = \tau_{{\rm d}} \;\ln A_{\rm p}.
\end{equation}
Then, the expected duration of the entire outburst is
defined as
\begin{equation}
T_{\rm exp} = T_{\rm r,exp} +T_{\rm d,exp} = (\tau_{\rm d} + \tau_{\rm r})
\; \ln A_{\rm p}.
\end{equation}
This quantity is still instrument dependent in the sense that some of the
amplitudes are only upper limits. Using the observed rise and decay
timescales (Table~\ref{tb:b}) and the calibrated outburst amplitudes
(Table~\ref{tb:E}), we calculate $T_{\rm r, exp}$, $T_{\rm d,exp}$ and
$T_{\rm exp}$, and the results are listed in Table~\ref{tb:dur}. We see
that $T_{\rm exp}$ is usually a factor of 2 or more longer than
$T_{\rm obs}$. 

\begin{table}[tbp]
\dummytable\label{tb:dur}
\end{table}

\begin{figure}
\psfig{figure=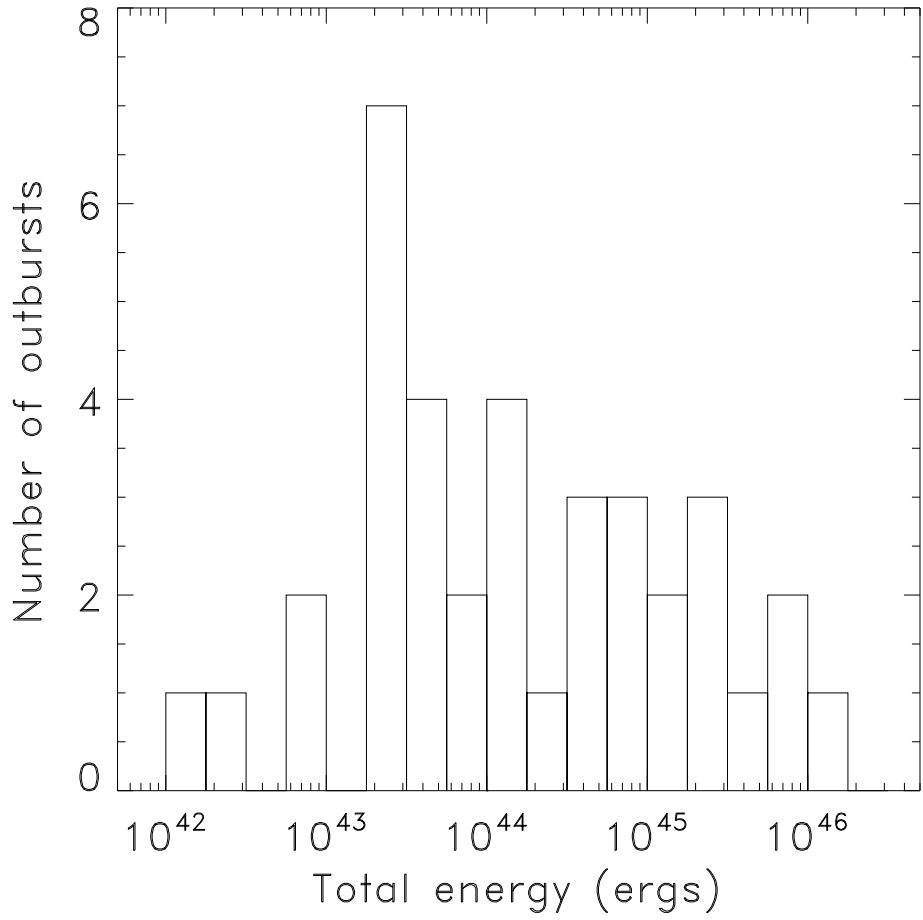}
\caption{ The distribution of the total energy
radiated in the 0.4-10 keV band during the outburst. \label{fg:E} }
\end{figure}

A measure of the energetics of an XN outburst is the total radiated energy,
i.e., the time-integrated X-ray luminosity over the entire outburst. For 
outbursts which follow an exponential profile during both the rising and
decay phase, the expected total energy of the outburst can be calculated as
\begin{eqnarray}
E_{\rm exp} & = & L_{\rm p} [ \tau_{\rm r}(1-e^{-T_{\rm r,exp}/\tau_{\rm r}})
                  + \tau_{\rm d}(1-e^{-T_{\rm d,exp} / \tau_{\rm d}} ) ]
 \nonumber \\ 
 & = & L_{\rm p} (\tau_{\rm r} + \tau_{\rm d}) ( 1 - A_{\rm p}^{-1}).
\end{eqnarray}
For $A_{\rm p} \gg 1$, we have $E_{\rm exp} \sim L_{\rm p} (\tau_{\rm r} +
\tau_{\rm d})$. This formula does not include the irregularities in the
light curves, nor does it take into account a possible spectral evolution
during the outburst as seen, for example, in the 1991 outburst of XN
1124--683 (\cite{ekea94}). From the observed light curves, however, we know
that the irregularities in general will not affect the derived total energy
by more than a factor of 2, and substantial spectral evolution occurs only
when the X-ray flux has decayed significantly. Therefore, the total energy
estimated here is in general accurate, to about a factor of 2-3. In
Table~\ref{tb:dur} we list the derived total energy for sources with
sufficient data. It is seen that this quantity covers 4 orders of magnitude
in dynamic range with $<$$\log(E_{\rm exp})$$> = 44.47 \pm 1.01$ which 
corresponds to the mean and (upper, lower) 1-$\sigma$ boundary for
$E_{\rm exp}$ of $2.92\, (29.8, 0.3) \times 10^{44}$ ergs. This corresponds
to (on average) a total amount of mass $\Delta M = E_{\rm exp} / \eta c^2$
of $1.64\, (16.7, 0.16) \times10^{-9} \Msun$ being accreted on to the
compact object during an XN outburst, where we assume a radiation
efficiency $\eta =10\%$.  The distribution of $E_{\rm exp}$ in
Fig.~\ref{fg:E} shows a broad hump centered at around $10^{44}$ ergs. 


\subsection{Additional Light Curve Features}
\label{ssc:f}

\subsubsection{Precursors}
\label{sssc:prec}

\begin{figure}
\psfig{figure=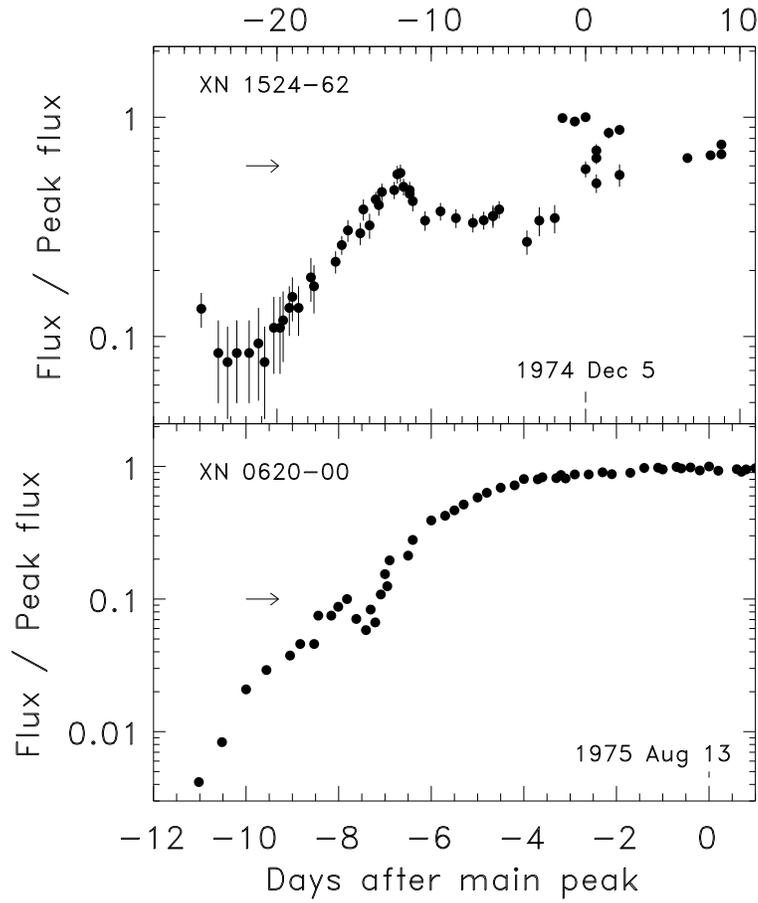}
\caption{ The precursor peaks seen in the light
curves of (a) XN 1524--62 (Kaluzienski et al.\ 1975) and (b) XN 0620--00
(Elvis et al.\ 1975). The {\it y}-axis is the ratio of the observed flux to
the peak flux. The precursors are marked by the horizontal arrows which
are at 60\% of the peak flux for XN 1524--62 and 10\% for XN 0620--00.
Notice that the time span of the {\it upper} panel is 3 times longer then
that of the {\it lower} panel. \label{fg:prec} }
\end{figure}

The rising phase of most XN outbursts is characterized by a fast, monotonic
flux increase. In a few cases, however, it exhibits a {\em precursor} peak
prior to the main peak. The best example is the 1974 outburst of XN
1524--62 recorded by {\it Ariel 5} (Fig.~\ref{fg:prec}a). Its 3-6 keV flux
reached a local maximum of $\sim80$\% of the main peak about two weeks
before the main peak. The rising to the precursor peak was slightly slower
than that to the main peak. After the precursor, the 3--6 keV flux declined
by a factor of 2 followed by a rapid rise to the main peak. A less
prominent precursor was seen in the 1975 outburst of XN 0620--00, which
crested at $\sim10$\% of the main peak about a week earlier, although this
is evident only in a linear light curve plot (Fig.~\ref{fg:prec}b). There
is also weak evidence for a precursor in the outbursts of XN 1608--522 in
1970, XN 1730--220, XN 1915+105, and XN 1908+005 in 1975. While there are
not enough cases for a statistical treatment, the lack of other clear
examples probably indicates that precursors are not a common phenomenon. 

\begin{figure}
\psfig{figure=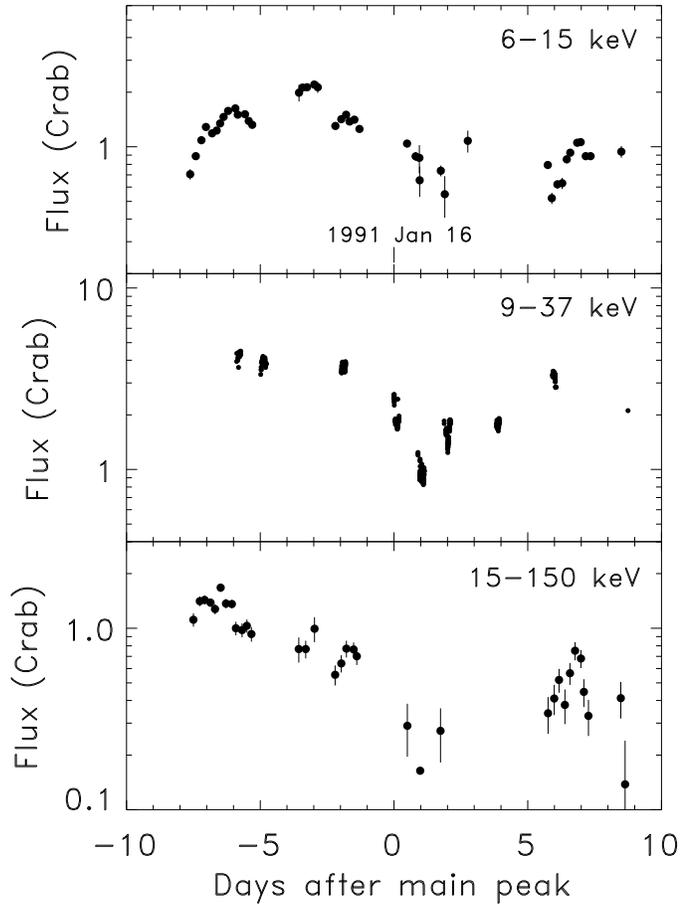}
\caption{ High energy light curves of XN
1124--683. The marked peak date at Day 0 is for the {\it Ginga} light curve
in 1-37 keV. The precursor(s) are seen only in (a) the {\it Granat}/WATCH
6-15 keV (Brandt et al.\ 1992) and (b) {\it Ginga}/LAC 9-37 keV light
curves (Ebisawa, Ogawa, \& Terada 1994), but not clearly in (c) the WATCH
15-150 keV light curve (Brandt et al.\ 1992). \label{fg:prec2} }
\end{figure}

The precursors we discussed above may be of a different physical origin
from the often more structured light curves seen in high energies, as
illustrated by the case of XN 1124--683. The {\it Ginga} 1--37 keV light
curve shows only a monotonic rise, but in 9--37 keV it clearly exhibits two
peaks (Fig.~\ref{fg:prec2}b): one 7 days before the main peak (in 1--37
keV), with 80\% of $F_{\rm p}$ and another 6 days after the main peak with
60\% of $F_{\rm p}$. The 6-15 keV light curve from {\it Granat}/WATCH, on
the other hand, has 3 peaks at --6 days ($\sim20\%$ of $F_{\rm p}$), --3
days ($\sim25\%$ of $F_{\rm p}$), and 7 days ($\sim12\%$ of $F_{\rm p}$)
(Fig.~\ref{fg:prec2}a). Finally, the WATCH 15--150 keV light curve also
shows 3 concurrent peaks with different flux ratios (Fig.~\ref{fg:prec2}c).
Such a complexity suggests that (1) an XN outburst may peak at different
times in different energies, (2) the hard X-rays usually rise and peak
earlier than the soft X-rays (while as noted in(\S~\ref{sssc:rise}) 
the optical may peak before both the hard and soft X-rays), and (3) the
precursor phenomenon is a rare feature in the low energy light curves,
while the multi-peak behavior is more common in high energy light curves.

\subsubsection{Plateaus}
\label{sssc:pla}

We have discussed the plateau events and classified them as one of the
major morphologies of the XN light curves, but this phenomenon warrants
some further discussion. In Table~\ref{tb:pla} we list the decay timescale
during the plateau phase ($\tau_{\rm plt}$), the duration of the
plateaus ($T_{\rm plt}$), and the decay timescale during the subsequent
decay phase following the plateau ($\tau_{\rm tail}$). For the plateau
events XN 1705--25 in 1977, XN 1716--249 in 1993, and XN 1915+105 in 1992,
the plateau decay constant $\tau_{\rm plt}$ is the same as $\tau_{\rm d}$
listed in Table~\ref{tb:b}. The events in BH systems are listed separately
from those in the NS systems. Generally speaking, a large spread in both
the decay timescale and duration, and in the subsequent decay behavior is
evident for both BHs and NSs.  After the plateau (long or short) phase, the
sources usually take on a normal exponential decay, but occasionally
undergo a sharp cutoff. This can happen in both BHs and NSs.

\begin{table}[tbp]
\dummytable\label{tb:pla}
\end{table} 

One thing is striking in Table~\ref{tb:pla} -- on average, all measures of
the plateau durations in BHs are longer than those in NSs by a factor
of 2 or more. For the BHs, even after excluding the extreme case of XN
1915+105, we get $<$$\tau_{\rm plt}$$>=338$ days, $<$$T_{\rm plt}$$>=45$
days, and $<$$\tau_{\rm tail}$$>=34$ days, respectively, which are still
substantially longer than those for the NSs. Individually, XN 1915+105 in
1992 (Fig.~\ref{fg:LP}c) holds the record in both the decay timescale
($\sim2400$ days) and plateau duration ($\sim300$ days), which makes it the
most energetic ($E_{\rm exp} \sim 2.8 \times 10^{46}$ ergs) XN event ever. 

With the exception of XN 1915+105, the longest plateau occurred in XN
1716--249 (Fig.~\ref{fg:LP}b), also a BHC. Both XN 1716 and XN 1915
underwent another, though much shorter outburst of similar peak flux about
200 days after the end of the plateau phase. In other words, these two
sources seem to be turning back on again very soon after their large,
initial outbursts. We do not know if these later events are entirely
separate outbursts or ``mini-outbursts" associated with the decay phase of
a previous outburst. It is nevertheless somewhat puzzling that these
sources have managed to accumulate enough material to trigger the
subsequent outbursts. We note that except for these light curve
similarities, the two sources are quite different: the peak X-ray spectrum
of XN 1716--249 has no soft component whereas XN 1915+105 has; XN 1915 is
an extremely slow riser whereas XN 1716 is fast; XN 1915 is a superluminal
jet source while XN 1716 is apparently not, although it has weak radio
emission and Hjellming et al.\ (1996) have suggested that its
X--ray--to--radio may be a low intensity analog of the superluminal sources
XN J1655 and XN 1905. 

\subsubsection{Secondary maxima}
\label{sssc:2ndmaxima}

Even in the canonical FRED light curves, the exponential decay may not
always be strictly monotonic. Instead, it is often interrupted by flux
increases by a factor of two or more, followed by resumption of the normal
decay; thus displaying {\em secondary maxima} (\cite{clg93}). 

\begin{figure}
\psfig{figure=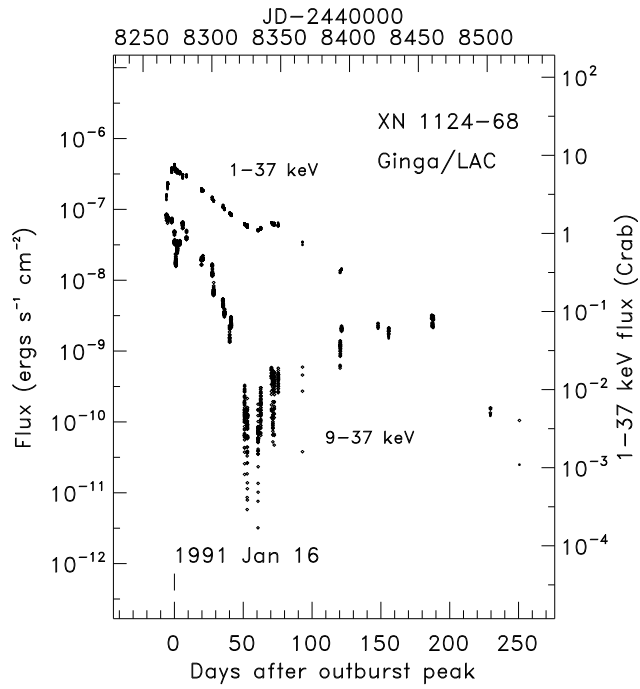}
\caption{ The {\it Ginga} 1-37 keV and 9-37 keV light
curves of the 1991 outburst of XN 1124--683 (Ebisawa et al.\ 1994).
\label{fg:1124} }
\end{figure}

\begin{figure}
\psfig{figure=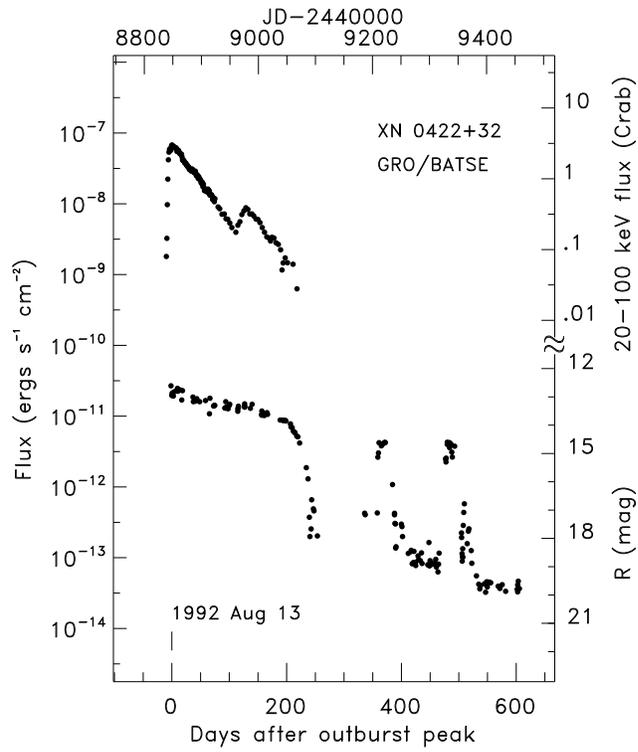}
\caption{ The {\it Compton}/BATSE 20-100 keV X-ray
and R-band optical light curves of the 1992 outburst of XN 0422+32
(Paciesas et al.\ 1995; Callanan et al.\ 1995). \label{fg:0422} }
\end{figure}

\begin{figure}
\psfig{figure=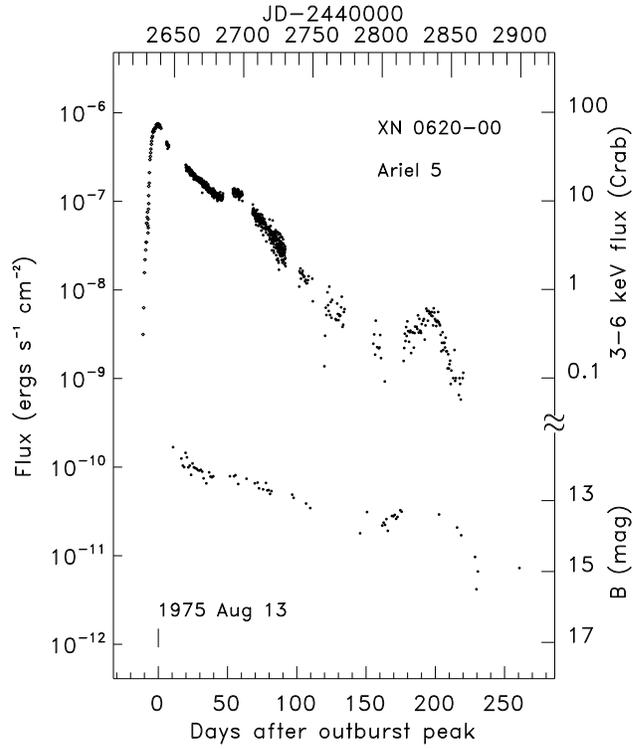}
\caption{ The {\it Ariel 5} 3-6 keV X-ray and B-band
optical light curves of the 1975 outburst of XN 0620--00 (Kaluzienski et
al.\ 1977; Elvis et al.\ 1975; Tsunemi et al.\ 1977). \label{fg:0620} }
\end{figure}

\begin{figure}
\psfig{figure=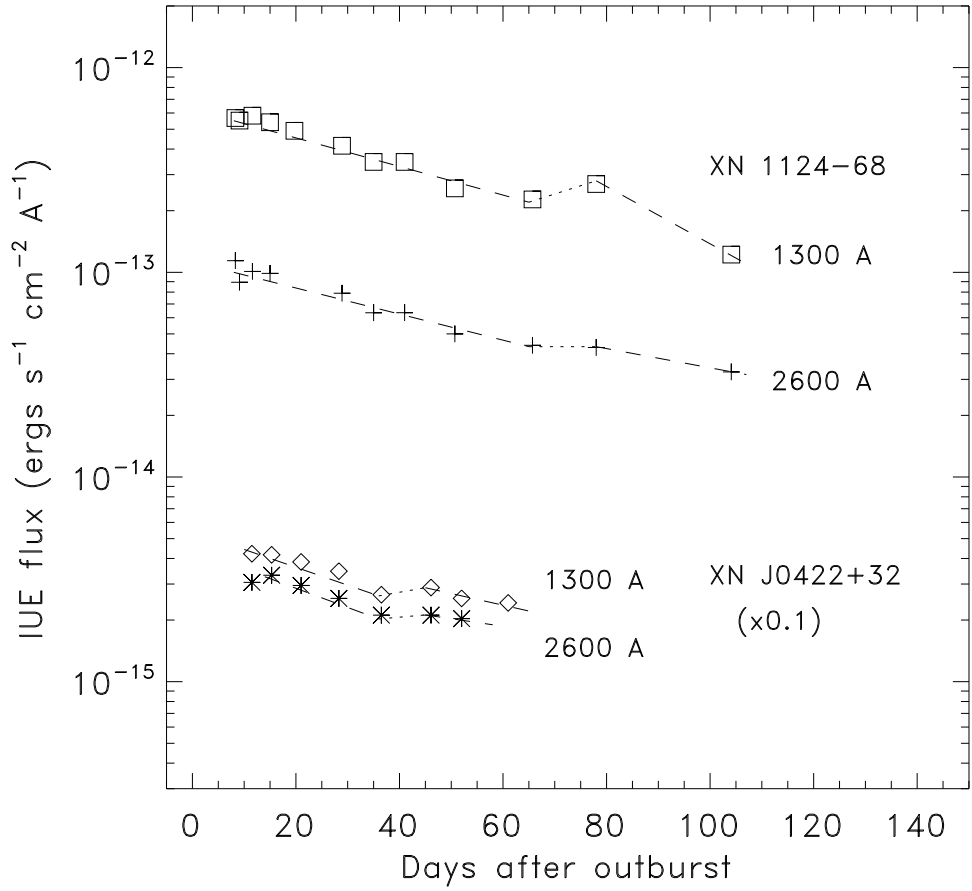}
\caption{ The {\it IUE} UV light curves of the 1991
outburst of XN 1124--683 and 1992 outburst of XN J0422+32 (Shrader \&
Gonzalez-Riestra 1991; Shrader et al.\ 1994). \label{fg:UV} }
\end{figure}

There are three morphological types of secondary maxima observed to date:
{\em glitches, bumps,} and {\em mini-outbursts}.\footnote{Following the
terminology adopted at the 1996 Aspen Winter Workshop on Black Hole X-Ray
Transients.} A glitch is an upward inflection superposed on a smooth
exponential decay, i.e., after the glitch the decay follows roughly the
same path as before, but is offset upwards by a factor of $\sim2$ or more.
Glitches are seen in the X-ray light curves of XN J0422+32
(Fig.~\ref{fg:0422}), XN 0620--00 (Fig.~\ref{fg:0620}), XN 1124--683
(Fig.~\ref{fg:1124}), XN 1543--47 (Fig~\ref{fg:F}b), and XN 2000+25
(Fig.~\ref{fg:F}a). We note that a glitch is always the {\em first}
secondary maximum to appear in a light curve within 100 days after the
peak, irrespective of how many additional features may occur subsequently.
Glitches have also appeared in the optical light curve of XN 0620--00
(Fig.~\ref{fg:0620}) and possibly of XN 1456--32 (Fig.~\ref{fg:opt}a), and
in the UV light curves of XN J0422+32 and XN 1124--683 (Fig.~\ref{fg:UV}),
at about the same time as in the X-rays.

\begin{figure}
\psfig{figure=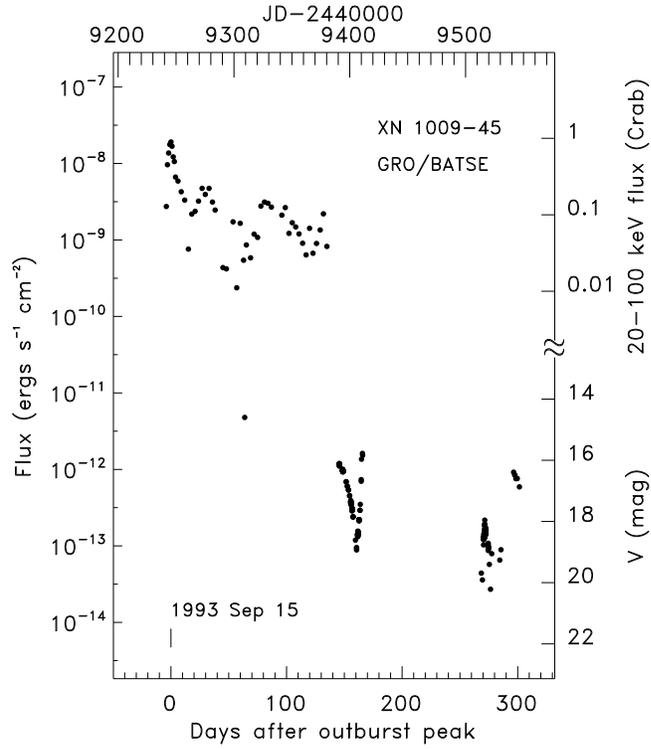}
\caption{ The {\it Compton}/BATSE 20-100 keV hard
X-ray and V-band optical light curves of the 1993 outburst of XN 1009--45
(Paciesas et al.\ 1995; Bailyn \& Orosz 1995). \label{fg:1009} }
\end{figure}

On the other hand, both ``bumps" and ``mini-outbursts" are small-amplitude
(relative to the primary peak) events superposed on a normal decay profile,
except that in some cases the mini-outbursts appear to be more like
outbursts by themselves, i.e., when the source is closer to quiescence.
Bumps have been seen both in X-rays from XN 0620 (Fig.~\ref{fg:0620}) and
XN 1124 (Fig.~\ref{fg:1124}) and in the optical from XN 0620
(Fig.~\ref{fg:0620}), while mini-outbursts have been seen only in the
optical from XN J0422 (Fig.~\ref{fg:0422}) and XN 1009
(Fig.~\ref{fg:1009}). XN J0422+32 provided the most dramatic example of
mini-outbursts approximately one year after its initial outburst
(\cite{swchn97}). Its R-band light curve is clearly divided into two phases
separated by a precipitous decrease of more than 4 magnitudes $\sim220$
days after the peak. It is remarkable that in both phases the baseline
decay rates are almost the same. Even more intriguing is the fact that the
mini-outbursts temporarily brought the flux back to a level consistent with
an extrapolation of the first-phase decay curve! Therefore, the occurrence
of the mini-outbursts may be causally related to the earlier flux
``free-fall''. 

Like other light curve properties, the secondary maxima are energy
dependent.  The glitch in the 1991 outburst of XN 1124--683 about 70 days
after the peak appears clearly in {\it Ginga}'s soft (1-5 keV) band, is
less apparent in the medium (5-9 keV) band, and is not seen at all in the
high energy ($>9$ keV) band (\cite{tmd91}; \cite{ekea94}). Therefore, it is
a soft X-ray feature, similar to the one seen in XN 0620--00 which has an
accompanying optical feature. On the other hand, the prominent glitch which
appeared in the 20--100 keV light curve of XN J0042+32 has no apparent
counterpart in the optical (Fig.~\ref{fg:0422}), suggesting that the
optical/UV glitches are decoupled from the hard X-ray glitches.
Furthermore, a glitch appeared in the UV light curves of both XN 1124--683
and XN J0422+32, at about 70 days and 45 days after the peak, respectively
(\cite{swhhs94}). It is striking that in both cases, the glitch is more
prominent in the 1300\AA\ than in the 2600\AA\ light curve, thus strongly
hinting at a common origin for the X-ray and UV behavior (\cite{swhhs94}).
We should also point out that while the UV glitch in XN 1124--683 coincides
with the one in soft X-rays, in XN J0422+32 there is no counterpart either
in the optical, although the data are noise limited, or in hard X-rays by
BATSE (20--100 keV), where there is at most a marginal deflection.

\subsubsection{Multi-peak light curves}
\label{sssc:p}

The multi-peak light curve occurs only in a few cases, and its puzzling
nature deserves some brief discussion in this section. First, XN J1655--40
is a truly unusual source, for its multi-peak light curve
(Fig.~\ref{fg:MP}), but more importantly for its superluminal jets. Like
the two long plateau sources XN 1716-249 and XN 1915+105 (the only other
Galactic superluminal jet source), XN 1655--40 has not been detected in the
last two decades and it became active only recently. In its 1994 outburst,
it underwent several consecutive outburst peaks with an interval of
$\sim120$ days (\cite{hbaea95}). Unlike XN 1915 or XN 1716, however, its
first peak is not the largest one in either flux or duration. Some of the
peaks are followed by radio flares but some are not (\cite{hr95}). The
light curves of the individual peaks are not FREDs but more or less
triangular with roughly equal rise and decay timescales. A similar
multi-peak light curve does not show in the 1996 outburst of XN 1655, which
is nevertheless characterized by an extremely slow and erratic rise and
decay.

The multi-peak light curve of XN 0042+32 in a weak (0.03 Crab) outburst in
1977 (\cite{wr78}), resembles the 1994 outburst of XN J1655--40
(Fig.~\ref{fg:MP}). Similarly, the light curves for individual peaks are
approximately triangular and the first peak is not the strongest. The only
apparent difference from XN 1655 is that the quasi-periodical peak
interval, in this case, is 11.6 days instead of 120 days. It is important
to note, however, that XN 0042 was detected in the 3--6 keV band, so that a
multi-peak light curve behavior occurs in both hard and soft X-rays. 

\section{DISTRIBUTION OF X-RAY NOVA SYSTEMS}
\label{sc:d}

\subsection{Frequency of XN Outbursts}
\label{ssc:rate}

Now we address the question: what is the overall rate of XN events in the
Galaxy? This is a nontrivial question because it depends on the sky
coverage, lifetime, limiting sensitivity and the energy band of X-ray
instruments launched at different times over the last 30 years. In our
attempt to quantify the sky coverage over this period of time, as a
fraction of the full sky over which an XN would be detected, a number of
simplifications were involved. First, we did not distinguish among
coverages in different energy bandpasses. {\it Compton}/BATSE for example,
even though with no sensitivity below 20 keV, has discovered three new XNs
and detected recurrent emission from several others during its initial 4
years of operation, while the peak XN emission is often thought to be in
the 2-10 keV band. Second, we have not attempted to calibrate the sky
coverage of various satellites according to their sensitivities, which
often differ considerably for the multiple experiments and/or multiple
modes a satellite may have operated on (e.g., HEAO 1 scanned during the
initial mission phase and was subsequently pointed). The limiting
sensitivities of the different instruments varied widely, but most were
able to detect transient sources to a level of $\sim0.1$ Crab. However, it
does not simply suffice for the peak intensity to reach (or slightly
exceed) this threshold for an event to be registered as an XN outburst.
Sufficient data are needed to allow a characterization of the transient,
thus a more realistic threshold is $\sim0.3-0.5$ Crab. It is also somewhat
subjective to attempt to identify what truly constitutes the detection
probability for a given source and instrument configuration and solid angle
coverage. Short-lived and relatively faint events could easily go
undetected by a state-of-the-art ASM like XTE if they happened to occur
within the nominal 10\% of the sky included in the solar or SAA zones of
avoidance. 

\begin{figure}
\psfig{figure=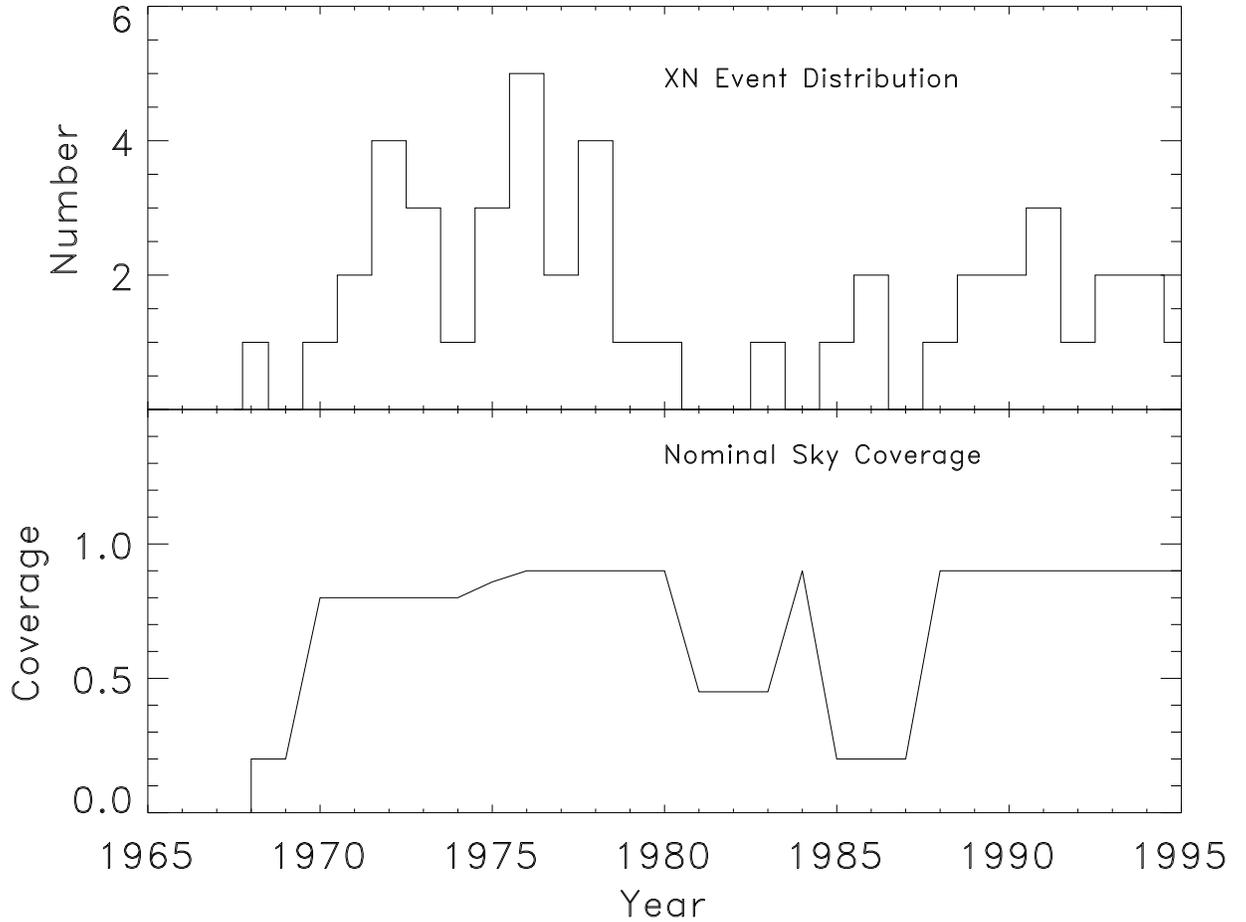}
\caption{ ({\it Upper panel}): Histogram of the XN
events per year from 1965 to 1995. ({\it Lower panel}): Nominal sky
coverage factor over the same period. Since we include only events which
lasted more than 10 days, any scanning instruments which can cover the
whole sky within a few days are considered to provide a sky coverage
fraction of close to unity. Notice the period around 1985 with very low sky
coverage and the corresponding low event rate. Notice also the low event
rate in early 1980's when the sky coverage was adequate. \label{fg:event} }
\end{figure}

Keeping all of these caveats in mind, we computed an approximate XN rate by
averaging over the nominal 30-year coverage baseline, and including {\em
all types} of events, BHC, NS, and sources of unknown nature. Not included
are the 11 ``Possible XNs'' in Table~\ref{tb:possible}, which would have
clearly shifted the estimated rates upwards. The upper panel of
Fig.~\ref{fg:event} is a histogram based on all the qualified 50 events.
The estimated sky-coverage factor is illustrated in the lower panel, this
was calculated by considering the instrumental capabilities and mission
lifetimes of 14 different satellites, as well as sounding rocket programs
of the late 1960's (Table~\ref{tb:mission}). In some cases, it was
necessary to apply some subjectivity; for example, we assigned sky-coverage
factors of $\sim10\%$ to scanning experiments such as {\it HEAO 1} and {\it
UHURU}.  All-sky monitors were assigned values of 80\%--90\%. On the basis
of all of the above, we estimate a rate of $\sim2.6$ XN outbursts per year.
The corresponding nominal rates for the NS and BH subgroups are $\sim1.1$
and $\sim1.5$ respectively. 

\begin{table}[tbp]
\dummytable\label{tb:mission}
\end{table}

\subsection{Recurrence Times and Properties of Recurrent Sources}
\label{ssc:recur}

Many XNs are known to have undergone more than one outburst. XN 0620--00
(XN Mon 1975) reached a similar optical brightness during an outburst in
1917 (\cite{ewl76}). XN 2023+338 (V404 Cyg) also erupted in 1938
(\cite{waa48}), 1956 and possibly 1979 (\cite{rga89}). The BHC XN 1630--472
has undergone quasi-periodic outbursts every $\sim600$ days since 1969
(\cite{paw95}). Thus, the reported recurrence time in different systems
varies from about 1 to 60 years.

\begin{figure}
\psfig{figure=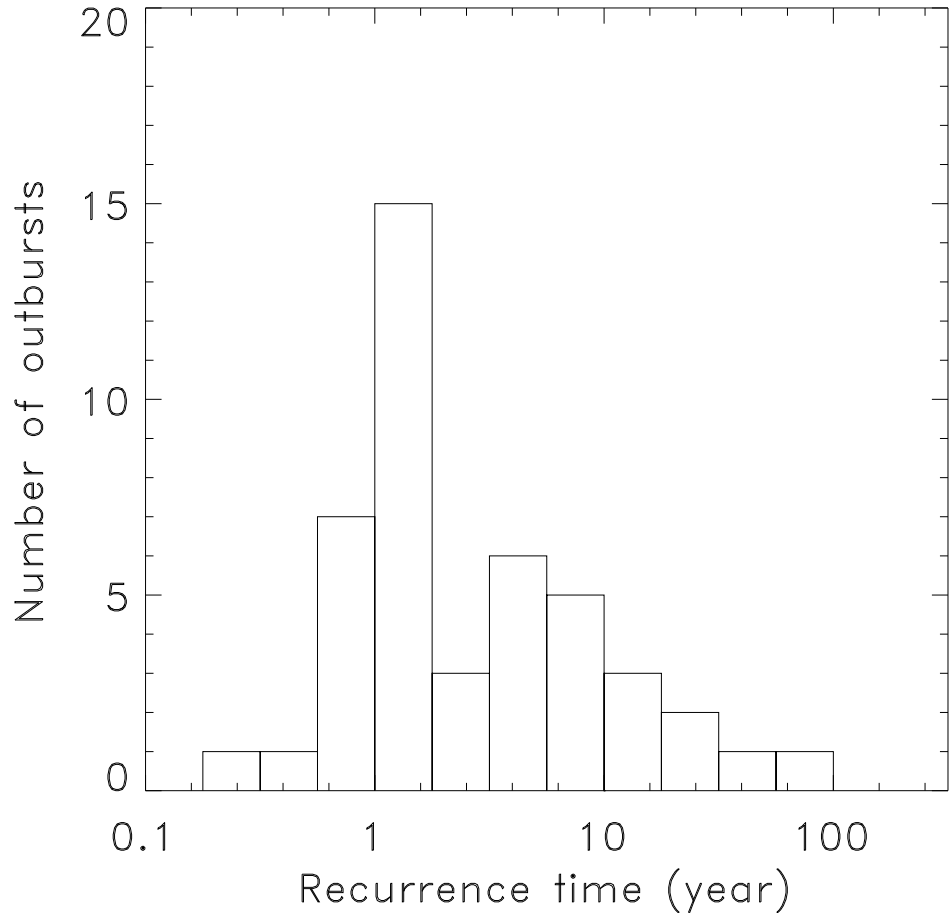}
\caption{ The distribution of the recurrence times
between outbursts. \label{fg:recur} }
\end{figure}

In Table~\ref{tb:b}, more than 50\% (13 of the 24) of the XN sources have
more than one outburst and 6 of them have at least 3 outbursts. If we take
all the recurrent events and ignore the possibility of missing events
between some of the recorded outbursts, we see that the maximum recurrence
time is 57.8 years (XN 0620) and the minimum is 0.25 years (XN 1608--522).
The mean recurrence time in logarithmic scale is $0.41\pm 0.54$ which
corresponds to an mean and 1-$\sigma$ (upper, lower) boundary for
$T_{\rm rec}$ of $2.6\, (8.9, 0.7)$ years. Fig.~\ref{fg:recur} shows the
distribution of the recurrence times. It has a sharp peak around 1 year and
a broad base. The true distribution is likely to be affected by two
factors. On one hand there must exist some missed outbursts between some of
the recorded events, which will increase the count of short recurrence
times. On the other hand, some of the sources which currently have only one
recorded outburst (and therefore are not included in Fig.~\ref{fg:recur})
may go into outburst in the future, thus increasing the count of long
recurrence times.

\begin{figure}
\psfig{figure=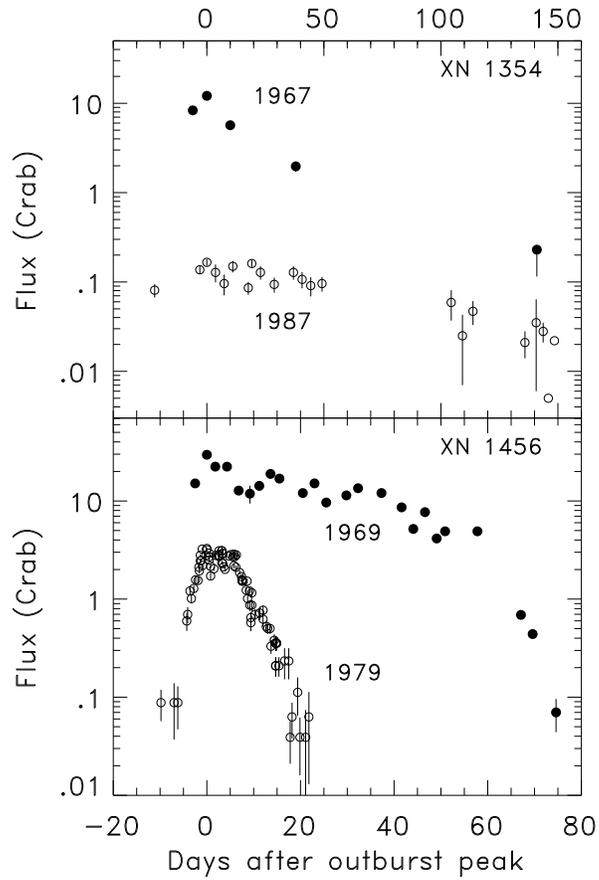}
\caption{ Comparison between the major and minor
outbursts in XN 1354--64 (Cen X-2, Chodil et al.\ 1968; Kitamoto et al.
1990) and XN 1456--32 (Cen X-4, Conner, Evans, \& Belian 1969; Evans,
Belian, \& Conner 1970; Kaluzienski et al.\ 1980). The two major outbursts
from the two sources are about the same peak intensity, but their two minor
ones differ by more than a factor of 10. Notice the changing of shapes and
durations from the major to the minor outbursts. Also notice that the time
span of the upper panel is 2 times that of the lower panel.
\label{fg:majmin} }
\end{figure}

For recurrent XNs, the peak flux, duration, and light curve morphology
often differ from outburst to outburst. The question is then: are there any
systematic differences between the strong and weak outbursts? For example,
in \S~\ref{sssc:decay} we have seen that the decay timescale $\tau_{\rm d}$
of the strong outbursts has, on average, a much narrower distribution
(around 30 days) than that of the weak outbursts. The question is, does
this property hold also for multiple outbursts from the same source? An
examination of Table~\ref{tb:b} for XN 1543--47, XN 1630--472 and XN
1908+005, reveals that this is not the case. For the two recurrent XNs in
the Centaurus region, XN 1354--64 (a BHC) and XN 1456--32 (a NS),
Fig.~\ref{fg:majmin} shows an opposite trend. The strong outburst of 24
Crab from Cen X-2 in 1967 (although its identification with XN 1354--64 is
only tentative) decayed twice as fast as the weak outburst in 1987 which
was about 100 times fainter. The strong outburst of $\sim30$ Crab of XN
1456--32 in 1969, on the other hand, decayed 5 times more slowly than the
weak outburst 10 years later which was 7 times fainter. Thus, it appears
that there is no correlation between the outburst magnitude, duration and
decay timescale.  We also do not find any clear correlation between the
recurrence time and the outburst luminosity.

\subsection{Spatial Distribution}
\label{ssc:spatial}

\begin{figure}
\psfig{figure=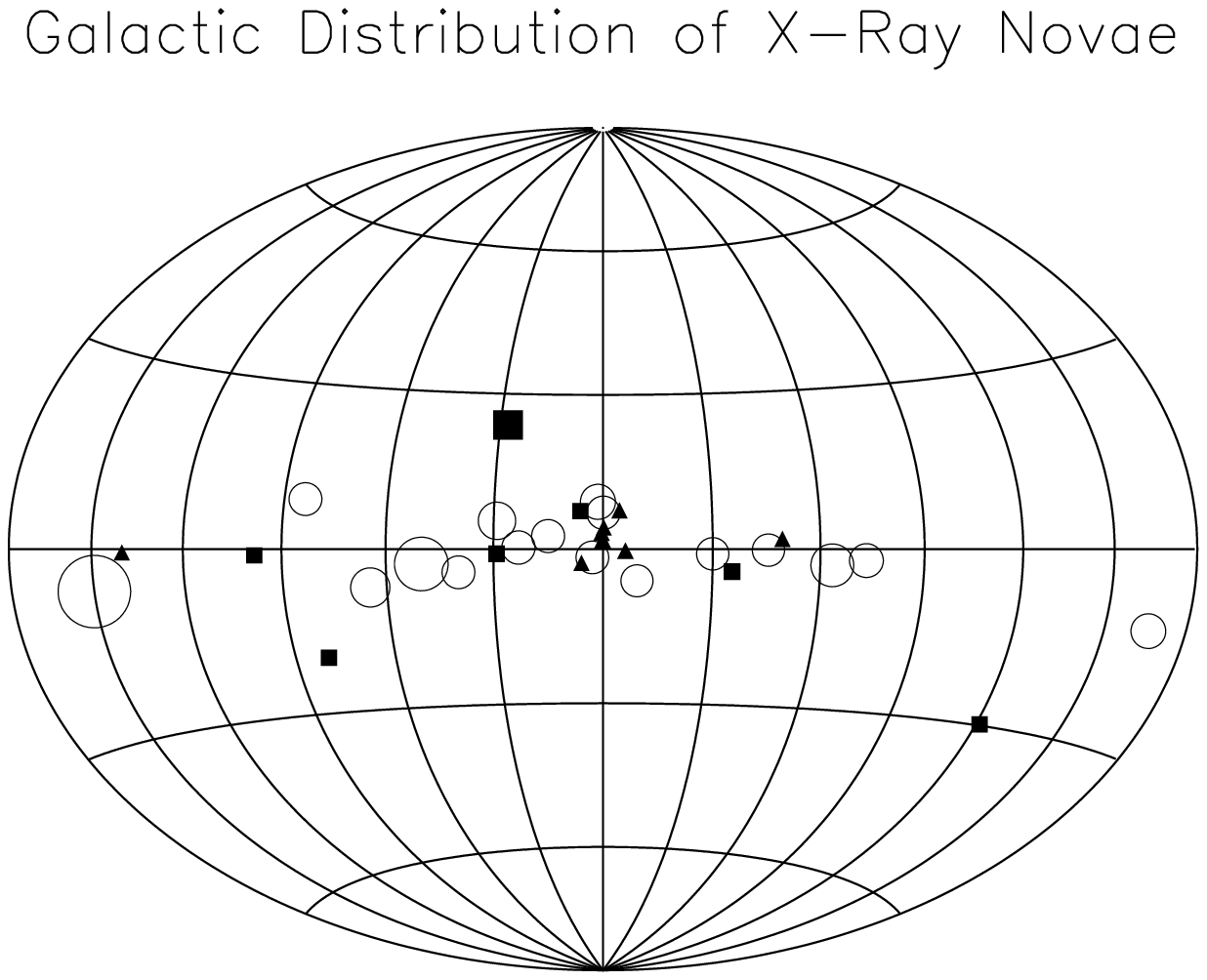}
\caption{ Distribution of X-ray nova systems in the
Galactic coordinate system. The symbols indicate the type of compact object
intrinsic to each system: {\em empty circles} are for positive and probable
BHCs, {\em filled squares} are for NS, and {\em filled triangles} are for
unknown types. Symbol sizes are proportional to XN peak intensities.
\label{fg:sky} }
\end{figure}

In Fig.~\ref{fg:sky} we have plotted the positions of 24 XNs on a
Galactic-Aitoff grid. The symbol sizes have been scaled in approximate
proportion to the peak intensities of the events. For multiple outbursts
from a single source, we have chosen the brightest outburst for
representation in this figure. The different symbols correspond to BH, NS,
and unknown systems as detailed in the figure caption. Clearly there is a
general concentration of sources along the Galactic plane, with an
additional enhancement associated with the bulge. Considering only the
probable neutron star systems, a more irregular distribution is seen with
several sources well out of the plane, but any conclusions are limited by
the small-number statistics. The BH systems are roughly uniformly
distributed along the plane, including several in the general anti-center
direction. The tendency for BHC systems to be less concentrated towards the
Galactic bulge, unlike the broader class of LMXBs which exhibit a distinct
concentration in the bulge region has been noted by White (1994) (see also 
\cite{wvp96}), who suggests that this may be indicative of a tendency for
BHC systems to be associated with a Population I distribution. One must be
cautious in accepting this interpretation however, since a large fraction
of the sources cannot be classified, and a number of the ones which are
assumed to be BHCs are at best tentatively classified. 

\begin{figure}
\psfig{figure=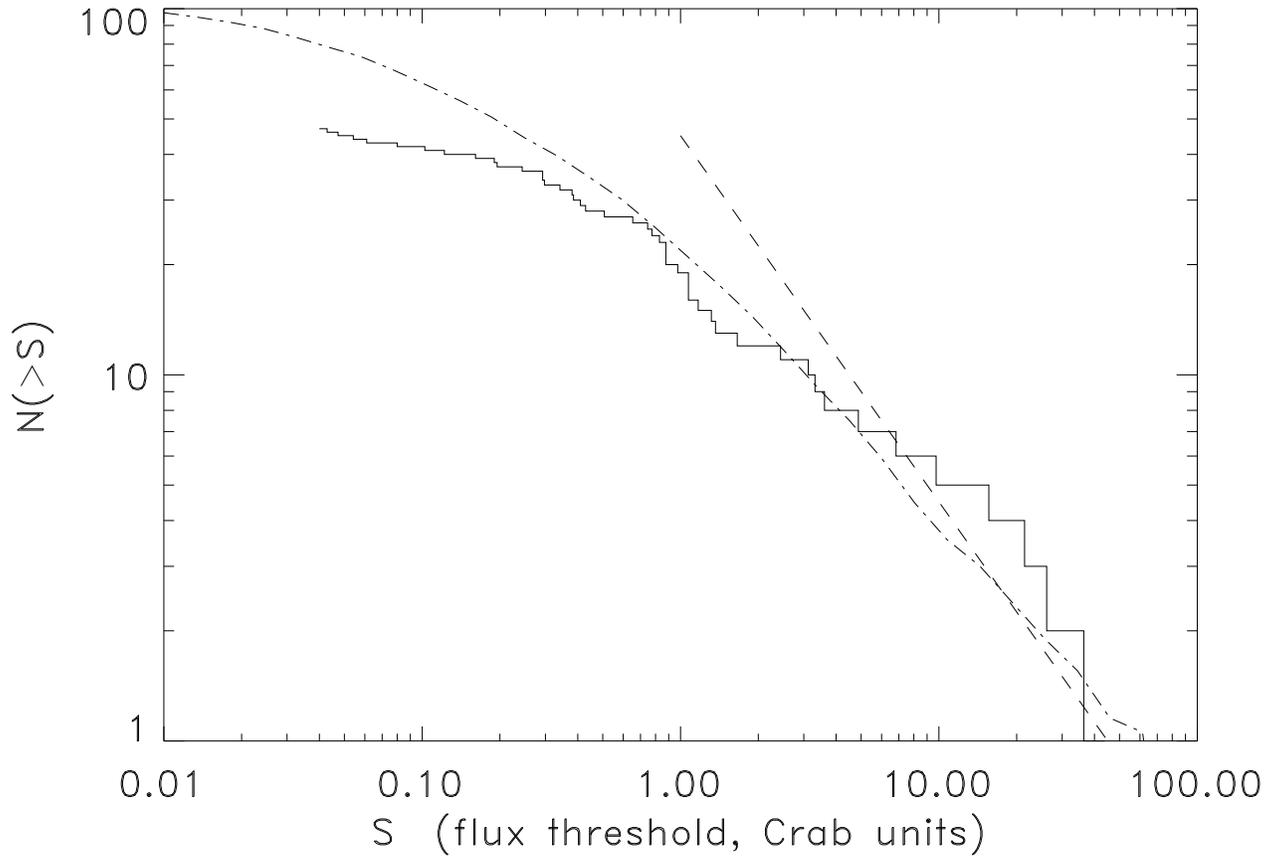}
\caption{ The $\log N - \log S$ distribution ({\it
solid line}) of the peak X-ray fluxes of all the XN outbursts. The {\it
dashed line\ } is for sources uniformly distributed on a disk of infinite
size, $N \propto S^{-1}$. The {\it dot-dashed line\ } is for a source
distribution model on a finite disk of 12 kpc in radius while the observer
is at 8.5 kpc from the center (see text). \label{fg:logNlogS} }
\end{figure}

The Galactic disk-bulge distribution of XNs can also be inferred from
the $\log(N)-\log(S)$ relationship of the XN outburst peak fluxes, shown as
a histogram in Fig.~\ref{fg:logNlogS}. We have treated recurrent outbursts
from the same sources as individual events in this analysis. The dashed
line shows a source distribution extending beyond the local range into a
disk of infinite size, so that it turns into a power law of $S^{-1}$ at
flux thresholds below $\sim30$ Crab. Intuitively, one might expect the
actual distribution of XN sources to fit such a distribution, but it is
clearly too steep when compared to the observations. We considered the case
in which the XN sources are distributed on the Galactic plane (12 kpc in
radius), and the observer is located at 8.5 kpc from the Galactic center;
this is represented by the dash-dotted line in Fig.~\ref{fg:logNlogS}. 

We can attempt to calculate $\log(N)-\log(S)$ numerically by making some
basic assumptions about the luminosity and spatial distributions. From
Table~\ref{tb:E}, we find that the range of X-ray luminosities among
events with distance estimates is $36.0 < \log L_{\rm x,p} < 39.3$, with
a mean value of 37.5. The distribution is approximately symmetric, rather
than exhibiting an excess towards the low-luminosity end. We attribute
this lack of asymmetry to detection inefficiency rather than to a real
physical effect, nonetheless we have approximated the distribution in
luminosity as a Gaussian. The spatial distribution of all XNs exhibits
an excess in the Galactic bulge region as noted above, although the BHC
subsample may be uniformly distributed over the disk. We have modeled this
spatial distribution as a constant surface density over the disk plus a
$r^{-3}$ power-law component to represent the apparent excess associated
with the bulge. We then performed a numerical integration over these
luminosity and space distributions, using again a geometry with the origin
offset from the disk center by 8.5 kpc. We derive a $\log N -\log S$ curve
which we can then scale, i.e., vary the mean surface density as a free
parameter, to approximately fit the observed $\log (N)-\log(S)$
distribution at the bright end, where it is presumably reasonably well
constrained by observations. The resulting distribution begins to flatten
at $10^{-2}$ Crab, and is flat by $10^{-3}$ Crab.

By constraining our model to overlay the bright portion of the observed
$\log N -\log S$ distribution and then extrapolating to a flux threshold of
$\sim 10^{-2}$ where it becomes flat, we can estimate the total number of
XNs in the galaxy. In this manner we derived a mean surface density of
about 0.25 kpc$^{-2}$ from which we estimate that there are a few hundred
XNs in the galaxy. This estimate is a lower limit in the sense that the
true mean recurrence time scale may be much larger than what we infer from
the observed subsample as discussed previously. This situation may not be
improved upon observationally for decades, although theory of binary star
formation and evolution holds promise for the nearer term.

\section{FROM THE LIGHT CURVES TO THE PHYSICS OF OUTBURSTS}
\label{sc:model}

In this section, we discuss briefly some of the theoretical implications of
the properties of XN light curves we have found so far. In \S~\ref{sc:m},
we saw that many XN light curves are FREDs or possible FREDs, and most
other light curve morphologies are of this basic profile but distorted or
stretched in a variety of ways. The question is, therefore, is there a
common underlying outburst mechanism which is responsible for at least the
majority of the outbursts? And, if so, what constraints do the statistical
properties and light curve features discussed in \S~\ref{sc:main} place on
such a model? Finally, if discrepancies exist, where should we look for
possible solutions?

There were originally two basic outburst mechanisms which have been
suggested. One is the mass transfer instability (MTI) model which
attributes the outbursts to a temporary increase in the mass transfer rate
from the companion as a result of the continuous X-ray heating of the
companion's surface in quiescence (e.g., \cite{hkl86}, 1990). The other is
the disk thermal instability model (DTI) which is based on the existence of
two thermal states at a given disk
surface density (due to an opacity jump around the hydrogen ionization
temperature, e.g., \cite{cgw82,flp83,mmh84,hw89,mw89}). More recently it
has been pointed out, however, that the MTI model has some severe
shortcomings which make it inapplicable to the XN outbursts (e.g.,
\cite{gh93,myi93}). The DTI model (also called the limit cycle model), on
the other hand, has been very successful in explaining dwarf nova outbursts
(e.g., \cite{cjk93} and references therein). Recently, Cannizzo, Chen, \&
Livio (1995, hereafter CCL95) embarked on a series of calculations that
enabled to clarify several important properties of limit cycles experienced
by disks around BHs. We will thus focus only on the DTI model in our
following discussions of each of the observed properties. A related model
for the accretion flow in BH systems that has attracted much attention
recently is the advection dominated accretion flow (ADAF) model, which
appears to be able to successfully explain the quiescent XN spectra and
luminosities (e.g., \cite{acklr95,nmy95}). ADAF models can be incorporated
into DTI models for the outburst, as we discuss below.

{\em Universal exponential decay and its timescale.\ } An important result
of CCL95 is that the DTI model cannot produce exponential decays if the
viscosity parameter $\alpha$ is taken either as a step function between the
two stable branches (\cite{hw89,mw89}) or as a function of radius of the
form $r^\epsilon$ (\cite{cjk93}). Robust exponential decays can be
reproduced {\it only} if $\alpha$ is a function of the local aspect ratio
$h/r$, of the form $\alpha = \alpha_0 (h/r)^n$ with $n=1.5$. Here $h$ is
the disk scale height at a given annulus of radius $r$, and $\alpha_0$ is a
constant. This result is the direct consequence of a more fundamental
finding that the width of the transition front in the disk, which
transforms the hot state into the cold state and vice versa, is the
geometrical mean of $h$ and $r$ (CCL95; \cite{vw96}). Therefore, the
universality of exponential decay is indeed the result of some fundamental
physical processes in the disk. CCL95 have further shown that the decay
timescale is proportional to $M/\alpha_0$ where $M$ is the mass of the
accreting object. This is interesting because under the assumption that
$\alpha_0$ is more or less universal (e.g., if it is determined by a MHD
dynamo in the disk), then this seems to be able to explain why the dwarf
novae usually have a decay constant of only about 3 days while the BHXNs
have a decay constant of $\sim 30$ days. It is quite clear, however, that
this relation is not precise, since the NSXNs have similar decay timescales
as the BHXNs even though the mass of NSs is typically smaller than that
of BHs by a factor of a few. 

{\em Fast rise\ }. In the disk instability model, an outburst can be
triggered either ``outside-in'' or ``inside-out'', depending on the history
of the last outburst and the mass accretion rate through the disk in
quiescence. Cannizzo (1996) found that the fast rises can be obtained only
if the outbursts are triggered at radii larger than about $10^{10}$ cm from
the BH (see also \cite{cwp86}). On the other hand, the outbursts are
always triggered near the inner edge of the disk ($\sim10^7$ cm for
standard disks in XN systems) if the mass transfer rate to {\it the inner
edge} is greater than about $10^{-19}\Msun$ yr$^{-1}$ (CCL95; see also
\cite{lny96}). This problem may be circumvented if we employ the ADAF model
for the quiescent state of the system. In this model, the inner part of the
accretion flow (of dimensions of the order of $10^4$ gravitational radii)
is nearly spherical and optically thin, and it joins at a transition
radius, $r_{\rm tr}$, to an outer standard accretion disk (e.g., Narayan et
al.\ 1996, 1997). Under these conditions, even outbursts that are triggered
at the inner edge of the standard disk (at $r_{\rm tr} \sim 10^{10}$ cm)
are in fact started at large radii, thus might be able to produce a fast
rise in X-rays.

A relative weakness of the ADAF model is the lack of a fully convincing
physical mechanism to remove the inner part of the disk at the end of the
outbursts and/or during quiescence. Therefore, the exact location of the
transition radius (the inner edge of the quiescent disk) is not determined
by the theory (although it can be constrained by the requirement of
stability). However, this may reflect the real situation in the sense that
the uncertainty in the location of the inner edge of the quiescent disk may
naturally lead to a broad distribution of the rise timescales in outbursts,
similar to what we see in Fig.~\ref{fg:rise}. An important corollary is
that the existence of outbursts with a very slow rise may imply that there
is no ADAF in operation in the quiescent state of these sources. We also
note that because of the requirement for the standard disk to first fill in
(by diffusion) the central parts occupied by the ADAF at quiescence, one
would expect naturally a delay in the rise of the X-ray flux, relatively to
the rise in the optical such as the one observed in XN 1655--40 (\cite{orbm97}).
Similar models, including a hole at the center of the disk to explain the
UV delay, have been proposed for dwarf novae (e.g., \cite{lp92}).

{\em The diversity of XN light curve morphologies.\ } It is presently
unclear if the variety of light curves observed in XN outbursts, such as,
plateaus and triangular profiles, can all be explained by a disk
instability model.  Nevertheless, it should be noted that a wide range of
outburst behaviors in dwarf novae have been successfully modeled by a disk
instability (\cite{cjk93}), and that the light curves that were obtained do
include both plateau and triangular outbursts.

Naively, one might think that multi-peak outbursts may be due to a
different physical mechanism than that responsible for the "normal" XN
outbursts. However, a strong piece of evidence against this notion may be
the fact that these and most other XNs included in our study all have spent
a long time in quiescence before their first eruption, which suggests that
the physical mechanism which keeps these sources dormant may be universal.
It is probably due to the external conditions, such as the binary
parameters, the nature of the secondary mass-donor star, and especially its
response to the X-ray irradiation during outburst, which are responsible
for the diverse nature of the XN light curve morphologies and features. 

{\em Secondary maxima.} There has been so far no single satisfactory
mechanism to explain the observed variety of secondary maxima. Chen, Livio,
\& Gehrels (1993) were the first to suggest that this phenomenon represents
an intrinsic physical process which characterizes the outbursts. They
proposed that the glitches and bumps are caused by mass transfer events
from the companion star, in response to the X-ray heating during the main
outburst. They correctly predicted the short orbital period of XN 0422+32
from its early onset of the glitch observed in the UV (Fig.~\ref{fg:UV}).
But their model cannot explain the series of mini-outbursts observed in XN
0422 (Fig.~\ref{fg:0422}). A similar model was suggested by Augusteijn,
Kuulkers, \& Shaham (1993), with a mathematical formulation for the
``echoing'' of the initial outburst. The latter model has been successful
in predicting the equal spacings between the onset of the mini-outbursts in
XN 0422, but it has failed to account for the large amplitude of these
events. Mineshige (1994), on the other hand, showed that the secondary
maxima could in principle be caused by a thermal instability in the disk,
if the disk is moderately irradiated by X-rays, although there is no
mechanism for the distinction between glitches, bumps and mini-outbursts,
nor for the time delay between these events and the main peak. Recently,
Kim et al.\ (see \cite{wjc96}) proposed that the secondary maxima are
due to the ``stagnation'' effect in an indirectly irradiated disk.

While none of the above models can give satisfactory explanations for all
the observed features of the secondary maxima, it is worth noting that all
the models involve X-ray irradiation of either the companion star or the
accretion disk. Therefore, a better understanding of the relationship
between the disk X-ray emission and the response of the disk and companion
star to irradiation is clearly needed.

One should also keep in mind that the bump shown in the 1--37 keV light
curve of XN 1124-683 (Fig.~\ref{fg:1124}) is completely due to an X-ray
spectral transition when the rise in the 9--37 keV flux starts to dominate
the total flux. This apparently has nothing to do with any of the models
just mentioned. The bump shown in the 3-6 keV light curve of XN 0620--00
(Fig.~\ref{fg:0620}), on the other hand, is probably the result of some
kind of X-ray irradiation because there is also a similar optical feature
(Fig.~\ref{fg:0620}) which appeared several days before the X-ray bump
(\cite{clg93}).

{\em Recurrence times.\ } In general, the DTI model predicts that the total
accreted mass during an outburst, $\Delta M$, is related to the outburst
recurrence time, $T_{\rm rec}$ and the mass overflow rate from the
companion star, \Mdot$_{\rm c}$, through a simple expression,
\begin{equation}
\dot M_{\rm c} \sim \frac{ \Delta M }{ T_{\rm rec} }.
\end{equation}
Therefore, using $\Delta M$ in Table~\ref{tb:dur} and $T_{\rm rec}$
inferred from Table~\ref{tb:b}, we can derive $\dot M_{\rm c}$ for
sources with sufficient data. The results are listed in Table~\ref{tb:mdc}.
These values agree reasonably well with those derived by van Paradijs
(1996). For comparison, for sources with known binary period and companion
mass, we also list the theoretical mass transfer rates predicted by King et
al.\ (1996) in their binary evolution theory. We see that in all the cases
except XN 1456--32 the two values agree reasonably well. An important point is
that according to van Paradijs (1996), for a given binary period the mass
transfer rate from the companion has to be smaller than some critical value
for the source to be a transient. The large mass transfer rate (partly due
to an uncertainty in its distance) inferred for XN 1630--472 thus probably
indicates a very unusual system with a very large orbital period
($>12$ days).

\begin{table}[tbp]
\dummytable\label{tb:mdc}
\end{table}

\section{SUMMARY AND CONCLUSIONS}
\label{sc:s}

In this work, we studied the properties of X-ray and optical light curves
of X-ray novae. Our collection of XN light curves, although not complete,
demonstrates both the striking universality and the rich variability of the
long term time profiles of the XN outbursts. First, we have attempted to
classify the morphologies observed in the light curves, and in particular
identified the following general classes: (1) fast rise and exponential
decay (FRED); (2) triangulars; (3) plateaus (long and short); (4) variable
decay, and (5) multi-peaks. Only very few outbursts cannot be classified in
any of these categories. In addition to the global morphologies, we have
identified other light curve features, such as precursors and secondary
maxima (the latter of which include ``glitches'', ``bumps'', and
mini-outbursts). 

We have presented the distributions of (1) rise timescales, (2) peak
luminosities, (3) outburst amplitudes, (4) decay timescales, and (5) total
energy radiated during an outburst. From these distributions we see that
the rise timescales have a much broader distribution than the decay
timescales. The latter are distributed in a narrow range around 30 days for
the brightest outbursts, although the distribution has a broad base for
fainter outbursts. We have also found that the distribution of peak
luminosities in Eddington units has a narrow Gaussian profile, centered
around $0.2 L_{\rm Edd}$. The outburst amplitude, which is defined as the
ratio between the peak luminosity and the quiescent luminosity in the same
energy band (0.4--10 keV in our case), was found to have a broad
distribution. Contrary to a recent claim by Narayan, Garcia, \& McClintock
(1997), there is significant overlap between the distribution of the
amplitudes of the BHXNs and those of the NSXNs, rather than a clear
division. While this in itself should not be taken as evidence against the
existence of a BH horizon, this quantity is clearly not the best parameter
to demonstrate this effect.

A Log(N) -- Log(S) distribution based on all of the events in our study is
consistent with a uniform distribution over the Galactic disk plus a
power-law component associated with the apparent excess in the Galactic
bulge region, as viewed from a vantage point displaced 8.5 kpc from the
center. Extrapolation to $\sim$ mCrab sensitivity threshold levels suggests
a few times $10^{2}$ XN in the Galaxy -- however, this is a lower limit
since we may be sampling only the short end of the recurrence-time
distribution.

We have argued that the basic properties of the light curves, and in
particular the universal exponential decay and the narrow range of its
decay constant can be explained in terms of the disk thermal instability
model. A few of the properties (e.g., the broad rise timescale
distribution) suggest that the inner part of the accretion flow in
quiescence may be represented by an ADAF, joining at some transition radius
to an outer standard disk. A full theoretical understanding of additional
light curve features, such as precursor and secondary maxima will probably
require further study of a variety of feedback processes. It seems very
plausible, for example, that the secondary maxima are related to X-ray
irradiation effects, either of the secondary star or of the accretion disk.
Some of the smaller ``bumps'' may simply represent the effects of spectral
evolution at certain phases of the outburst.

Finally, we would like to note that future, more sensitive X-ray sky
monitors will be able to both discover XN outbursts earlier, and follow
them for longer periods of time. This will certainly lead to an improved
characterization of the various light curve features and eventually to more
meaningful constraints on theoretical models.

\acknowledgements{We sincerely thank John Cannizzo, Josh Grindlay, Jan
van Paradijs, Jeff McClintock, Jerome Orosz, Craig Wheeler for informative
conversations and comments on the manuscript, Paul Callanan, Eugene
Churazov, Ken Ebisawa, Marat Gilfanov, Alan Harmon, Louis Kaluzienski,
Shunji Kitamoto, Jim Lochner, Jerome Orosz, William Paciesas, Kentaro
Terada, and Mark Wagner for providing the original light curve data, and
David Palmer for providing the freeware DataThief by Kees Huyser \& Jan
van der Laan which greatly eased the pain of digitizing data from
publications. We are grateful to the referee for pointing out some errors
in the original manuscript and for helping us to improve the presentation
of the paper. ML acknowledges support from NASA grants NAGW-2678 and GO
4377. This research has made use of data obtained through the High Energy
Astrophysics Science Research Archive Center (HEASARC) Online Services,
provided by the NASA-Goddard Space Flight Center, and the Simbad database,
operated at CDS, Strasbourg, France.}

\newpage

\clearpage

\clearpage

\setcounter{table}{0}

\begin{deluxetable}{c@{}llr@{}lr@{}lr@{}llr@{}ll}
\tablenum{1}
\tablewidth{0pt}
\tablecaption{Black Hole X-Ray Nova Systems}
\tablehead{
\multicolumn{2}{c}{Source} & \colhead{X-ray Nova}  & \multicolumn{2}{c}{$f_M$}   & \multicolumn{2}{c}{$i$}
 & \multicolumn{2}{c}{$M_{\rm BH}$} & \colhead{Star} & \multicolumn{2}{c}{$M_{\rm c}$} & \colhead{Ref.} \\ 
\multicolumn{2}{c}{Name} & \colhead{Designation} & \multicolumn{2}{c}{(\Msun)} & \multicolumn{2}{c}{(deg)}
 & \multicolumn{2}{c}{(\Msun)}      & \colhead{Sp}   & \multicolumn{2}{c}{(\Msun)}     &      \\
\multicolumn{2}{c}{(1)}  & \colhead{(2)}   & \multicolumn{2}{c}{(3)}  & \multicolumn{2}{c}{(4)}  
 & \multicolumn{2}{c}{(5)}         & \colhead{(6)}  & \multicolumn{2}{c}{(7)}     & \colhead{(8)} 
}
\startdata
A     & 0620--00  & XN Mon 1975 & 2.91 & $\pm0.08$ & $66^\circ$ & $\pm4^\circ$  & 4.9  & $-$10    & K4 V  & 0.4 & $-$0.7 & 1--3 \nl
H     & 1705--250 & XN Oph 1977 & 4.0  & $\pm0.8$  & $70^\circ$ & $\pm10^\circ$ & 4.9  & $\pm1.3$ & K3 V  & 0.7 &        & 4--6 \nl 
GS    & 2000+25   & XN Vul 1988 & 4.97 & $\pm0.10$ & $65^\circ$ & $\pm9^\circ$  & 8.5  & $\pm1.5$ & K2-K7 V & 0.4 & $-$0.9 & 7 \nl
GS    & 2023+338  & XN Cyg 1989 & 6.08 & $\pm0.06$ & $56^\circ$ & $\pm2^\circ$  & 12.3 & $\pm0.3$ & K0 IV & 0.9 &        & 8--12 \nl
GRS   & 1124--683 & XN Mus 1991 & 3.01 & $\pm0.15$ & $54^\circ$ & $-65^\circ$   & 5.0  & $-$7.5   & K2 V  & 0.8 &        & 13 \nl
GRO J & 0422+32   & XN Per 1992 & 1.21 & $\pm0.06$ & $48^\circ$ & $\pm3^\circ$  & 3.57 & $\pm0.34$ & M2 V & 0.4 &        & 14--16 \nl
GRO J & 1655--40  & XN Sco 1994 & 3.16 & $\pm0.15$ & $69^\circ.5$ & $\pm0^\circ.08$ & 7.02 & $\pm0.22$ & F3-F6 IV & 1.2 & $-$1.5 & 17 \nl 
\enddata
\tablecomments{Col.~(3) Mass function; Col.~(4) Orbital inclination angle; Col.~(5) Mass of the
black hole; Col.~(6) Spectral type of the secondary star; Col.~(7) Mass of the secondary star}
\tablerefs{(1) McClintock \& Remillard 1986; (2) Haswell et al.\ 1994; (3) Shahbaz, Naylor, \& 
 Charles 1994; (4) Remillard et al.\ 1994; (5) Martin et al.\ 1995; (6) Remillard et al.\ 1996; 
 (7) Callanan et al.\ 1996; (8) Casares, Charles, \& Naylor 1992; (9) Wagner et al.\ 1992; (10) 
 Casares \& Charles 1994; (11) Shahbaz et al.\ 1994; (12) Sanwal et al.\ 1996; (13) Orosz et
 al.\ 1996; (14) Orosz \& Bailyn 1995; (15) Casares et al. 1995; (16) Filippenko, Matheson, \&
 Ho 1995; (17) Orosz \& Bailyn 1997.}
\end{deluxetable}

\begin{deluxetable}{r@{\hspace{1pt}}r@{}lllll}
\tablenum{2}
\tablewidth{0pt}
\tablecaption{X-ray Nova Sources}
\tablehead{
\multicolumn{3}{c}{Source} & \colhead{XN} & \colhead{X-ray} & \colhead{Variable} &
\colhead{Other} \\
 & & & \colhead{Designation} & \colhead{Source} & \colhead{Star} & \colhead{Name(s)}
}
\startdata
3U  &   & 0042+32   &             &           &           & 2A 0042+323 \nl
GRO & J & 0422+32   & XN Per 1992 &           & V518 Per  & GRS 0417+335 \nl
A   &   & 0620--00  & XN Mon 1975 & Mon X-1   & V616 Mon  & Nova Mon 1917 \nl
EXO &   & 0748--676 &             &           & UY Vol    & 2E 0748.4--6737 \nl
MX  &   & 0836--425 &             &           &           & 4U 0836--42, GS 0836--429 \nl
GRS &   & 1009--45  & XN Vel 1993 & \nl
GS  &   & 1124--683 & XN Mus 1991 &           & GU Mus    & GRS 1121--68 \nl
GS  &   & 1354--64  & XN Cen 1967 & Cen X-2\tnm{a} &      & MX 1353--644 \nl
4U  &   & 1456--32  & XN Cen 1969 & Cen X-4   & V822 Cen  & 2E 1455.2--3127 \nl
A   &   & 1524--62  &             & TrA X-1   & KY TrA    \nl
4U  &   & 1543--47  &             &           &           & MX 1543--475 \nl
4U  &   & 1608--522 &             & GX 331--1 & QX Nor    & 1ES 1609--52.2 \nl
4U  &   & 1630--472 &             & Nor X-1   &           & GX 337+00, A 1630--472 \nl
GRO & J & 1655--40  & XN Sco 1994 & \nl
H   &   & 1705--250 & XN Oph 1977 &           & V2107 Oph & A 1705--250 \nl
GRS &   & 1716--249 & XN Oph 1993 &           &           & GRO J1719--24 \nl
4U  &   & 1730--220 & \nl
A   &   & 1742--289 &             & GX .2--.2 &           & A 1743--288 \nl
EXO &   & 1846--031 & \nl
4U  &   & 1908+005  &             & Aql X-1   & V1333 Aql & A 1908+005 \nl
GRS &   & 1915+105  & \nl
4U  &   & 1918+146  &             &           &           & A 1918+14, H 1922+154 \nl
GS  &   & 2000+25   & XN Vul 1988 &           & QZ Vul    \nl
GS  &   & 2023+338  & XN Cyg 1989 &           & V404 Cyg  & Nova Cyg 1938 \nl
\enddata
\tablenotetext{a}{Identification of GS 1354--64 with Cen X-2 is uncertain (Kitamoto
et al.\ 1990).}
\end{deluxetable}

\begin{deluxetable}{r@{\hspace{1pt}}llll}
\tablenum{3}
\tablewidth{0pt}
\tablecaption{Possible X-ray Nova Sources}
\tablehead{
\multicolumn{2}{c}{Source} & \colhead{Other Name(s)} & \colhead{Note}
}
\startdata
MX  & 0656--072 & 3A 0656--072                 & \nl
4U  & 1516--56  & Cir X-1, BR Cir              & A high-mass system? \nl
A   & 1658--298 & V2134 Oph, H 1658--298, MX 1659--29 \nl
4U  & 1659--487 & GX 339--4, V821 Ara, A 1659--487 \nl
MX  & 1730--335 & Rapid Burster, 4U 1730--333 & \nl
KS  & 1731--260 & IRAS 17311--2604            & \nl
KS  & 1732--273 & \nl
GS  & 1734--275 & \nl
4U  & 1735--28  & GX 359+2, H 1735--285 \nl
X   & 1740--294 & GC X-4                       & Not in SIMBAD database. \nl
GRS & 1741--288 & \nl
KS  & 1741--293 & MXB 1743--29                & \nl
H   & 1741--322 & \nl
X   & 1742--294 & GC X-2                       & Not in SIMBAD database. \nl
A   & 1744--36  & \nl
MX  & 1746--203 & 4U 1743--19, H 1745--203 \nl
EXO & 1747--214 & GPS 1747--212               & \nl
A   & 1749--285 & GX+1.1,--1.0                & Not in SIMBAD database. \nl
MX  & 1803--245 & \nl
GS  & 1826--238 & \nl
\enddata
\end{deluxetable}

\begin{deluxetable}{r@{}ll@{}llr@{  }r@{ }rll@{}l@{}l@{}l}
\tablenum{4}
\tablewidth{0pt}
\tablecaption{General Information of X-Ray Nova Systems\tnm{a}}
\tablehead{
\multicolumn{2}{c}{Source}  & \colhead{Type\tnm{b}} & \colhead{RA} & \colhead{Dec} & \colhead{$l$}
 & \colhead{$b$} & \colhead{$D$} & \colhead{Star} & \colhead{$m_V$} & \colhead{$A_V$}
 & \colhead{$P_{\rm orbit}$} & \colhead{Ref.} \\
\colhead{} & \colhead{} & \colhead{} & \colhead{(2000)} & \colhead{(2000)} & {(deg)}
 & {(deg)} & {(kpc)} & \colhead{Sp} & {(mag)} & {(mag)}
 & {(day)} \\
\multicolumn{2}{c}{(1)} & \colhead{(2)} & \colhead{(3)} & \colhead{(4)} & \colhead{(5)}
 & \colhead{(6)} & \colhead{(7)} & \colhead{(8)} & \colhead{(9)} & \colhead{(10)}
 & \colhead{(11)} & \colhead{(12)} 
}
\startdata
  & 0042+32   & BH? & 00 44 50.4  & +33 01 17     & 121.3 & $-$29.8 & 7?   & G?    & 19.29 & 0.6  & 11.6  & 1,2 \nl
J & 0422+32   & BH  & 04 21 42.75 & +32 54 26.8   & 165.9 & $-$11.9 & 2.2  & M0 V  & 22.24 & 1.2  & 0.212 & 1,2 \nl
  & 0620--00  & BH  & 06 22 44.51 & $-$00 20 44.5 & 210.0 & $-$6.5  & 0.87 & K4 V  & 18.4  & 1.2  & 0.323 & 1,2 \nl
  & 0748--676 & NS  & 07 48 33.8  & $-$67 45 08.6 & 280.0 & $-$19.8 & 2.1  &       & $>$23 & 1.26 & 0.159 & 1,2 \nl
  & 0836--429 & NS  & 08 37 23    & $-$42 53.1    & 261.9 & $-$1.1  & 10?  &       & $>$23 & 11   &       & 1,2 \nl
  & 1009--45  & BH? & 10 11 32    & $-$44 49 41   & 276.2 & +9.0    & 3?   &       & $>$21 & 5    &       & 3,12 \nl
  & 1124--683 & BH  & 11 26 26.64 & $-$68 40 32.5 & 295.3 & $-$7.1  & 5.5  & K5 V  & 20.5  & 0.87 & 0.433 & 1,9 \nl
  & 1354--64  & BH? & 13 58 09.68 & $-$64 44 04.9 & 310.0 & $-$2.8  & 10?  &       & 22?   & 3    &       & 1,2 \nl
  & 1456--32  & NS  & 14 58 22.0  & $-$31 40 07   & 332.2 & +23.9   & 1.2  & K3 V  & 18.3  & 0.3  & 0.629 & 1,2 \nl
  & 1524--62  & BH? & 15 28 17.1  & $-$61 52 58   & 320.3 & $-$4.4  & 4.4  &       & $>$21 & 2.4  &       & 1,2 \nl
  & 1543--47  & BH? & 15 47 08.5  & $-$47 40 10   & 330.9 & +5.4    & 4    & A2 V? & 16.7? & 2.1  & 0.6?  & 1,2 \nl
  & 1608--522 & NS  & 16 12 42.9  & $-$52 25 23   & 330.9 & $-$0.9  & 3.3  &       & $>$20 & 5.2  &       & 1,2 \nl
  & 1630--472 & BH? & 16 34 00.4  & $-$47 23 39   & 336.9 & +0.3    & 10?  &       &       & 27   &       & 1,2 \nl
J & 1655--41  & BH  & 16 54 00.00 & $-$39 50 44.0 & 345.0 & +2.5    & 3.2  & F5-G2 & 17.3  & 3.45 & 2.62  & 4,5,11 \nl
  & 1705--250 & BH  & 17 08 14.5  & $-$25 05 29   & 358.6 & +9.1    & 4.3  & K3 V  & 21.3  & 4.5  & 0.521 & 1,2 \nl
  & 1716--249 & BH? & 17 19 36.87 & $-$24 01 03.4 & 0.1   & +7.0    & 2.4  &       & $>$21 & 2.4  &       & 6   \nl
  & 1730--220 & NS? & 17 33 56.5  & $-$22 02 07   & 4.5   & +5.9    &      &       &       &      &       & 1,2 \nl
  & 1742--289 & NS  & 17 45 37.0  & $-$29 01 07   & 359.9 & $-$0.0  & 8.5  &       &       & 100  &       & 1,2 \nl
  & 1846--031 & BH? & 18 49 17.01 & $-$03 03 43.1 & 30.0  & $-$0.9  & 7?   &       &       & 18   &       & 1,2 \nl
  & 1908+005  & NS  & 19 11 15.95 & +00 35 06     & 35.4  & $-$4.3  & 2.5  & K0 V  & 19.2  & 1.2  & 0.792 & 1,2 \nl
  & 1915+105  & BH? & 19 15 11.49 & +10 56 44.9   & 45.4  & $-$0.2  & 12.5 &       & $>$19 & 25   &       & 7,8 \nl
  & 1918+146  & BH? & 19 20 17    & +14 42 19     & 49.3  & +0.4    &      &       &       &      &       & 1,2 \nl
  & 2000+25   & BH  & 20 02 49.52 & +25 14 11.3   & 63.4  & $-$3.0  & 2    & K5 V  & 21.2  & 5.4  & 0.344 & 1,10 \nl
  & 2023+338  & BH  & 20 24 03.8  & +33 52 04     & 73.1  & $-$2.1  & 3.5  & K0 IV & 19    & 3.0  & 6.46  & 1 \nl
\enddata
\tablenotetext{a}{Most data are from vP95 which also contains exhaustive references for each source.}
\tablenotetext{b}{We use ``BH'' or ``NS'' for firm BH or NS systems, ``BH?'' or ``NS?'' for BHC or NS
 candidates.}
\tablerefs{(1) van Paradijs 1995; (2) Bradt \& McClintock 1983; (3) Bailyn \& Orosz 1995; (4) Bailyn
 et al.\ 1995; (5) Harmon et al.\ 1995; (6) Della Valle, Mirabel, \& Rodriguez 1994; (7) Mirabel et
 al.\ 1994; (8) Mirabel \& Rodriguez 1994; (9) Orosz et al.\ 1996; (10) Callanan et al.\ 1996; (11)
 Orosz \& Bailyn 1997; (12) Della Valle \& Benetti 1993.
}
\end{deluxetable}

\begin{deluxetable}{lllrcl}
\tablenum{5}
\tablewidth{0pt}
\tablecaption{Morphological Types of X-Ray Light Curves}
\tablehead{
\multicolumn{2}{c}{Type} & \colhead{Code} & \multicolumn{2}{c}{Number} & \colhead{Definition}
 \\ \cline{4-5}
\colhead{} &  &  & \colhead{X-ray} & \colhead{Optical}
}
\startdata
1a & FRED           & F  & 14 &   & Fast rise followed by exponential decay. \nl
1b & Possible FRED  & F' & 6  & 9 & Exponential decay with no rise phase data. \nl 
2  & Triangle       & T  & 7  &   & Rise timescales $>1/3$ decay timescale. \nl  
3a & Short plateau  & Ps & 4  &   & Short ($<30$ days) flat top. \nl
3b & Long plateau   & Pl & 4  &   & Long ($>30$ days) flat top. \nl
4  & Variable decay & V  & 4  &   & Several decay stages or wiggles. \nl
5  & Multiple peak  & M  & 4  &   & Multiple peaks of similar strength. \nl
   & Uncertain      & U  & 2  & 1 & Uncertain shape.
\enddata
\end{deluxetable}

\begin{deluxetable}{r@{}lllrcrrrc}
\tablewidth{0pt}
\tablenum{6}
\tablecaption{Quiescent X-Ray Emission}
\tablehead{
\multicolumn{2}{c}{Source} & \colhead{Date} & \colhead{$\Delta$ Mon}
 & \colhead{S.I/Band}              & \colhead{$F_{\rm q,obs}$}
 & \colhead{$\log(F_{\rm q})$}     & \colhead{$\log(L_{\rm q})$}
 & \colhead{$\log({\dot M}_{\rm q})$} & \colhead{Ref.} \\
\multicolumn{2}{c}{(1)} & \colhead{(2)} & \colhead{(3)} &
 \colhead{(4)} & \colhead{(5)} & \colhead{(6)} & \colhead{(7)}
 & \colhead{(8)} & \colhead{(9)}
}
\startdata
  & 0042+32   & 1978?   & $\ge10$ & AS?/2-11 & $<0.41$  & $<-10.82$ & $<35.03$ & $<-10.72$ & 1 \nl 
J & 0422+32   & 1992/02 &  --6 & RP/0.01-2  & $<0.0088$ & $<-13.11$ & $<31.90$ & $<-13.86$ & 9 \nl
  & 0620--00  & 1992/03 & +199 & RP/0.4-1.4 & 0.0043    &  --13.63  &   30.79  &  --14.96  & 2 \nl 
  & 0748--676 & 1980/05 & --57 & EI/0.2-3.5 & 0.030     &  --11.94  &   33.04  &  --12.72  & 10 \nl 
  & 0836--429 & 1991/05 & +244 & RP/1-2.4   & 2.26      &   --9.46  &   36.82  &   --8.93  & 11 \nl
  & 1009--45  & 1991-?  & $\ge10$ & CB/20-100 & $<5$    &  $<-8.80$ & $<36.69$ &  $<-9.07$ & 18 \nl
  & 1124--683 & 1992/03 &  +14 & RP/0.3-2.4 & $<0.0039$ & $<-13.12$ & $<32.66$ & $<-13.10$ & 3 \nl 
  & 1354--64  & 1975?   & $\ge92$ & AS/2-10 & 3.4       &  --10.03  &   36.14  &   --9.61  & 1 \nl 
  & 1456--32  & 1994/02 & +177 & A/0.5-4.5  & 0.070     &  --11.89  &   32.46  &  --13.29  & 4,13 \nl 
  & 1524--62  & 1991/02 &   +6 & RP/0.1-2.4 & $<0.029$  & $<-12.24$ & $<33.40$ & $<-12.36$ & 5 \nl 
  & 1543--47  & 1990/08 & --20 & RP/0.4-2.4  & $<0.013$  & $<-12.61$ & $<32.88$ & $<-12.87$ & 12 \nl 
  & 1608--522 & 1993/08 &      & A/0.5-10   & 0.0134    &  --12.49  &   33.23  &  --12.53  & 13 \nl 
  & 1630--472 & 1993/03 &   +6 & RP/0.1-2.4 & $<0.0046$ & $<-11.11$ & $<35.55$ & $<-10.20$ & 6 \nl 
J & 1655--40  & 1996/03 &  --2 & A/2-10     & 0.0061    &  --12.82  &   32.52  &  --13.23  & 14 \nl 
  & 1705--250 & 1991/03 & +163 & RH/0.4-2.4 & $<0.0165$ & $<-12.32$ & $<32.88$ & $<-12.88$ & 7 \nl 
  & 1716--249 & 1991-94 & $>-4$ & CB/20-100 & $<5$      &  $<-9.89$ & $<35.14$ & $<-10.61$ & 15,18 \nl
  & 1730--220 & 1970-74 & $>-24$ & U/2-6    & $<4.0$    &  $<-9.94$ & $<35.19$ & $<-10.57$ & 16 \nl 
  & 1742--289 & 1974?   & $>-2$ & AR/3-10   & $<8.66$   &  $<-9.74$ & $<36.93$ &  $<-8.82$ & 1 \nl 
  & 1846--031 & 1991-95 & $>20$ & CB/20-100 & $<5$      & $<-10.57$ & $<35.44$ & $<-10.32$ & 1 \nl
  & 1908+005  & 1992/03 & ?    & RP/0.4-2.4 & 0.0372    &  --12.29  &   32.81  &  --12.95  & 7 \nl 
  & 1915+105  & 1991-92 & $>-4$ & CB/20-100 & $<5$      & $<-10.12$ & $<36.62$ &   --9.14  & 17,18 \nl 
  & 1918+146  & 1973?   & $>6$ & U/2-10     & $<5.0$    &  $<-9.76$ & $<35.48$ & $<-10.28$ & 1 \nl 
  & 2000+25   & 1992/05 &  +48 & RP/0.4-2.4 & $<0.00036$ & $<-13.88$ & $<31.47$ & $<-14.28$ & 7 \nl 
  & 2023+338  & 1992/11 &  +42 & RP/0.2-2.4 & 0.02      &  --12.49  &   33.84  &  --11.91  & 8 \nl 
\enddata
\tablecomments
{
Col.\ (2) Observation year/month; \\
Col.\ (3) Months after (+) or before ($-$) the nearest outburst peak; \\
Col.\ (4) Satellite-Instrument/Band (keV). Satellite-instrument code: 
 A--{\it ASCA}, AR--{\it Ariel 5}/RMP, AS--{\it Ariel 5}/SSI, CB--{\it 
 Compton}/BATSE, EI--{\it Einstein}/IPC, RH--{\it ROSAT}/HRI, RP--{\it 
 ROSAT}/PSPC, U--{\it UHURU}; \\
Col.\ (5) Observed quiescent flux or flux upper limit in units of mCrab in the observed band; \\
Col.\ (6) Logarithmic quiescent 0.4-10 keV flux in units of ergs s$^{-1}$ cm$^{-2}$; \\
Col.\ (7) Logarithmic quiescent 0.4-10 keV luminosity in units of ergs s$^{-1}$; \\
Col.\ (8) Logarithmic quiescent mass accretion rate in units of $M_\odot$ yr$^{-1}$.
}
\tablerefs{1.~van Paradijs 1995; 2.~McClintock, Horne, \& Remillard 1995; 
3.~Greiner et al.\ 1994; 4.~van Paradijs et al.\ 1987; 5.~Barret et al.\ 1995;
6.~Parmar, Angenili, \& White 1995; 7.~Verbunt et al.\ 1994; 8.~Wagner et al.\
1995; 9.~ Callanan et al.\ 1996; 10.~ Parmar et al.\ 1986; 11.~Belloni et al.\
1993; 12.~Greiner et al.\ 1994; 13.~Asai et al.\ 1996; 14.~Robinson et al.\
1996; 15.~Harmon \& Paciesas 1993; 16.~Cominsky et al.\ 1978; 17.~Paciesas
et al.\ 1996; 18.\ S.N.\ Zhang, 1997 (private communication).}
\end{deluxetable}

\begin{deluxetable}{r@{}l@{}c@{}l@{}l@{}c@{}c@{}c@{}c@{}c@{}rl@{}l@{}l@{}l@{}r@{}rl}
\tablenum{7}
\tablewidth{0pt}
\tablecaption{Basic Properties of X-Ray Nova Outbursts}
\tablehead{
 &  &  & \multicolumn{7}{c}{X-ray} && \multicolumn{6}{c}{Optical} 
 \\ \cline{4-10}\cline{12-17}
 &  & \colhead{Outburst} & {Mor.} & \colhead{S.I./Band} & {$F_{\rm p,obs}$}
 & {$\log(L_{\rm obs})$}
 & \colhead{$\tau_{\rm r}$} & \colhead{$\tau_{\rm d}$} & \colhead{$T_{\rm obs}$}
 && {Mor.} & \multicolumn{2}{c}{$m_{\rm v,p}$}  & {$\Delta m_{\rm v}$}
 & \colhead{$\tau_{\rm d,o}$} & {$T_{\rm obs,o}$} \\ 
\multicolumn{2}{c}{Source} & \colhead{yr/mon} & {type} & \colhead{(keV)} & {(Crab)} & {(erg/s)}
 & {(day)} & {(day)} & {(day)} && {type} & \multicolumn{2}{c}{(mag)} & {(mag)} & {(day)}
 & {(day)} & \colhead{Ref.} \\
\multicolumn{2}{c}{(1)} & \colhead{(2)} & {(3)} & \colhead{(4)}
 & \colhead{(5)} & \colhead{(6)} & \colhead{(7)} & \colhead{(8)} & \colhead{(9)}
 && {(10)} & \multicolumn{2}{c}{(11)} & {(12)} & {(13)}
 & {(14)} & \colhead{(15)}
}
\startdata
  & 0042+32   & 1970/02 &      &  B/25-300 & 0.2   & 37.75 &       &       &     &&      &   &      &     &       &     & 1     \nl
  &           & 1977/02 & M    & AS/1-20   & 0.044 & 37.65 & 1.1   & 2.8   & 58  &&      &   &      &     &       &     & 2     \nl
J & 0422+32   & 1992/08 & F    & CB/20-100 & 3     & 37.38 & 1.3   & 40.4  & 228 && F$'$ &   & 13   & 8.0 & 197.1 & 607 & 3,4   \nl
  & 0620--00  & 1917/11 &      &           &       &       &       &       &     && F$'$ & B & 12   & 6.4 &  90.5 &  94 & 5     \nl
  &           & 1975/08 & F    & AA/3-6    & 45    & 38.10 & 1.6   & 26.3  & 231 && F$'$ & B & 11.2 & 7.2 &  79.4 & 250 & 6-9   \nl
  & 0748--676 & 1985/02 & M?   & EM/0.1-20 & 0.038 & 35.92 &       & 20    & 30  &&      &   & 17.2 & $>$5.8 & 45 &  30 & 10    \nl
  & 0836--429 & 1971/01 & T    &  U/2-6    & 0.047 & 37.10 &       & 17.0  & 82  &&      &   &      &     &       &     & 11    \nl
  &           & 1990/11 & U    & GA/1-20   & 0.032 & 36.82 & 5?    &       & 70  &&      &   &      &     &       &     & 12,75 \nl
  & 1009--45  & 1993/09 & V    & GW/8-20   & 0.9   & 36.96 & 1.2   & 88.2  & 139 && U    &   & 13.8 & $>$6.4 & 58.1 & 494 & 13-16 \nl
  & 1124--683 & 1991/01 & F    & GA/1-20   & 8     & 38.73 & 0.9   & 28.3  & 261 && F$'$ &   & 13.6 & 6.9 & 48.5  &     & 59,77 \nl
  & 1354--64  & 1967/04 & F    &  R/3-8    & 24    & 39.17 & 16    & 30.5  & 44  &&      &   &      &     &       &     & 17    \nl
  &           & 1971?   &      &  O/2-6    & 0.057 & 36.91 &       &       &     &&      &   &      &     &       &     & 11,18 \nl
  &           & 1987/02 & Pl   & GA/1-20   & 0.087 & 37.67 & 30.9  & 46.4  & 199 &&      &   & 16.9 & 5.1 &       &     & 18    \nl
  & 1456--32  & 1969/07 & F$'$ &  V/3-12   & 29.6  & 38.02 & 0.4   & 36.9  & 80  &&      &   &      &     &       &     & 19,20 \nl
  &           & 1979/05 & Ps   & AA/3-6    & 4     & 36.80 & 2.0   & 4.8   & 33  && F$'$ &   & 12.8 & 5.5 & 7.8   & 35  & 21-23 \nl
  & 1524--62  & 1974/11 & F    & AA/3-6    & 0.83  & 37.30 & 2.9   & 57.4  & 148 &&      &   & 17.5 & $>$3.5 & 117 &    & 60    \nl
  &           & 1990/08 &      & RP/0.1-2.4 & 0.08 & 37.40 &       &       &     &&      &   &      &     &       &     & 24,61 \nl
  & 1543--47  & 1971/08 & F    &  U/2-6    & 2.1   & 37.79 & 1.2   & 42.7  & 507 &&      &   &      &     &       &     & 25,26 \nl
  &           & 1983/08 & F$'$ &  T/1.5-3.8 & 10   & 38.44 & 0.6   & $<$17.9 &185 &&     &   & 14.9 & 1.8 &       &     & 27,62 \nl
  &           & 1992/04 & T    & CB/20-300 & 1.04  & 37.85 & 0.7   & 1.6   & 9   &&      &   &      &     &       &     & 28    \nl
  & 1608--522 & 1970/04 & M?   & VB/3-12   & 0.93  & 37.43 &       &       &     &&      &   &      &     &       &     & 29    \nl
  &           & 1970/09 & T    & VB/3-12   & 0.78  & 37.35 & 22.3  & 15.2  & 78  &&      &   &      &     &       &     & 29    \nl
  &           & 1971/06 & F    & VB/3-12   & 0.71  & 37.31 & 1.4   & 13.1  & 42  &&      &   &      &     &       &     & 29    \nl
  &           & 1971/09 & F    & VB/3-12   & 0.80  & 37.36 & 9.6   & 17.0  & 53  &&      &   &      &     &       &     & 29    \nl
  &           & 1975/11 &      & VB/3-12   & 0.21  & 36.78 &       &       &     &&      &   &      &     &       &     & 29-31 \nl
  &           & 1977/07 & Pl   &  S/1-10   & 1.4   & 37.61 & 2.6   & 24.5  & 198 &&      &   & 18.2 & $>$1.8 &    &     & 29    \nl
  &           & 1979/02 & T    & AA/3-6    & 0.6   & 37.24 & 8.1   & 20.4  & 37  &&      &   &      &     &       &     & 29    \nl
  &           & 1979/04 & Ps   & AA/3-6    & 0.5   & 37.16 & 9.8   & 13.4  & 43  &&      &   &      &     &       &     & 29    \nl
  &           & 1983/04 &      &  T/2-20   & 0.53  & 37.35 &       &       &     &&      &   &      &     &       &     & 65 \nl
  &           & 1991/04 &      & CB/20-100 & 0.14  & 36.61 &       &       &     &&      &   &      &     &       &     & 76 \nl
  & 1630--47  & 1971/02 & T    &  U/2-6    & 0.15  & 38.06 & 62.5? & 41.8  & 930 &&      &   &      &     &       &     & 32,64 \nl
  &           & 1972/10 & V?   &  U/2-6    & 0.084 & 37.81 & 1.3   & $<$44.6 & 119 &&    &   &      &     &       &     & 32    \nl
  &           & 1974/04 & F    & AS/2-6    & 0.23  & 38.25 & $<$11 & 95.5  & 250 &&      &   &      &     &       &     & 32    \nl
  &           & 1976/06 & F$'$ & AS/2-6    & 0.29  & 38.35 & $<$14.9 & 130 & 239 &&      &   &      &     &       &     & 32    \nl
  &           & 1977/11 & F$'$ & AA/3-6    & 1.4   & 38.56 & 9.4   & 24.8  & 117 &&      &   &      &     &       &     & 33,34 \nl
  &           & 1979/03 &      &  E/1-50   & 0.0094 & 37.48 &      &       &     &&      &   &      &     &       &     & 66    \nl
  &           & 1984/04 & F$'$ & EM/1-50   & 0.49  & 38.45 &       & 19.4  & 114 &&      &   & 15.3 &     &       &     & 35,36 \nl
  &           & 1987/10 &      & GW/1-10   & 0.0004 & 36.24 &      &       &     &&      &   &      &     &       &     & 67 \nl
  &           & 1989/03 &      & MT/2-28   & 0.16  & 38.14 &       &       &     &&      &   &      &     &       &     & 67,68 \nl
  &           & 1992/09 &      & RP/2-10   & 0.0044 & 36.62 &      &       &     &&      &   &      &     &       &     & 66    \tablebreak
J & 1655--40  & 1994/08 & M    & CB/1-200  & 1.7   & 38.12 & 2.9   & 5.1   & 166 &&      &   & 14.2 & $>$3.8 &    &     & 37,38 \nl
  &           & 1995/08 &      & CB/1-200  & 1.28  & 37.99 &       &       &     &&      &   &      &     &       &     & 78 \nl
  &           & 1996/07 &      & XA/2-11   & 3.5   & 38.04 &       &       &     &&      &   &      &     &       &     & 79,80 \nl
  & 1705--250 & 1977/08 & Ps   & H1/2-200  & 2.2   & 38.96 & 0.5   & 334.8 & 37  &&      &   & 15.9 & 5.4 &       &     & 39-41 \nl
  & 1716--249 & 1993/10 & Pl   & GW/0.1-100 & 5.5  & 38.72 & 0.8   & 387.4 & 84  &&      &   & 16.6 & $>$4.4 &    &     & 42,69 \nl
  & 1730--220 & 1972/08 & V    &  U/2-6    & 0.12  & 36.35 & $<$18.0 & 82.4 & 230 &&     &   &      &     &       &     & 63    \nl
  & 1742--289 & 1975/02 & V    & AA/3-10   & 1.9   & 38.85 & 3.6   & 99.6  & 213 &&      &   &      &     &       &     & 43,44 \nl
  & 1846--031 & 1985/04 & F    & EM/1-25   & 0.22  & 38.21 &       & 80.4  & 157 &&      &   &      &     &       &     & 45    \nl
  & 1908+005  & 1969/12 &      & VB/3-6    & 0.81  & 36.80 &       &       &     &&      &   &      &     &       &     & 70    \nl
  &           & 1970/08 &      & VB/3-6    & 0.41  & 36.50 &       &       &     &&      &   &      &     &       &     & 70    \nl
  &           & 1971/09 & T?   &  U/3-6    & 0.85  & 36.82 & 14.2? & 17.2  & 279 &&      &   &      &     &       &     & 46,70 \nl
  &           & 1972/04 &      & VB/3-6    & 0.49  & 36.58 &       &       &     &&      &   &      &     &       &     & 70    \nl
  &           & 1973/01 & T?   &  U/3-6    & 0.82  & 36.81 & 52.4? & 5.6   & 54  &&      &   &      &     &       &     & 46,70 \nl
  &           & 1974/04 &      & VB/3-6    & 1.67  & 37.11 &       &       &     &&      &   &      &     &       &     & 70    \nl
  &           & 1975/06 & F    & AA/3-6    & 0.99  & 36.89 & 4.2   & 23.6  & 45  &&      &   & 16.5 & 2.7 &       &     & 46,47 \nl
  &           & 1976/06 & Ps   & AA/3-6    & 0.76  & 36.77 & 1.6   & 29.7  & 54  &&      &   & 16.4 & 2.8 &       &     & 46,47 \nl
  &           & 1978/06 & F    & AS/3-6    & 1.1   & 36.94 & 1.5   & 22.3  & 66  && F$'$ &   & 15.5 & 3.7 & 33.2  & 40  & 48    \nl
  &           & 1979/03 &      &  E/3-6    & 0.5   & 36.59 &       &       &     &&      &   &      &     &       &     & 71,72 \nl
  &           & 1980/05 &      &  T/1-22   & 0.04  & 36.19 &       &       &     &&      &   &      &     &       &     & 73    \nl
  &           & 1987/03 &      & GA/1-6    & 1.6   & 37.55 &       &       &     &&      &   &      &     &       &     & 74    \nl
  &           & 1988/10 &      & GA/1-6    & 0.65  & 37.16 &       &       &     &&      &   &      &     &       &     & 74    \nl
  &           & 1989/09 &      & GA/1-6    & 0.33  & 36.86 &       &       &     &&      &   &      &     &       &     & 74    \tablebreak
  & 1915+105  & 1992/08 & Pl   & CB/20-100 & 0.3   & 37.80 & 19.5  & 2370  & 542 &&      &   &      &     &       &     & 28,49 \nl
  & 1918+146  & 1972/07 & F$'$ &  U/2-6    & 0.05  & 35.94 & 3.5   & 8.4   & 28  &&      &   &      &     &       &     & 63    \nl
  & 2000+25   & 1988/05 & F    & GA/3-6    & 12    & 37.99 & 0.3   & 30.1  & 169 && F$'$ &   & 18.1 & 3.1 & 102.5 & 212 & 50-52 \nl
  & 2023+338  & 1938/10 &      &           &       &       &       &       &     && F$'$ &   & 12.5 & 6.5 & 25.2  & 65  & 53   \nl
  &           & 1956/08 &      &           &       &       &       &       &     &&      &   & 14.1 &     &       &     & 54    \nl
  &           & 1989/05 & F    & GL/1.7-37 & 21    & 39.31 & 5.5   & 38.8  & 178 && F$'$ &   & 17.9 & 1.1 & 7.3   & 36  & 55-58 \nl
\enddata
\tablecomments
{Col.\ (3) and (9) The light curve morphology types. \newline
Col.\ (4) Observing satellite/instrument code: AA: {\it Ariel~5}/ASM; AS: {\it Ariel~5}/SSE; 
 B: Balloon experiments; C: {\it Copernicus}; CB: {\it Compton}/BATSE; E: {\it Einstein};
 EM: {\it EXOSAT}/ME; GA: {\it Ginga}/ASM; GL: {\it Ginga}/LAC; GS: {\it Granat}/Sigma;
 GW: {\it Granat}/Watch; H: {\it Hakucho}; H1: {\it HEAO-1}; MH: {\it Mir-Kvant}/HEXE;
 MT: {\it Mir-Kvant}/TTM; O: {\it OSO-7}; R: Rocket experiments; RP: {\it Rosat}/PSPC;
 S: {\it SAS-3}; T: {\it Tenma}; VA: {\it Vela~5A}; VB: {\it Vela~5B}; U: {\it UHURU}; XA: {\it RXTE}/ASM.}
\tablerefs{(1) Laros \& Wheaton 1980; (2) Watson \& Ricketts 1978; (3) Paciesas et al.\ 1995; (4) 
Callanan et al.\ 1995; (5) Eachus, Wright, \& Liller 1976; (6) Elvis et al.\ 1975; (7) Kaluzienski 
et al.\ 1977; (8) Tsunemi et al.\ 1977; (9) Lloyd, Noble, \& Penston 1977; (10) Parmar et al.\ 1986; 
(11) Markert et al.\ 1977; (12) Aoki et al.\ 1992; (13) Paciesas et al.\ 1995; (14) Bailyn \& Orosz 
1995; (15) Kaniovsky et al.\ 1993; (16) Della Valle et al.\ 1997; (17) Chodil et al.\ 1968; 
(18) Kitamoto et al.\ 1990; (19) Conner, Evans, \& Belian 1969; (20) Evans, Belian, \& Conner 1970; 
(21) Kaluzienski et al.\ 1980; (22) Canizares et al.\ 1980; (23) Matsuoka et al.\ 1980; (24) 
Barret et al.\ 1992; (25) Matilsky et al.\ 1972; (26) Li, Sprott, \& Clark 1976; (27) Kitamoto et 
al.\ 1984; (28) Harmon et al.\ 1994; (29) Lochner \& Roussel-Dupr\'e 1994; (30) Kaluzienski \& Holt
1975; (31) Kaluzienski 1977; (32) Jones et al.\ 1976; (33) Kaluzienski \& Holt 1978; 
(34) Share et al., 1978; (35) Parmar, Stella, \& White 1986; (36) Tanaka 1984; (37) Harmon et al.\ 
1995; (38) Bailyn et al.\ 1995; (39) Watson, Ricketts, \& Griffiths 1978; (40) Wilson \& Rothschild 
1983; (41) Griffiths et al.\ 1978; (42) Paciesas et al.\ 1995; (43) Eyles et al.\ 1975; (44) 
Branduardi et al.\ 1976; (45) Parmar et al.\ 1993; (46) Kaluzienski et al.\ 1977; (47) Thorstensen, 
Charles, \& Bowyer 1978; (48) Charles et al.\ 1980; (49) Paciesas et al.\ 1996; (50) Tsunemi et al.\ 
1989; (51) Tanaka, Makino, \& Dotani 1991; (52) Chevalier \& Ilovaisky 1990; (53) Casares et al.\ 
1991; (54) Richter 1987; (55) Tanaka 1989; (56) Kitamoto et al.\ 1989; (57) Wagner et al.\ 1991; (58)
Terada et al.\ 1994; (59) Ebisawa et al.\ 1994; (60) Kaluzienski et al.\ 1975; (61) Barret et al.\
1995; (62) van der Woerd \& White 1989; (63) Cominsky et al.\ 1978; (64) Forman, Jones, \& Tananbaum 1976;
(65) Mitsuda et al.\ 1989; (66) Parmar, Angelini, \& White 1995; (67) Parmar et al.\ 1997; (68) in 't
Zand 1992; (69) Della Valle et al.\ 1994; (70) Priedhorsky \& Terrell 1984; (71) Czerny, Czerny, \&
Grindlay 1987; (72) Holt \& Kaluzienski 1979; (73) Koyama et al.\ 1981; (74) Kitamoto et al.\ 1993;
(75) Belloni et al.\ 1993; (76) Zhang et al.\ 1996; (77) Della Valle, Jarvis, \& West 1991; (78) Zhang
et al.\ 1997; (79) Remillard 1997; (80) Cui 1997 (priv. comm.).
}
\end{deluxetable}

\begin{deluxetable}{r@{}lccccc}
\tablenum{8}
\tablewidth{0pt}
\tablecaption{Energy Output of XN Outbursts}
\tablehead{
\multicolumn{2}{c}{Source} & \colhead{Outburst} & {log($F_{\rm p}$)}
 & {log($L_{\rm p}$)} & {$\log(\frac{L_{\rm p}}{L_{\rm q}})$}
 & {$\log(\frac{L_{\rm p}}{L_{\rm Edd}})$} \\
\multicolumn{2}{c}{(1)} & \colhead{(2)} & \colhead{(3)} & \colhead{(4)}
 & \colhead{(5)} & \colhead{(6)}  
}
\startdata
  & 0042+32   & 1970/02 & --9.20 & 36.60 & $>1.57$ & --2.51 \nl
  &           & 1977/02 & --8.81 & 36.99 & $>1.96$ & --2.13 \nl
J & 0422+32   & 1992/08 & --7.46 & 37.29 & $>5.39$ & --1.37 \nl
  & 0620--00  & 1975/08 & --6.00 & 38.09 &  7.30   & --0.87 \nl
  & 0748--676  & 1985/02 & --8.85 & 35.93 &  2.89   & --2.33 \nl
  & 0836--429 & 1971/01 & --8.80 & 37.49 &  0.67   & --1.62 \nl
  &           & 1990/11 & --9.32 & 36.97 &  0.15   & --2.15 \nl
  & 1009--45  & 1993/09 & --7.73 & 37.55 & $>0.86$ & --1.57 \nl
  & 1124--683 & 1991/01 & --6.77 & 38.86 & $>6.21$ & --0.08 \nl
  & 1354--64  & 1967/04 & --6.54 & 39.58 &  3.44   &   0.46 \nl
  &           & 1971?   & --8.97 & 37.43 &  1.29   & --1.69 \nl
  &           & 1987/02 & --8.54 & 37.85 &  1.71   & --1.26 \nl
  & 1456--32  & 1969/07 & --5.87 & 38.40 &  5.94   &   0.14 \nl
  &           & 1979/05 & --7.11 & 37.29 &  4.83   & --0.97 \nl
  & 1524--62  & 1974/11 & --8.53 & 37.17 & $>3.77$ & --1.95 \nl
  &           & 1990/08 & --8.60 & 37.10 & $>3.70$ & --2.02 \nl
  & 1543--47  & 1971/08 & --7.24 & 38.28 & $>5.40$ & --0.83 \nl
  &           & 1983/08 & --6.65 & 38.85 & $>5.97$ & --0.26 \nl
  &           & 1992/04 & --7.54 & 37.93 & $>5.05$ & --1.19 \nl
  & 1608--522 & 1970/04 & --7.65 & 37.75 &  4.52   & --0.51 \nl
  &           & 1970/09 & --7.73 & 37.67 &  4.44   & --0.59 \nl
  &           & 1971/06 & --7.76 & 37.63 &  4.40   & --0.63 \nl
  &           & 1971/09 & --7.71 & 37.68 &  4.45   & --0.58 \nl
  &           & 1975/11 & --7.97 & 37.43 &  4.20   & --0.83 \nl
  &           & 1977/07 & --7.47 & 37.92 &  4.69   & --0.34 \nl
  &           & 1979/02 & --7.83 & 37.56 &  4.33   & --0.70 \nl
  &           & 1979/04 & --8.03 & 37.36 &  4.13   & --0.90 \nl
  &           & 1983/04 & --7.81 & 37.73 &  4.50   & --0.53 \nl
  &           & 1991/04 & --9.11 & 36.10 &  2.87   & --2.16 \nl
  & 1630--47  & 1971/02 & --8.41 & 38.46 & $>2.91$ & --0.65 \nl
  &           & 1972/10 & --8.66 & 38.21 & $>2.66$ & --0.90 \nl
  &           & 1974/04 & --8.21 & 38.66 & $>3.11$ & --0.45 \nl
  &           & 1976/06 & --8.12 & 38.75 & $>3.20$ & --0.36 \nl
  &           & 1977/11 & --7.64 & 39.23 & $>3.68$ &   0.11 \nl
  &           & 1979/03 & --9.39 & 37.92 & $>2.37$ & --1.19 \nl
  &           & 1984/04 & --8.51 & 38.60 & $>3.05$ & --0.52 \nl
  &           & 1989/03 & --8.32 & 38.39 & $>2.84$ & --0.72 \nl
  &           & 1992/09 & --9.91 & 37.20 & $>1.65$ & --1.92 \nl
J & 1655--40  & 1994/08 & --7.35 & 37.86 &  5.34   & --1.10 \nl
  &           & 1995/08 & --7.45 & 37.77 &  5.25   & --1.19 \nl
  &           & 1996/07 & --7.16 & 38.02 &  5.50   & --0.94 \tablebreak
  & 1705--250 & 1977/08 & --7.24 & 38.04 & $>5.16$ & --0.76 \nl
  & 1716--249 & 1993/10 & --6.72 & 38.39 & $>3.25$ & --0.73 \nl
  & 1730--220 & 1972/08 & --8.44 & 36.68 & $>1.49$ & --1.58 \nl
  & 1742--289 & 1975/02 & --7.39 & 39.28 & $>2.35$ &   0.16 \nl
  & 1846--031 & 1985/04 & --8.07 & 38.06 & $>2.62$ & --1.06 \nl 
  & 1908+005  & 1969/12 & --7.64 & 37.29 &  4.48   & --0.97 \nl
  &           & 1970/08 & --7.94 & 36.99 &  4.18   & --1.27 \nl
  &           & 1971/09 & --7.62 & 37.31 &  4.50   & --0.95 \nl
  &           & 1972/04 & --7.86 & 37.07 &  4.26   & --1.19 \nl
  &           & 1973/01 & --7.64 & 37.30 &  4.49   & --0.96 \nl
  &           & 1974/04 & --7.33 & 37.60 &  4.79   & --0.66 \nl
  &           & 1975/06 & --7.56 & 37.37 &  4.56   & --0.89 \nl
  &           & 1976/06 & --7.67 & 37.26 &  4.45   & --1.00 \nl
  &           & 1978/06 & --7.50 & 37.42 &  4.61   & --0.84 \nl
  &           & 1979/03 & --7.85 & 37.08 &  4.27   & --1.18 \nl
  &           & 1980/05 & --8.78 & 36.15 &  3.34   & --2.11 \nl
  &           & 1987/03 & --7.20 & 37.74 &  4.93   & --0.52 \nl
  &           & 1988/10 & --7.59 & 37.35 &  4.54   & --0.91 \nl
  &           & 1989/09 & --7.89 & 37.05 &  4.24   & --1.21 \nl
  & 1915+105  & 1992/08 & --7.47 & 39.00 & $>2.38$ & --0.42 \nl
  & 1918+146  & 1972/07 & --8.76 & 36.48 & $>1.00$ & --2.64 \nl
  & 2000+25   & 1988/05 & --6.71 & 38.60 & $>7.13$ & --0.45 \nl
  & 2023+338  & 1989/10 & --6.24 & 39.14 &  5.30   & --0.06 \nl                 
\enddata
\tablecomments{
Col.\ (3) Calibrated peak flux in 0.4--10 keV band in units of ergs s$^{-1}$
 cm$^{-2}$. \\
Col.\ (4) Logarithmic luminosity in 0.4-10 keV band. For unknown distance, we
 assume 3 kpc. \\
Col.\ (5) Logarithmic outburst amplitude (or lower limit) in 0.4--10 keV band
 as the ratio of the peak luminosity to the quiescent luminosity (or upper
 limit). \\
Col.\ (6) Logarithmic peak 0.4-10 keV luminosity in Eddington units. For
 unknown masses, we assume $5 \Msun$ for BHCs and $1.4 \Msun$ for NS.
}
\end{deluxetable}

\begin{deluxetable}{r@{}llrrrrcc}
\tablenum{9}
\tablewidth{0pt}
\tablecaption{Duration and Total Energy }
\tablehead{
\multicolumn{2}{c}{Source} & \colhead{Year} &
 \colhead{$T_{\rm obs}$} &
 \colhead{$T_{\rm exp}$} &
 \colhead{$T_{\rm r,exp}$} &
 \colhead{$T_{\rm d,exp}$} &
 \colhead{$\log(E)$} &
 \colhead{$\log(\Delta M)$} \\
\colhead{} & \colhead{} & \colhead{} & \colhead{(day)} & \colhead{(day)} & 
\colhead{(day)} & \colhead{(day)} & \colhead{(ergs)} & \colhead{($M_\odot$)}
}
\startdata
  & 0042+32  & 77/02\tnm{*} &  58 & $>$17 & $>$4 & $>$12 & 42.51 & --10.74\nl
J & 0422+32  & 92/08 & 228   & $>$517 & $>$16  & $>$501 & 43.85 &  --9.40 \nl
  & 0620--00 & 75/08 & 231   &    468 &    26  &    442 & 44.47 &  --8.78 \nl
  & 1009--45 & 93/09 & 139   & $>$177 &  $>$2  & $>$174 & 44.37 &  --8.88 \nl
  & 1124-683 & 91/01 & 261   & $>$417 & $>$12  & $>$404 & 45.26 &  --7.99 \nl
  & 1354-64  & 67/04 & 44    &    292 &   126  &    165 & 46.08 &  --7.17 \nl
  &          & 87/02 & 199   &    304 &   121  &    182 & 44.67 &  --8.58 \nl
  & 1456--32 & 69/07 & 80    &    510 &     5  &    504 & 44.91 &  --8.35 \nl
  &          & 79/05 & 33    &    75 &     22  &     53 & 43.06 & --10.19 \nl
  & 1524--62 & 74/11 & 148   & $>$523 & $>$25  & $>$498 & 43.88 &  --9.37 \nl
  & 1543--47 & 71/08 & 507   & $>$545 & $>$14  & $>$530 & 44.86 &  --8.39 \nl
  &          & 83/08 & 185   & $>$254 &   $>$8 & $>$246 & 45.06 &  --8.19 \nl
  &          & 92/04 & 9     &  $>$26 &  $>$8  &  $>$18 & 43.23 & --10.03 \nl
  & 1608--52 & 70/09 &  78   &    383 &    227 &    155 & 44.18 &  --9.07 \nl
  &          & 71/06 &    42 &    146 &     14 &    132 & 43.73 &  --9.52 \nl
  &          & 71/09 &  53   &    272 &     98 &    174 & 44.04 &  --9.21 \nl
  &          & 77/07 &  198  &    292 &     28 &    264 & 44.29 &  --8.96 \nl
  &          & 79/02 &  37   &    284 &     80 &    203 & 43.95 & --9.30 \nl
  &          & 79/04 &  43   &    220 &     93 &    127 & 43.66 &  --9.59 \nl
  & 1630--47 & 71/02 &  93   & $>$698 & $>$418 & $>$280 & 45.42 &  --7.83 \nl
  &          & 72/10 &  119  & $>$281 &   $>$7 & $>$273 & 44.81 &  --8.44 \nl
  &          & 74/04 &       & $>$762 &  $>$78 & $>$683 & 45.62 &  --7.63 \nl
  &          & 76/06\tnm{\dag} & 239 & $>$1067 & $>$109 & $>$957 & 45.85 & --7.40 \nl
  &          & 77/11 &  117  & $>$289 &  $>$79 & $>$210 & 45.70 &  --7.56 \nl
J & 1655-40  & 94/08\tnm{*}& 166 & 98 &     35 &     62 & 43.70 &  --9.56 \nl
  & 1705-25  & 77/08\tnm{\dag} & 37 & $>$3983 & $>$5 & $>$3977 & 44.48 & --8.78 \nl
  & 1716--25 & 93/10\tnm{\dag} &       & $>$2905 & $>$5 & $>$2899 & 45.18 & --8.07 \nl
  & 1730-22  & 72/08 &  230  & $>$344 &  $>$61 & $>$282 & 43.60 &  --9.65 \nl
  & 1742-29  & 75/02 &  213  & $>$715 &  $>$24 & $>$690 & 46.23 &  --7.03 \nl
  & 1908+00  & 71/09 &  279  &    325 &    147 &    178 & 43.75 &  --9.51 \nl
  &          & 75/06 &  45   &    291 &     44 &    247 & 43.76 &  --9.50 \nl
  &          & 76/06 &  53   &    320 &     16 &    304 & 43.69 &  --9.56 \nl
  &          & 78/06 &   66  &    252 &     15 &    236 & 43.74 &  --9.51 \nl
  & 1915+10  & 92/08\tnm{\dag} & 542 & $>$13094 & $>$106 & $>$12987 & 46.45 & --6.80 \nl
  & 1918+15  & 72/07 &       & $>$27 &   $>$8  &  $>$19 & 42.44 & --10.81 \nl
  & 2000+25  & 88/05 &  163  & $>$499 &   $>$4 & $>$494 & 45.02 &  --8.23 \nl
  & 2023+34  & 89/05 &  178  &    590 &     73 &    517 & 45.73 &  --7.53 \nl
\enddata
\tablenotetext{*}{Observed light curve contains multiple peaks, the expected
duration is for only one peak.}
\tablenotetext{\dag}{Long plateau outburst with cutoff, the expected duration
is for an extended plateau phase. The energy fluence is calculated using the
observed plateau duration instead of the decay timescale.}
\end{deluxetable}

\begin{deluxetable}{r@{}llccc}
\tablenum{10}
\tablewidth{0pt}
\tablecaption{Outbursts with Plateau}
\tablehead{
\multicolumn{2}{c}{Source} & \colhead{Year} &
 \colhead{$\tau_{\rm plt}$} &
 \colhead{$T_{\rm plt}$} &
 \colhead{$\tau_{\rm tail}$} \\
 &  &  & \colhead{(day)} & \colhead{(day)} & \colhead{(day)} 
}
\startdata
  &             &      & BHs &     &     \nl \hline
J & 0422+32     & 1992 & 70  & 15  & 42  \nl
  & 1354$-$64   & 1987 & 360\tnm{*} & 57 & 44 \nl
  & 1543$-$47   & 1971 & 540 & 52  & 47  \nl
  & 1705$-$250  & 1977 & 335\tnm{*} & 31 & ?  \nl
  & 1716$-$249  & 1993 & 387 & 72  & 1.4 \nl
  & 1915+105    & 1992 & 2370\tnm{*} & 310 & 20 \nl
  &             & Mean & 677 & 90  & 31  \nl \hline
  &             &      & NSs &     &     \nl \hline
  & 1456$-$32   & 1979 & 78  & 9   & 3.6 \nl
  & 1608$-$52   & 1977 & 860\tnm{\dagger} & 51 & 37 \nl
  &             & 1979 & 250\tnm{\dagger} & 18 & 12 \nl
  & 1908+005    & 1975 & 71\tnm{*}  & 20 & 14 \nl
  &             & 1976 & 150\tnm{*} & 16 & 21 \nl
  &             & 1978 & 110\tnm{*} & 15 & 28 \nl
  &             & Mean & 258 & 22  & 19 \nl
\enddata
\tablenotetext{*}{Flux is highly variable, $\tau_{\rm plt}$ is for the overall shape.}
\tablenotetext{\dagger}{This value is highly uncertain due to sparse data.}
\end{deluxetable}

\begin{deluxetable}{lccccr}
\tablenum{11}
\tablewidth{0pt}
\tablecaption{X-ray Nova related X-ray Missions\tablenotemark{*}
}
\tablehead{
\colhead{Observatory} & \colhead{Start} & \colhead{Stop} &
\colhead{Coverage} & \colhead{FOV} & \colhead{Energy} \\
\colhead{} & \colhead{Year} & \colhead{Year} & \colhead{factor} &
\colhead{} & \colhead{Band} 
}
\startdata
Sounding Rockets & 1960's & 1960's & $\ll1$ & vary         & 3$-$12\phn   \nl
Vela         & 1969.4 & 1979.6 & 0.80 & $4\pi$             & 3$-$12\phn   \nl
Ariel-5      & 1974.9 & 1980.3 & 0.90 & $4\pi$             & 3$-$6\pnn    \nl
SAS-3        & 1975.4 & 1979.5 & 0.60 & $3\x(1\dg\x32\dg)$ & 1.5$-$60\phn \nl
Tenma        & 1983.5 & 1984.5 & 0.90 & 45\dg              & 1$-$60\phn   \nl
Hakucho      & 1979.2 & 1984.2 & 0.45 & $50\dg\x360\dg$    & 1.5$-$30\phn \nl
Ginga        & 1987.1 & 1991.8 & 0.90 & $4\pi$             & 1$-$20\phn   \nl
GRANAT/Watch & 1989.9 & 1992.8 & 0.80 & $3\dg\x60\dg$      & 6$-$180      \nl
CGRO/BATSE   & 1991.3 & 1995.9 & 0.90 & $4\pi$             & 20$-$300     \nl
Eureka/Watch & 1992.8 & 1993.5 & 0.10 & 60\dg              & 6$-$180      \nl
 & & & & \nl
OSO 3/Copernicus & 1967.2 & 1982.3 & 0.05 & 40\dg            & 8$-$210    \nl
OSO 4            & 1969.1 & 1975.6 & 0.30 & $40\dg\x360\dg$  & 14$-$200   \nl
OSO 5            & 1972.2 & 1980.8 & 0.40 & $2\x(23\dg\x360\dg)$ & 8$-$200 \nl
EXOSAT           & 1983.5 & 1986.3 & 0.05 & $0.8\dg\x0.8\dg$  & 1$-$20\phn \nl
HEAO-1           & 1977.5 & 1989.5 & 0.08 & $8\dg\x360\dg$    & 0.2$-$60\phn \nl
Uhuru            & 1970.9 & 1973.3 & 0.15 & $12.7\dg\x360\dg$ & 2$-$20\phn \nl
\enddata
\tablenotetext{*}{All of this information was retrieved from the online database
 at HEARSAC.}
\end{deluxetable}

\begin{deluxetable}{ccccc}
\tablenum{12}
\tablewidth{0pt}
\tablecaption{Mass Transfer Rate from the Companion
}
\tablehead{
\colhead{Source} & \colhead{yr/mon} & {$T_{\rm rec}$}
 & {$\log(\dot M_{\rm c})$} & {$\log(\dot M_{\rm c,K})$\tablenotemark{*}} \\
 & & \colhead{(yr)} & {($M_\odot$/yr)} & {($M_\odot$/yr)}
}
\startdata
0042+32  & 77/02 &  7.00 & --11.59 & \nl
0620--00 & 75/08 & 57.75 & --10.54 & --10.08 \nl
1354--64 & 71/12 &  4.67 &  --9.25 & \nl
1456--32 & 79/05 &  9.83 & --11.18 & --9.81 \nl
1543--47 & 83/08 & 12.00 &  --9.27 & \nl
         & 92/04 &  8.67 & --10.97 & \nl
1608--52 & 70/09 &  0.42 &  --8.69 & \nl
         & 71/06 &  0.75 &  --9.40 & \nl
         & 71/09 &  0.25 &  --8.61 & \nl
         & 77/07 &  1.67 &  --9.18 & \nl
         & 79/02 &  1.58 &  --9.50 & \nl
1630--47 & 72/10 &  1.67 &  --8.66 & \nl
         & 74/04 &  1.50 &  --7.81 & \nl
         & 76/06 &  2.17 &  --7.74 & \nl
         & 77/11 &  1.42 &  --7.71 & \nl
1908+00  & 71/09 &  1.08 &  --9.54 & --9.56 \nl
         & 75/06 &  1.17 &  --9.57 & \nl
         & 76/06 &  1.00 &  --9.56 & \nl
         & 78/06 &  2.00 &  --9.81 & \nl
2023+34  & 89/05 & 32.75 &  --9.05 & --8.71 \nl
\enddata
\tablenotetext{*}{Theoretical mass transfer rate from the companion according to
King et al.\ (1996).}
\end{deluxetable}

\end{document}